\documentstyle[12pt]{article}
\input epsf.sty

\catcode`@=11
\def\chkspace{%
  \relax   % Calm down any expanding \if's.
  \begingroup\ifhmode\aftergroup\dochksp@ce\fi\endgroup}
\def\dochksp@ce{%
  \unskip              % Remove any preceding horizontal glue
  \futurelet\chkspct@k\d@chkspc  % Grab the next token and look ahead
}
\def\d@chkspc{%
  \let\nxtsp@ce=\relax
  \ifx\chkspct@k.\else     % Don't put spaces before punctuation ...
    \ifx\chkspct@k,\else
      \ifx\chkspct@k;\else
        \ifx\chkspct@k!\else
          \ifx\chkspct@k?\else
            \ifx\chkspct@k:\else
              \ifx\chkspct@k)\else
              \ifx\chkspct@k(\else
                \ifx\chkspct@k]\else
                  \ifx\chkspct@k-\else
                    \ifx\chkspct@k\egroup\else  % or close groups.
                      \let\nxtsp@ce=\put@space  % Put a space everywhere else.
                    \fi
                  \fi
                \fi
              \fi
              \fi
            \fi
          \fi
        \fi
      \fi
    \fi
  \fi
  \nxtsp@ce
}
\def\put@space{$\;$}
\catcode`@=12

\def\Pade{Pad$\acute{\rm e}$\chkspace}

\def\ra{{$\rightarrow$}\chkspace}
\def\etal{{\it et al.}\chkspace}

\def\adhoc{{\it ad hoc}\chkspace}
\def\ie{{\it i.e.}\chkspace}

\def\eg{{\it eg.}\chkspace}
\def\etc{{\it etc.}\chkspace}

\def\apriori{{\it a priori}\chkspace}

\def\ep{{e$^+$e$^-$}\chkspace}
\def\epa{{e$^+$e$^-$ annihilation}\chkspace}

\def\qu{\quad}

\def\gluino{\relax\ifmmode \tilde{g} \else $\tilde{g}$ \fi\chkspace}

\def\qq{q$\overline{\rm q}$\chkspace}

\def\qbar{$\overline{\rm q}$\chkspace}

\def\pbar{$\overline{\rm p}$\chkspace}

\def\bb{\relax\ifmmode {\rm b}\bar{\rm b}
       \else ${\rm b}\bar{\rm b}$ \fi\chkspace}
\def\cc{\relax\ifmmode {\rm c}\bar{\rm c}
       \else ${\rm c}\bar{\rm c}$ \fi\chkspace}
\def\tt{\relax\ifmmode {\rm t}\bar{\rm t}
       \else ${\rm t}\bar{\rm t}$ \fi\chkspace}
\def\ss{\relax\ifmmode {\rm s}\bar{\rm s}
       \else ${\rm s}\bar{\rm s}$ \fi\chkspace}

\def\qqg{\relax\ifmmode {\rm q}\overline{\rm q}{\rm g}
\else q$\overline{\rm q}$g \fi\chkspace}

\def\ttg{{t$\overline{\rm t}$g}\chkspace}

\def\afb{\relax\ifmmode A_{FB} \else
{{$A_{FB}$}}\fi\chkspace}
\def\afbb{\relax\ifmmode A_{FB}^b \else
{{$A_{FB}^b$}}\fi\chkspace}
\def\pafb{\relax\ifmmode \tilde{A}_{FB} \else
{{$\tilde{A}_{FB}$}}\fi\chkspace}
\def\pafbb{\relax\ifmmode \tilde{A}_{FB}^b \else
{{$\tilde{A}_{FB}^b$}}\fi\chkspace}

\def\pafbzo{\relax\ifmmode \tilde{A}_{FB}|_{O(0)} \else
{{$\tilde{A}_{FB}|_{O(0)}$}}\fi\chkspace}
\def\pafbfo{\relax\ifmmode \tilde{A}_{FB}|_{\oalp} \else
{{$\tilde{A}_{FB}|_{\oalp}$}}\fi\chkspace}
\def\pafbso{\relax\ifmmode \tilde{A}_{FB}|_{\oalpsq} \else
{{$\tilde{A}_{FB}|_{\oalpsq}$}}\fi\chkspace}
\def\pafbto{\relax\ifmmode \tilde{A}_{FB}|_{\oalpc} \else
{{$\tilde{A}_{FB}|_{\oalpc}$}}\fi\chkspace}

\def\pafbbzo{\relax\ifmmode \tilde{A}_{FB}^b|_{O(0)} \else
{{$\tilde{A}_{FB}^b|_{O(0)}$}}\fi\chkspace}
\def\pafbbfo{\relax\ifmmode \tilde{A}_{FB}^b|_{\oalp} \else
{{$\tilde{A}_{FB}^b|_{\oalp}$}}\fi\chkspace}
\def\pafbbso{\relax\ifmmode \tilde{A}_{FB}^b|_{\oalpsq} \else
{{$\tilde{A}_{FB}^b|_{\oalpsq}$}}\fi\chkspace}
\def\pafbbto{\relax\ifmmode \tilde{A}_{FB}^b|_{\oalpc} \else
{{$\tilde{A}_{FB}^b|_{\oalpc}$}}\fi\chkspace}

\def\afbo0{\tilde{A}_{FB}|_{O(0)}}
\def\afbo1{\tilde{A}_{FB}|_{\oalp}}
\def\afbo2{\tilde{A}_{FB}|_{\oalpsq}}
\def\afbo3{\tilde{A}_{FB}|_{\oalpc}}

\def\lam{\relax\ifmmode \Lambda_{\overline{MS}}
       \else {{$\Lambda_{\overline{MS}}$}}\fi\chkspace}
\def\lamuds{\relax\ifmmode \Lambda^{(3)}_{\overline{MS}}
       \else {{$\Lambda^{(3)}_{\overline{MS}}$}}\fi\chkspace}
\def\lamudsc{\relax\ifmmode \Lambda^{(4)}_{\overline{MS}}
       \else $\Lambda^{(4)}_{\overline{MS}}$\fi\chkspace}
\def\lamudscb{\relax\ifmmode \Lambda^{(5)}_{\overline{MS}}
       \else $\Lambda^{(5)}_{\overline{MS}}$\fi\chkspace}

\def\alp{\relax\ifmmode \alpha_s\else $\alpha_s$\fi\chkspace}
\def\alpbar{\relax\ifmmode \overline{\alpha_s}
       \else $\overline{\alpha_s}$\fi\chkspace}
\def\alpmz{\relax\ifmmode \alpha_s(M_Z)\else $\alpha_s(M_Z)$\fi\chkspace}
\def\alpmzsq{\relax\ifmmode \alpha_s(M_Z^2)
       \else $\alpha_s(M_Z^2)$\fi\chkspace}

\def\oalp{\relax\ifmmode O(\alpha_s)\else{{O($\alpha_s$)}}\fi\chkspace}
\def\oalpsq{\relax\ifmmode O(\alpha_s^2)
           \else{{O($\alpha_s^2$)}}\fi\chkspace}
\def\oalpc{\relax\ifmmode O(\alpha_s^3)
           \else{{O($\alpha_s^3$)}}\fi\chkspace}
\def\oalpf{\relax\ifmmode O(\alpha_s^4)
           \else{{O($\alpha_s^4$)}}\fi\chkspace}

\def\plb{Phys. Lett.\chkspace}
\def\npb{Nucl. Phys.\chkspace}
\def\rmp{Rev. Mod. Phys.\chkspace}
\def\prl{Phys. Rev. Lett.\chkspace}
\def\prd{Phys. Rev.\chkspace}
\def\zpc{Z. Phys.\chkspace}

\def\z0{{$Z^0$}\chkspace}
\def\Dst{\relax\ifmmode {\rm D}^* \else {D$^*$}\fi\chkspace}
\def\Dpl{\relax\ifmmode {\rm D}^+ \else {D$^+$}\fi\chkspace}
\def\D0{\relax\ifmmode {\rm D}^0 \else {D$^0$}\fi\chkspace}
\def\Kst{\relax\ifmmode {\rm K}^* \else {K$^*$}\fi\chkspace}
\def\K0{\relax\ifmmode {\rm K}^0_s \else {K$^0_s$}\fi\chkspace}
\def\Kpl{\relax\ifmmode {\rm K}^+ \else {K$^+$}\fi\chkspace}
\def\Kstz{\relax\ifmmode {\rm K}^{*0} \else {K$^{*0}$}\fi\chkspace}

\renewcommand{\baselinestretch}{1.5}

 1

\catcode`\@=11 % This allows us to modify PLAIN macros.
\makeatletter
\def\@seccntformat#1{\csname the#1\endcsname.\hskip 1em}
\makeatother
\begin{document}

%\pagestyle{empty}
 
%%% bugger eps figure defaults
\renewcommand{\topfraction}{.99}
\renewcommand{\bottomfraction}{.99}
\renewcommand{\textfraction}{.01}
\renewcommand{\floatpagefraction}{.98}
%%% 

\hfill{SLAC--PUB--7434}
 
\hfill{MIT-LNS-97-267}
 
\vskip .1truecm

\hfill{March 1997}
 
\vskip 1truecm

\centerline{\Large \bf PRECISE TESTS OF QCD}
\centerline{\Large \bf IN}
\centerline{\Large \bf e$^+$e$^-$ ANNIHILATION$^*$}
  
\vskip 1.5truecm
 
%\centerline{\bf P.N. Burrows}
\centerline{\bf P.N. Burrows$^{**}$}
 
\vskip .5truecm
 
%\centerline{\it Laboratory for Nuclear Science}
%\centerline{\it Massachusetts Institute of Technology}
%\centerline{\it Cambridge, MA 02139, USA}
\centerline{\it Stanford Linear Accelerator Center}
\centerline{\it Stanford University, Stanford, CA94309, USA}
 
\vskip .4truecm
 
\centerline{burrows@slac.stanford.edu}
  
\vskip 1.4truecm
  
\centerline{ABSTRACT}

\noindent
A pedagogical review is given of precise tests of QCD in 
electron-positron annihilation. Emphasis is placed on measurements that
have served to establish QCD as the correct theory of strong interactions,
as well as measurements of the coupling parameter \alp. An outlook is given
for future important tests at a high-energy \ep collider.

\centerline{\it Lectures given at the SLAC Summer Institute,} 
\centerline{\it August 19-30, 1996}
 
\vfill
 
\noindent
$\star$ {Work supported by Department of Energy contracts
DE--FC02--94ER40818 (MIT)
and   DE--AC03--76SF00515 (SLAC).}
 
\noindent
$**$ {Permanent address: Lab. for Nuclear Science,
M.I.T., Cambridge, MA 02139, USA.}
 
\eject

\noindent{\large \bf 1. Introduction - \ep Colliders}

\vskip .5truecm

\noindent
The production of hadronic final-states by a variety of
interactions is illustrated in Fig.~1. In electron-positron annihilation 
hadronic activity is, by construction, limited to the final state, making
the study of hadronic events cleaner and simpler relative to lepton-hadron
and hadron-hadron collisions, from both the experimental and
theoretical points-of-view. On the experimental side there are no remnants of
the beam particles to add confusion to the interpretation of hadronic
structures, and, apart from initial and final-state photon radiation effects,
the hadronic centre-of-mass frame coincides with the laboratory frame.
On the theoretical side the absence of hadrons in the incoming beams
removes dependence on the limited knowledge of the parton density 
functions of hadrons, as well as rendering QCD calculations at a
given order of perturbation theory easier to perform because there are
generally fewer strong-interaction Feynman diagrams to consider.
Electron-positron annihilation thus provides an ideal environment for precise 
tests of QCD. 
 
%% figure 1 
\begin{figure} [tbh]
% \hspace*{5cm}
   \epsfxsize=4.0in
   \epsfysize=7.0in
   \begin{center}
   \mbox{\epsffile{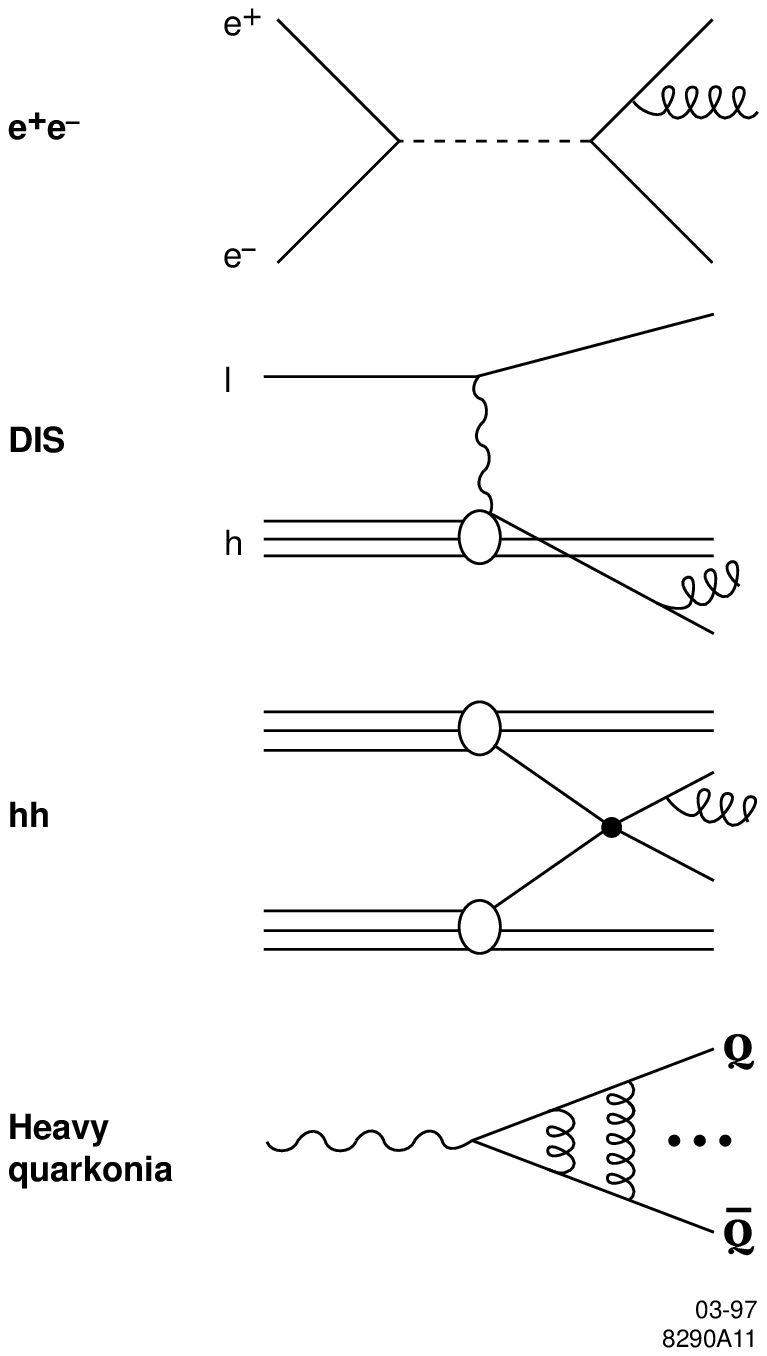}}
   \end{center}
   \caption[]{
Schematic of the production of hadronic final-states by different interactions:
\ep annihilation, deep inelastic lepton-hadron scattering, hadron-hadron
collisions, and production and decay of heavy quarkonia.
  }
\label{fig1}
\end{figure}

A large number of \ep colliders have been constructed over the past 25 years;
these are listed in Table~1. The range of c.m. energies $Q$ extends from  
a few GeV at the very first colliders up to almost 
200 GeV at the CERN LEP-II collider. The first generation of colliders
was built on speculation of allowing exciting high-energy physics studies.
They did not disappoint, the J/$\Psi$ being discovered at SPEAR, the gluon 
being observed
at PETRA, and a wealth of strong- and electroweak-interaction studies being
performed at PETRA, PEP, and TRISTAN, all of which served to establish the 
validity of
the Standard Model. A second generation of colliders has been designed to
serve as particle `factories': DA$\Phi$NE in the vicinity of the $\Phi$
resonance, BEPC near the charmonium threshold, DORIS, CESR, PEP-II and KEKB
around the bottomonium resonances, SLC and LEP at the \z0 resonance, and 
LEP-II at the W$^+$W$^-$ threshold. 

\clearpage

\begin{table}
\vskip .5truecm
\centering
\begin{tabular}{|ll|l|c|} \hline
& Collider $\quad\quad$ \hfill & Location $\quad\quad$
& c.m. energy $Q$ (GeV)                               \\ \hline
 & ADONE \hfill & Frascati \hfill  & 1 $-$ 3          \\
& DCI   \hfill & Orsay \hfill      & 1 $-$ 2.4        \\
$\bullet$ &
VEPP-2M \hfill & Novosibirsk \hfill      & 1 $-$ 1.5  \\
$\bullet$ &
DA$\Phi$NE \hfill & Frascati \hfill      & 1 $-$ 1.5  \\
& SPEAR \hfill & SLAC \hfill      & 2 $-$ 8           \\
$\bullet$ &
BEPC  \hfill & Beijing \hfill   & 2 $-$ 3             \\
& DORIS \hfill & DESY \hfill      & 3 $-$ 11          \\
$\bullet$ &
VEPP-4M \hfill & Novosibirsk \hfill      & 10 $-$ 12  \\
$\bullet$ &
CESR  \hfill & Cornell \hfill   & 10 $-$ 11           \\
$\bullet$ &
PEP-II \hfill & SLAC    \hfill   & $9\otimes 3.1$      \\
$\bullet$ &
KEKB   \hfill & KEK     \hfill   & $8\otimes 3.5$      \\
& PETRA \hfill & DESY \hfill      & 12 $-$ 47         \\
& PEP   \hfill & SLAC \hfill      & 29                \\
& TRISTAN \hfill & KEK \hfill     & 50 $-$ 64         \\
$\bullet$ &
SLC \hfill   & SLAC \hfill      & 88 $-$ 93           \\
& LEP \hfill  & CERN \hfill      & 88 $-$ 93          \\
$\bullet$ &
LEP-II \hfill  & CERN \hfill      & 130 $-$ 192       \\
& XLC  \hfill  & ???? \hfill      & 500 $-$ 1500      \\  \hline
\end{tabular}
\vskip .5truecm
\caption{\ep colliders 1972 - 200? $\bullet$ denotes running/under 
construction.}
\end{table}

With the exception of the SLC, all of
these colliders have been of the storage ring type, the largest, LEP-II
with a circumference of 27km, probably marking the limit of the energy that
can be achieved with current storage ring technology for an acceptable cost. 
The SLC
is the first example of a high-energy {\it linear} \ep collider; it achieves
the same collision energy as LEP, but has an effective length of about 3 miles
and was considerably cheaper to construct. Because of their intrinsically lower
cost/GeV, linear colliders represent the obvious path towards construction of
higher-energy \ep colliders with current acceleration technology. A number of
proposals for such an accelerator are represented by `XLC' in Table~1; they all
aim to achieve c.m. energies between 500 and 1500 GeV, which is believed to
cover the interesting range for study of electroweak-symmetry-breaking
processes. Some examples of QCD tests that could be made at the XLC will
be given towards the end of these lectures.

It would require a semester-long lecture series to do full justice to QCD
studies in \epa, so some hard choices have been made as to the material to
be covered here; I apologise well in advance for all that has been omitted.
No attempt has been made to give a complete review of all of the experimental
results in any of the areas covered; usually one or two results or figures are
shown as examples. For this purpose I have drawn heavily on material from
TASSO and SLD, the two experiments with which I have been involved since 1985;
no disrespect is intended to the many other experiments whose results may not
be shown. Tests of QCD in hadron-hadron and lepton-hadron collisions will not
be discussed here as they are covered in other lectures\cite{albrow,smith}
at this Institute. 
 
In the interests of pedagogy I shall review the fundamental properties
of QCD and the important experimental measurements from \epa that have been
key historically to establishing the theory. Having verified QCD, in a
qualitative sense, as being the only viable theory for describing strong 
interactions, I shall then review quantitative tests in the form of
measurements of \alp, the single parameter of the theory, and put the 
\ep measurements into context with determinations from other processes.
I shall focus on measures of the event topology, especially on 
jet definition and the relation between the
jets observed in detectors and the `true' underlying parton-jet structure.
This will introduce the problem of hadronisation, as well as the difficulty
of relating finite-order perturbative QCD calculations to the `all-orders'
data. I shall conclude by looking forward to the precise QCD tests that could be
made at a high-energy \ep collider and to the qualitatively new \ttg system
accessible at such a facility.

\vskip 1 truecm

\noindent{\large \bf 2. Our Theory of Strong Interactions - QCD}
 
\vskip .5truecm

\noindent
Quantum Chromodynamics (QCD)~\cite{qcd}
is our theory of the strong interaction between
quarks and gluons. It is a non-Abelian Yang-Mills gauge theory that describes 
the interactions of a triplet of spin-1/2 quarks possessing the colour quantum
number ($c$ = r,b,g) via an octet of vector gluons. The spinor quark fields
$q_c(x)$ transform as the fundamental representation of the SU(3) group,
whilst the gluon fields $A_{\mu}^a(x)$ ($a$ = 1,2,$\ldots$,8) transform
according to the adjoint representation. The SU(3) colour transformations
are generated by the $3\times3$ matrices $T^a$ = $\lambda^a/2$, where 
$\lambda^a$ are the Gell-Mann matrices~\cite{GM} which obey the commutation
relations:
\begin{eqnarray}
[T^a,T^b]\quad=\quad i\;f^{abc}\;T^c
\end{eqnarray}
and $f^{abc}$ are the structure constants of SU(3).
The Lagrangian has the form:
\begin{eqnarray}
{\cal{L}}\quad=\quad -{1\over {4}} F^a_{\mu\nu} F^{\mu\nu\;a}\;+\;
\overline{q}\;(i\gamma_{\mu} D^{\mu}\;-\;m) \;q
\end{eqnarray}
where $F_{\mu\nu}$ is the field strength tensor:
\begin{eqnarray}
F_{\mu\nu}^a\quad=\quad {\partial}_{\mu} A_{\nu}^a - 
{\partial}_{\nu} A_{\mu}^a + g f^{abc} A_{\mu}^b A_{\nu}^c
\end{eqnarray}
and $D_{\mu}$ is the covariant derivative:
\begin{eqnarray}
D_{\mu}\quad=\quad {\partial}_{\mu} - i g T^a A_{\mu}^a(x),
\end{eqnarray}
$g$ is the bare coupling of the theory, $m$ the bare mass of the quark field
and the gluons are massless.

Following~\cite{kramer}, the `essential features' of QCD may be summarised
as:
 
\vskip .3truecm
 
$\bullet$
quarks with spin 1/2 exist as colour triplets
 
$\bullet$
gluons with spin 1 exist as colour octets
 
$\bullet$
the coupling \qqg exists
 
$\bullet$
the couplings ggg and gggg exist
 
$\bullet$
the couplings are equal
 
$\bullet$
the coupling decreases as 1/ln$Q^2$

\noindent
For most of the first lecture I shall review the evidence, from \epa alone,
that QCD {\it is} the correct theory of strong interactions.

\vskip 1 truecm

\noindent{\large \bf 3. Establishing the QCD Lagrangian}

\vskip .75truecm
 
\noindent{\large \bf 3.1 Two-Jet Events and q$\overline{\rm \bf q}$ Production}

\vskip .5truecm

\noindent
Quarks were first postulated in 1964 by Gell-Mann and Zweig~\cite{GMZ} as a
calculational device to explain the rich spectroscopy of recently-discovered 
mesons and baryons in terms of bound \qq and qqq (or {\qbar}{\qbar}{\qbar}) 
states.
The first direct evidence for quarks came from the observations at SLAC in the
late 1960s that in electron-nucleon scattering experiments at high $Q^2$ the
electron scatters from quasi-free pointlike particles. In \epa a convincing
demonstration of the existence of quarks was provided 
by the observation of jets in the Mark I experiment at SPEAR in 
1975~\cite{mk1jet}. This analysis represents the first use of an
{\it event shape observable} which, as will be discussed later, is a key 
component in the study of hadronic final states, so I shall briefly describe 
it. 

By eye the spatial distribution of particles in hadronic events recorded 
in the Mark I detector operating at c.m. energies between 3.0 and 7.4 GeV
looked more-or-less isotropic, and it was hard to distinguish any clear jet
structure. The quantity sphericity,
\begin{eqnarray}
S\quad=\quad  
\frac{{\rm Min} \left(\Sigma_i p_{\perp i}^2\right)}{\Sigma_i \vec{p_i}^2},
\end{eqnarray}
where $\vec{p_i}$ represents the momentum of particle $i$ and the sums
run over all particles in each event,
was invented~\cite{sphericity} to characterise the degree of isotropy in the
particle flow. In each event an axis, the sphericity axis, is defined so as to 
minimise the quantity in brackets in the numerator; eq.~(5) then defines
the sphericity of the
event. A completely isotropic distribution of particles, or spherical event,
would yield $S$ $\sim$ 1, whilst a perfectly-collimated back-to-back two-jet
event would have $S$ = 0. Sphericity distributions from Mark I are shown in
Fig.~2 for data taken at several different c.m. energies. As the energy was
raised from 3.0 to 7.4 GeV a clear change in the sphericity distribution was
observed, the distribution shifting to lower values at higher energies. This
was interpreted in terms of an increasing degree of collimation of particle
production with c.m. energy, namely the onset of the production of two
back-to-back jets of hadrons. At higher energies the jet structure is much
more apparent by eye, as indicated in the \z0 decay event from SLD shown in
Fig.~3, and is striking evidence for the production of a back-to-back quark
and antiquark in \epa.

The Mark I analysis was also able to establish the nature of the spin of the
quark and antiquark. Shown in Fig.~4 is the distribution of the 
azimuthal-angle, $\phi$, of the
sphericity axis w.r.t. the beamline, at two c.m. energies. At $Q$ = 7.4 GeV the 
electron and positron beams in the SPEAR ring built up a degree of transverse
polarisation $P$ via the Sokolov-Ternov synchrotron radiation effect~\cite{sok}
and a clear modulation in $\phi$ is visible. This is in contrast to the flat
$\phi$ distribution at $Q$ = 6.2 GeV which corresponds to a beam-depolarising 
resonance ($P$ = 0) in the accelerator. A fit of the function:
\begin{eqnarray}
\frac {dN}{d\Omega}\quad \propto \quad 1\;+\;\alpha \cos^2\theta\;+\; P^2 \alpha
\sin^2\theta\cos2\phi
\end{eqnarray}
to the 7.4 GeV data yielded $\alpha$ = $0.78\pm0.12$; this is close to
unity, which is expected for production of two spin-1/2 
particles~\cite{spinhalf}. 
 
%% figure 2 
\begin{figure} [tbh]
 \hspace*{5cm}
   \epsfxsize=4.0in
   \epsfysize=6.0in
    \setlength{\baselineskip}{13pt}
   \begin{center}
    \mbox{\epsffile{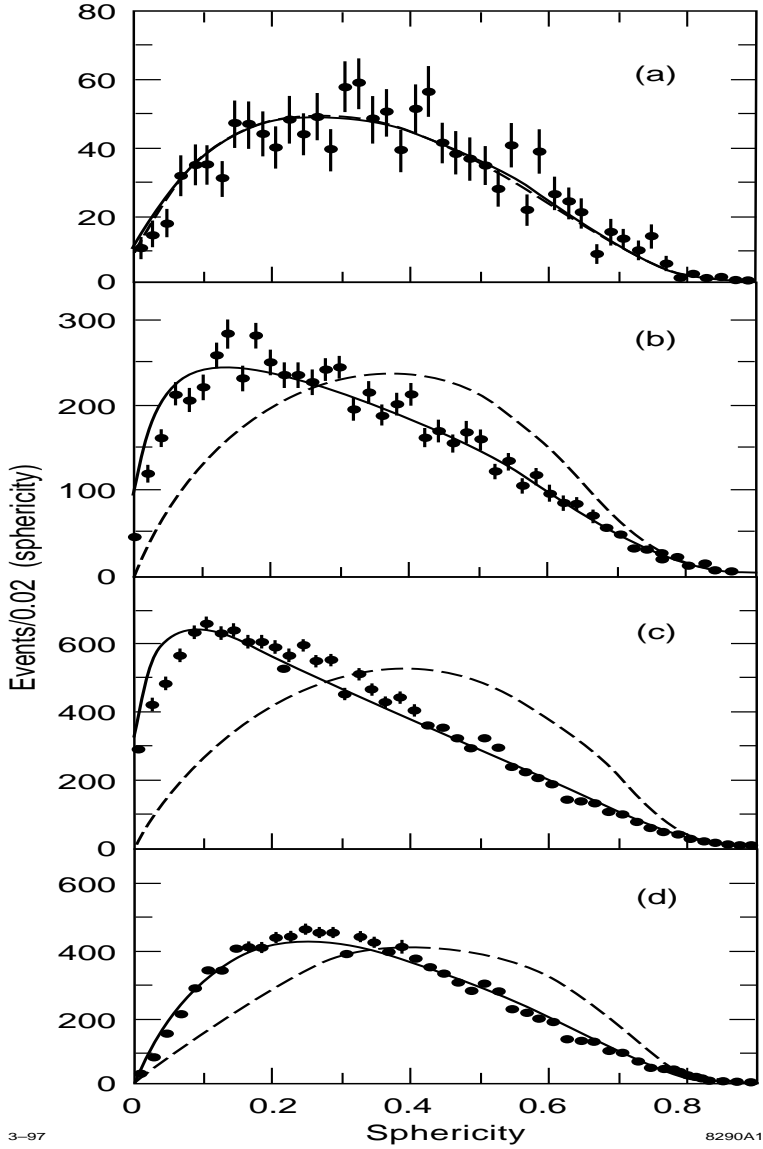}}
   \end{center}
   \caption[]{
Sphericity distributions measured by the Mark I experiment at 
SPEAR~\cite{mk1jet} at c.m. energies of (a) 3.0, (b) 6.2, (c) 7.4 GeV. 
The
narrowing of the distribution, and the trend towards smaller values as the c.m.
energy is raised, represent evidence for collimated production of hadrons in
\epa. The dashed line represents the expectation from a `phase-space model' of
hadron production. (d) As (c) but for a subset of events containing particles
with scaled momentum, $2p/Q$, less than 0.4.
  }
\end{figure}

\clearpage

%% figure 3 
\begin{figure} [tbh]
 \hspace*{5cm}
   \epsfxsize=5.2in
   \epsfysize=4.0in
    \setlength{\baselineskip}{13pt}
   \begin{center}
    \mbox{\epsffile{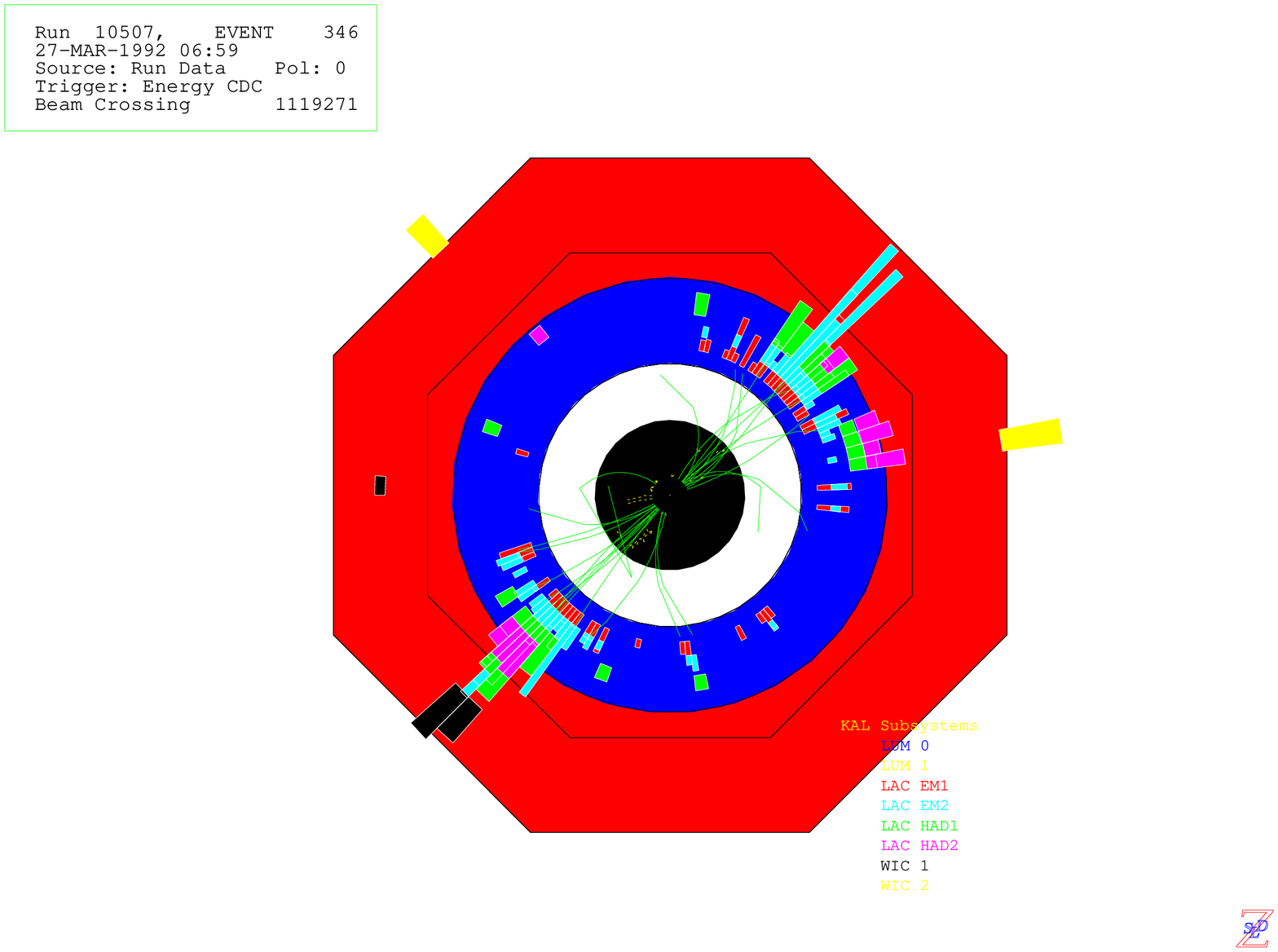}}
\end{center}
   \caption[]{
A contemporary two-jet event recorded by SLD: the process \ep \ra \z0 \ra \qq.
  }
\end{figure}

%% figure 4 
\begin{figure} [tbh]
 \hspace*{5cm}
   \epsfxsize=4.0in
   \epsfysize=5.0in
    \setlength{\baselineskip}{13pt}
   \begin{center}
    \mbox{\epsffile{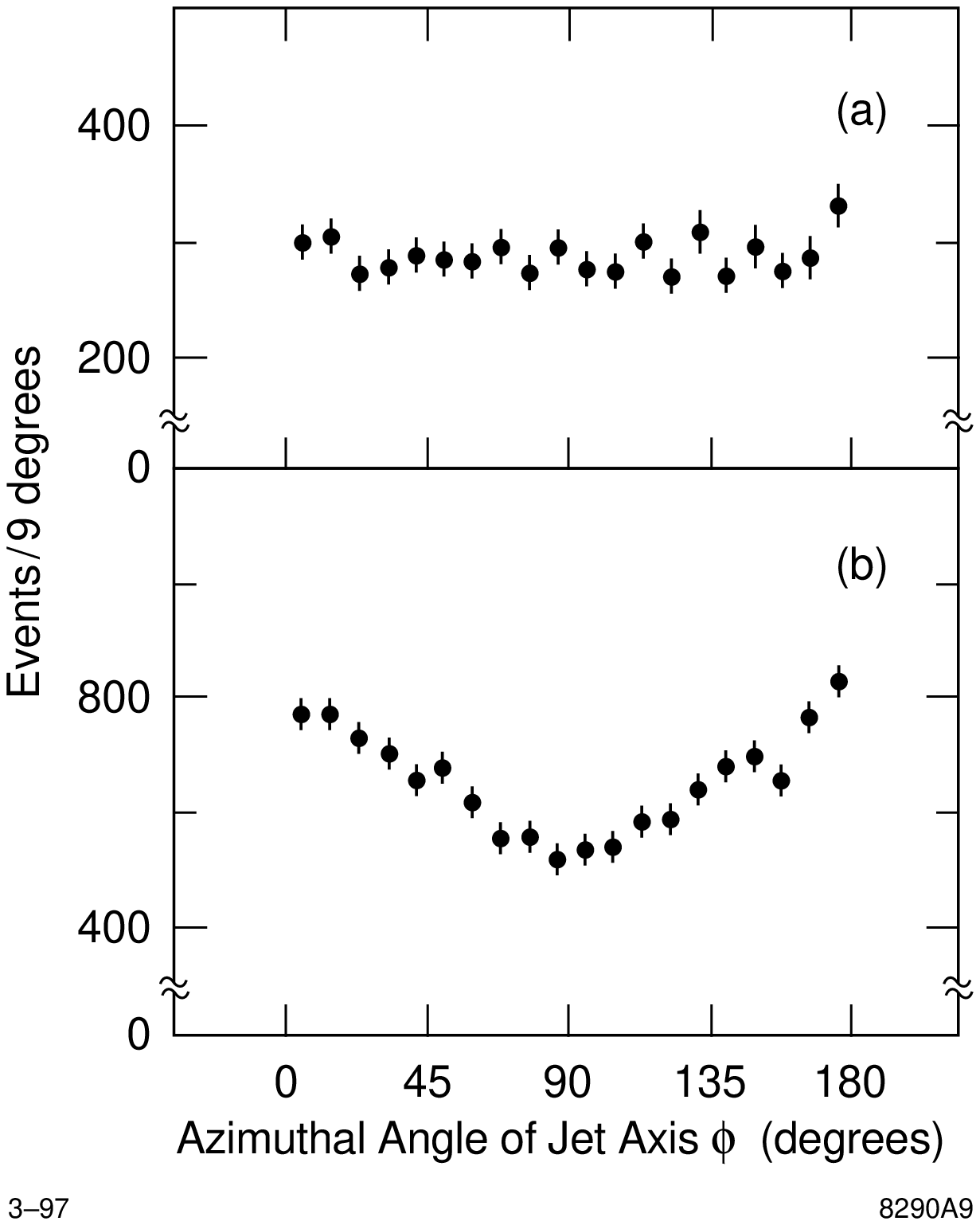}}
\end{center}
   \caption[]{
Azimuthal-angle distribution of the sphericity axis from Mark I~\cite{mk1jet}. 
At $Q$ = 6.2 GeV (a) the SPEAR beams were unpolarised and at $Q$ = 7.4 GeV 
(b) the 
polarisation-product was 0.47; the modulation in (b) is
characteristic of the production of two spin-1/2 particles in \epa.
  }
\end{figure}

These studies were subsequently extended at the higher-energy PETRA collider,
and examples from TASSO~\cite{tassph}
are shown in Fig.~5. Here the distribution of the 
polar-angle ($\theta_S$) of the sphericity axis is shown at c.m. energies of
14, 22 and 35 GeV. A fit to the functional form:
\begin{eqnarray}
\frac{dN}{d\cos\theta}\quad\propto\quad 1\;+\;a_{S,T}\cos^2\theta_{S,T}
\end{eqnarray}
yields, at 35 GeV for example, $a_S$ = $1.03\pm0.07$, again characteristic of
the production of two spin-1/2 particles in the \epa. Also shown in Fig.~5 is
our second example of an event shape observable in the form of the 
thrust-axis~\cite{thrust} polar-angle ($\theta_T$) 
distribution. Thrust will be discussed
later; it is qualitatively similar to sphericity in that it can be used to
quantify the degree of collimation of particle production, 
although it has properties that make it more attractive theoretically. 
The thrust-axis polar-angle distribution
in Fig.~5 was fitted to obtain, at 35 GeV for example, $a_T$ = $1.01\pm0.06$,
in good agreement with the result using the sphericity axis.

%% figure 5 
\begin{figure} [tbh]
 \hspace*{5cm}
   \epsfxsize=5.0in
   \epsfysize=5.0in
    \setlength{\baselineskip}{13pt}
   \begin{center}
    \mbox{\epsffile{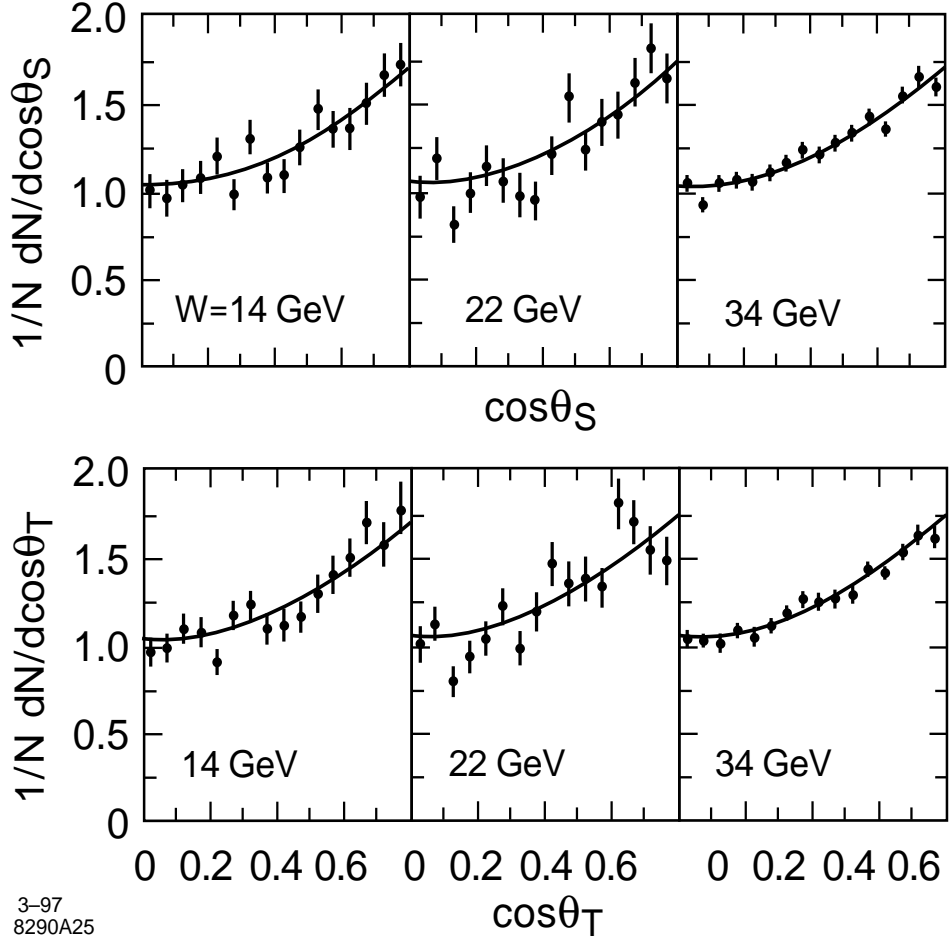}}
\end{center}
   \caption[]{
Polar-angle distributions of the sphericity and thrust axes from 
TASSO~\cite{tassph}. 
The $1+\cos^2\theta$ form is characteristic of the production of two spin-1/2
particles in the \epa. 
  }
\end{figure}

So far spin-1/2 quarks and antiquarks would appear to be well
established, and their colour-triplet nature, $N_C$ = 3, is required
in the quark-parton model (QPM) of hadrons to explain the existence of 
spin-3/2 baryon states 
such as the $\Delta^{++}$ (u$\uparrow$u$\uparrow$u$\uparrow$) and $\Omega^-$  
(s$\uparrow$s$\uparrow$s$\uparrow$), which would otherwise contain three
identical fermions in the same quantum state, in violation of the Pauli 
exclusion
principle. In \epa evidence for $N_C = 3$ is provided by the quantity:
$$
R \quad \equiv \quad \frac{\sigma(e^+e^-\;\rightarrow\;{\rm hadrons})}
{\sigma_{QED}(e^+e^-\;\rightarrow\;\mu^+\mu^-)}
$$
which, according to QED and the QPM, should be equal to 
$N_C\;\Sigma_f\;q_f^2$, where $q_f$ is the charge of the quark of flavour $f$
and the sum runs over all active flavours at a given c.m. energy.

%% figure 6 
\begin{figure} [tbh]
% \hspace*{5cm}
   \epsfxsize=6.0in
   \epsfysize=6.0in
   \begin{center}
    \mbox{\epsffile{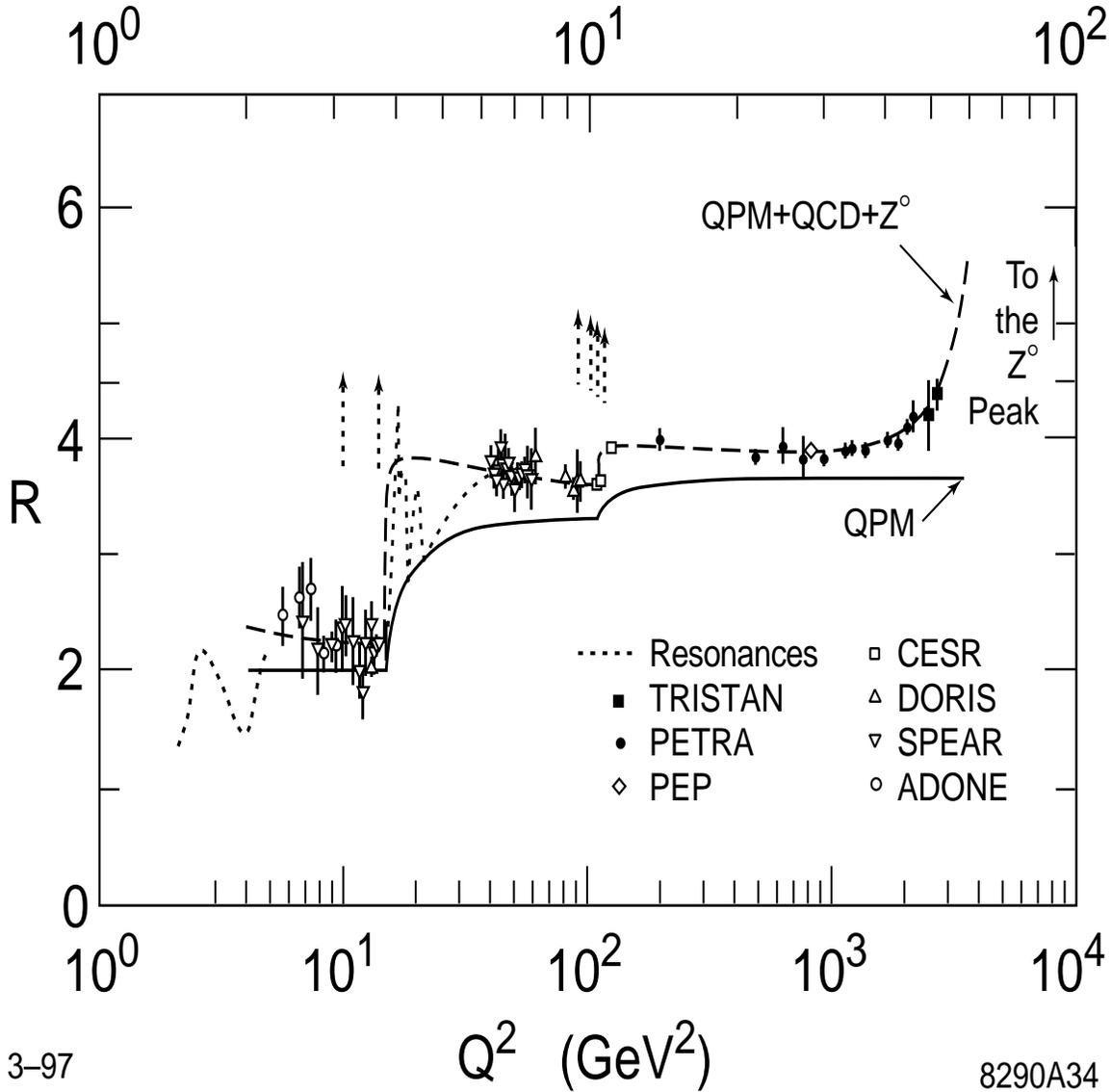}}
\end{center}
   \caption[]{
The $R$ ratio as a function of c.m. energy~\cite{marshall}. 
The expectation for $N_C$ = 3 is shown as the solid line.
  }
\end{figure}

A summary of $R$ measurements made up to 1988, 
as a function of c.m. energy, is presented in 
Fig.~6~\cite{marshall}. This is a tremendously information-rich figure. 
First, increases
in $R$ just above $Q^2$ = 10 and 100 GeV$^2$ represent the \cc and 
\bb production thresholds - further evidence, were it needed, for the
existence of quarks. Secondly, the QED + QPM prediction comes close to the
data only if the quarks are assigned fractional charges {\it and} the number of 
colours $N_C$ = 3 is used; the `colour singlet' expectation ($N_C$ = 1)
is simply too low by a factor of about
three! Thirdly, above $Q^2$ = 1000 GeV$^2$ the data points rise as $Q^2$ 
increases, representing the onset of contributions to \epa from \z0 exchange.
Finally, in regions between quark flavour thresholds and below the tail of the
\z0 resonance, there is a residual excess in the data relative to the
QED + QPM expectation, and the excess appears to decrease as $Q^2$ increases. 
In other words,
some mechanism causes an increase in the `phase-space' for hadron production
beyond QED + QPM, but at a rate that decreases with $Q^2$. In the language of
the 1990s we know that the extra contribution is due to gluon emission in the 
final state, and that the probability for this process, \alp, decreases 
roughly logarithmically with $Q^2$. The $R$-ratio thus provides indirect
evidence for the existence of the gluon, as well as for the non-Abelian
`running' of the strong coupling.

\vskip 1truecm
 
\noindent{\large \bf 3.2 Three-Jet Events and the Gluon}

\vskip .5truecm

\noindent
In \ep annihilation events containing {\it three} distinct jets of hadrons
were first observed in 1979 at the PETRA storage ring \cite{threejet}
at c.m. energies around 20 GeV.
Such events were interpreted~\cite{ellis} 
in terms of the fundamental process \ep \ra \qqg,
providing direct evidence for the existence of the gluon
and its coupling to quarks.  
A modern example of a three-jet event,
in fact the very event used to advertise this Summer Institute, is shown in
Fig.~7.

%% figure 7 
\begin{figure} [tbh]
% \hspace*{5cm}
   \epsfxsize=5.2in
   \epsfysize=4.0in
   \begin{center}
    \mbox{\epsffile{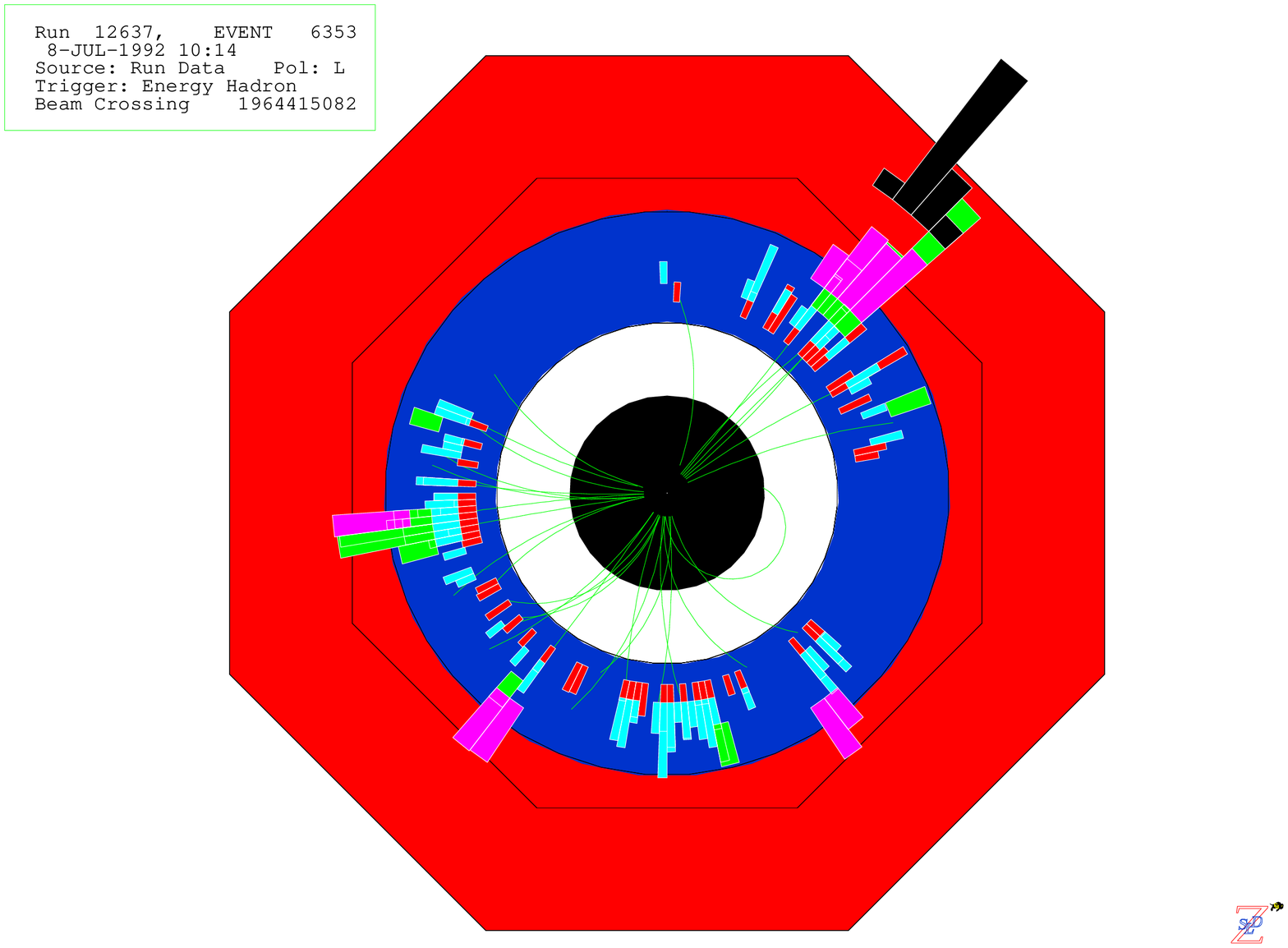}}
\end{center}
   \caption[]{
A contemporary three-jet event recorded by SLD: the process \ep \ra \z0 \ra 
\qqg.
  }
\end{figure}

Counting the number of jets per event, and then comparing the numbers
of two- and three-jet events, it was found~\cite{petalph} 
that around $Q$ = 20 GeV
\begin{eqnarray}
\frac{\# {\rm 3-jet\; events}}{\# {\rm 2-jet\; events}}\quad \approx \quad 0.15.
\end{eqnarray}                 
Since, at lowest order in
perturbative QCD, this ratio is simply the probability for gluon emission,
or \alp, the strong coupling parameter, simple event counting indicated
that the strong coupling at 20 GeV was around 0.15, \ie about ten times
larger than the electromagnetic coupling $\alpha$. More systematic
determinations of \alp will be discussed later.

Having observed the gluon directly in three-jet events one still needs to know
whether it is the gluon of QCD, namely a colour-octet vector particle.
Many studies of the nature of the gluon spin were performed at the PETRA 
and PEP storage rings and involved analysis of
the partition of energy among the three jets.
Ordering the three jets in \ep \ra \qqg $\;$ according to their energies
$E_1>E_2>E_3$, and normalising by the c.m. energy $Q$, we
obtain the scaled jet energies 
\begin{eqnarray}
x_i\quad=\quad {2 E_i\over Q}\quad\quad\quad(i=1,2,3),
\end{eqnarray}                 
represented in Fig.~8, where $ x_1 +  x_2 +  x_3 = 2$.
Making a Lorentz boost of the event into the rest frame of
jets 2 and 3 the historically-important 
Ellis-Karliner angle $\theta_{EK}$ is defined \cite{ek}
to be the angle between jets 1 and 2 in
this frame. For massless partons at tree-level:
\begin{eqnarray}
cos\theta_{EK}={{ x_2- x_3}\over{ x_1}}.
\end{eqnarray}
The results of an early study by TASSO~\cite{tassek} are shown in Fig.~9,
where the Ellis-Karliner angle distribution is compared, for data taken at $Q$
$\sim$ 30 GeV, with the prediction of QCD. 
One can also consider alternative `toy' models of
strong interactions, for example a model incorporating spin-0 (scalar) 
gluons~\cite{scalar}. From Fig.~9 the scalar-gluon model is clearly excluded.

%% figure 8 
\begin{figure} [tbh]
% \hspace*{5cm}
   \epsfxsize=3.0in
   \epsfysize=4.0in
   \begin{center}
    \mbox{\epsffile{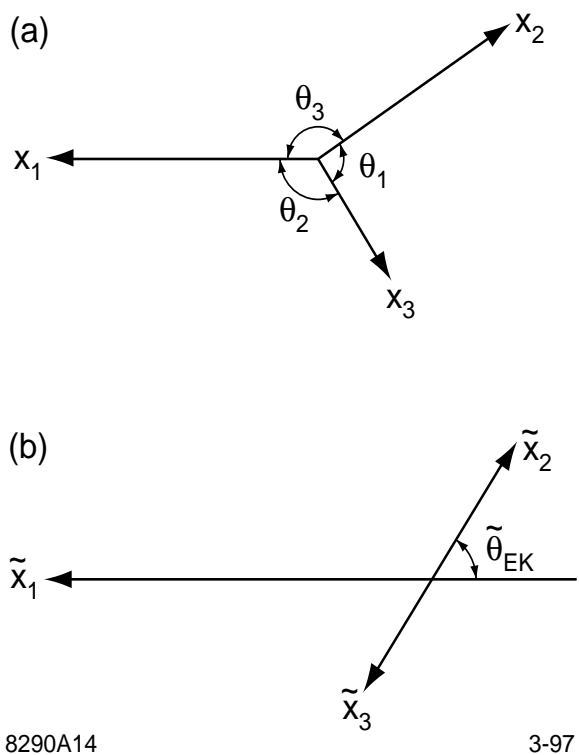}}
\end{center}
   \caption[]{
(a) Representation of the momentum vectors in a three-jet event, and (b)
definition of the Ellis-Karliner angle.
  }
\end{figure}

%% figure 9 
\begin{figure} [tbh]
% \hspace*{5cm}
   \epsfxsize=5.0in
   \epsfysize=4.0in
   \begin{center}
    \mbox{\epsffile{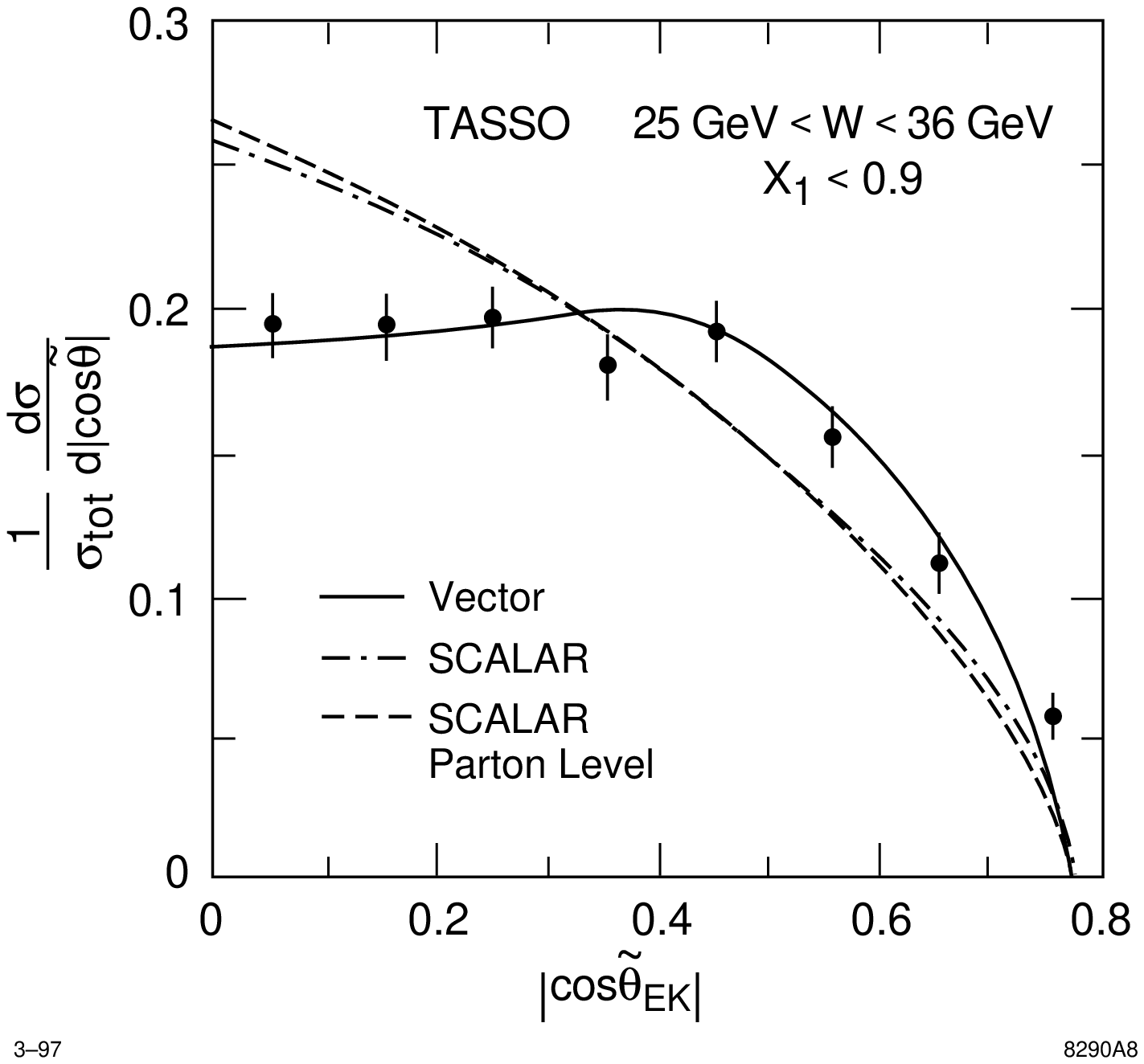}}
\end{center}
   \caption[]{
The Ellis-Karliner angle distribution of three-jet events recorded by TASSO at
$Q$ $\sim$ 30 GeV~\cite{tassek}; 
the data favour spin-1 (vector) gluons.
  }
\end{figure}

Similar studies have been extended at the \z0 resonance by the LEP and SLC
experiments. In this case
the inclusive differential cross sections, calculated at leading order
and assuming massless partons, can be written for vector gluons~\cite{kramer}:

\begin{eqnarray}
 {1\over\sigma}{d^2\sigma^V\over d x_1d x_2} \propto
{{ x_1^3+ x_2^3+(2- x_1- x_2)^3}\over{(1- x_1)(1- x_2)( x_1+ x_2-1)}},
\label{eqvec}
\end{eqnarray}

\noindent
for scalar gluons~\cite{scalar}:

\begin{eqnarray}
 {1\over\sigma}{d^2\sigma^S\over d x_1d x_2} \propto
\biggl[{{ x_1^2(1- x_1)+ x_2^2(1- x_2)+(2- x_1- x_2)^2( x_1+ x_2-1)}
\over  {(1- x_1)(1- x_2)( x_1+ x_2-1)}}-R \biggr],
\label{eqscal}
\end{eqnarray}

\noindent
where
\begin{eqnarray}
 {R}={10\;{\Sigma_j a_j^2}\over{\Sigma_j (v_j^2+a_j^2)}}
\end{eqnarray}
and $a_j$ and $v_j$ are the axial and vector couplings, respectively,
of quark flavor $j$ to the \z0, and
for a model of strong interactions incorporating spin-2 (tensor) 
gluons~\cite{rizzo,sldthree}: 
\begin{eqnarray}
 {1\over\sigma}{d^2\sigma^T\over d x_1d x_2} \propto
{{(x_1+x_2-1)^3 + (1-x_1)^3 + (1-x_2)^3}
\over{(1- x_1)(1- x_2)( x_1+ x_2-1)}}.
\label{eqten}
\end{eqnarray}
Singly-differential cross sections for $x_1$, $x_2$, $x_3$
or cos$\theta_{EK}$ can be obtained by numerical
integrations of Eqs.~(11), (12) and (14) 
and are compared with SLD data~\cite{sldthree} in Fig.~10.
The shapes are different for the vector, scalar and tensor gluon cases and
only the vector case describes the data.

%% figure 10 
\begin{figure} [tbh]
% \hspace*{5cm}
   \epsfxsize=5.0in
   \epsfysize=4.0in
   \begin{center}
    \mbox{\epsffile{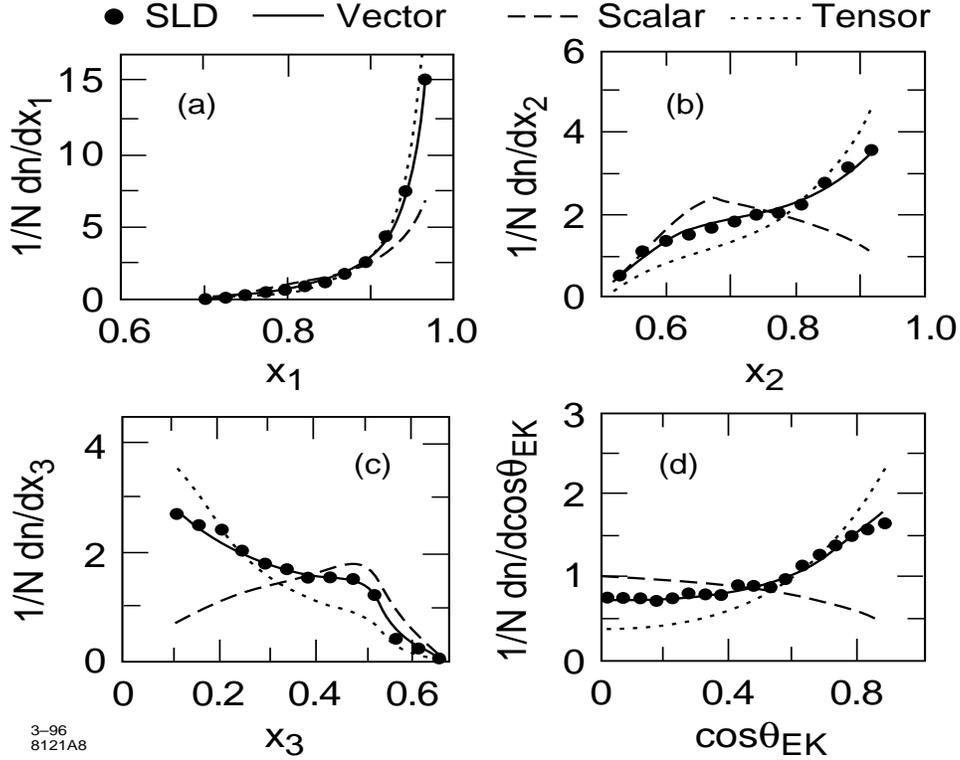}}
\end{center}
   \caption[]{
Comparison of leading-order scalar-, vector-, and tensor-gluon models with
SLD data~\cite{sldthree}; 
the scalar and tensor hypotheses are clearly excluded.
  }
\end{figure}
 
An additional interesting observable in three-jet events is the
orientation of the event plane w.r.t. the beam direction, which can be
described by three Euler angles (Fig.~11). These angular distributions were
studied first by TASSO \cite{TASSO}, and more recently by
L3 \cite{LTHREE} and DELPHI \cite{DELPHI}.
Again, the data were compared with the predictions of perturbative
QCD and a scalar gluon model, but the Euler angles
are less sensitive than the jet energy distributions
to the differences between the two cases \cite{LTHREE}.
One can parametrise the angular distributions in the form:
\begin{eqnarray}
{d\sigma\over d{\rm cos}\theta}\quad\propto\quad 1\;+\;\alpha(T)
{\rm cos}^2\theta
\end{eqnarray}
\begin{eqnarray}
{d\sigma\over d{\rm cos}\theta_N}\quad\propto\quad 1\;+\;\alpha_N(T)
{\rm cos}^2\theta_N
\label{eqthn}
\end{eqnarray}
\begin{eqnarray}
{d\sigma\over d\chi}\quad\propto\quad 1\;+\;\beta(T)
{\rm cos}2\chi
\end{eqnarray}
where $T$ is the thrust value \cite{thrust} of the event. 
As an example, fits of eq.~(16) to SLD distributions of
cos$\theta_N$ are shown in Fig.~12~\cite{sldthree} in four bins of thrust. 
The coefficients $\alpha(T)$,
$\alpha_N(T)$ and $\beta(T)$ depend on the gluon spin; they are
shown in Fig.~13 for leading-order calculations incorporating vector,
scalar and tensor gluons. 
The measured
$\alpha(T)$, $\alpha_N(T)$ and $\beta(T)$ are also shown in Fig.~13 and confirm 
that only vector gluons are compatible with the data.

%% figure 11 
\begin{figure} [tbh]
% \hspace*{5cm}
   \epsfxsize=3.0in
   \epsfysize=3.0in
   \begin{center}
    \mbox{\epsffile{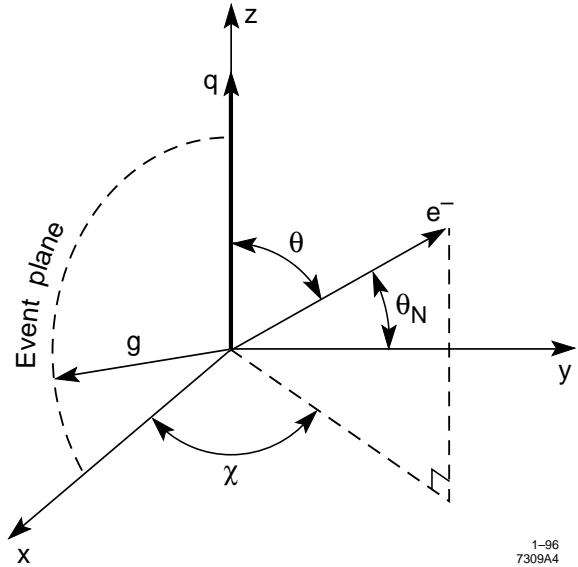}}
\end{center}
   \caption[]{
Definition of event-plane orientation angles.
  }
\end{figure}

%% figure 12 
\begin{figure} [tbh]
% \hspace*{5cm}
   \epsfxsize=5.0in
   \epsfysize=4.0in
   \begin{center}
    \mbox{\epsffile{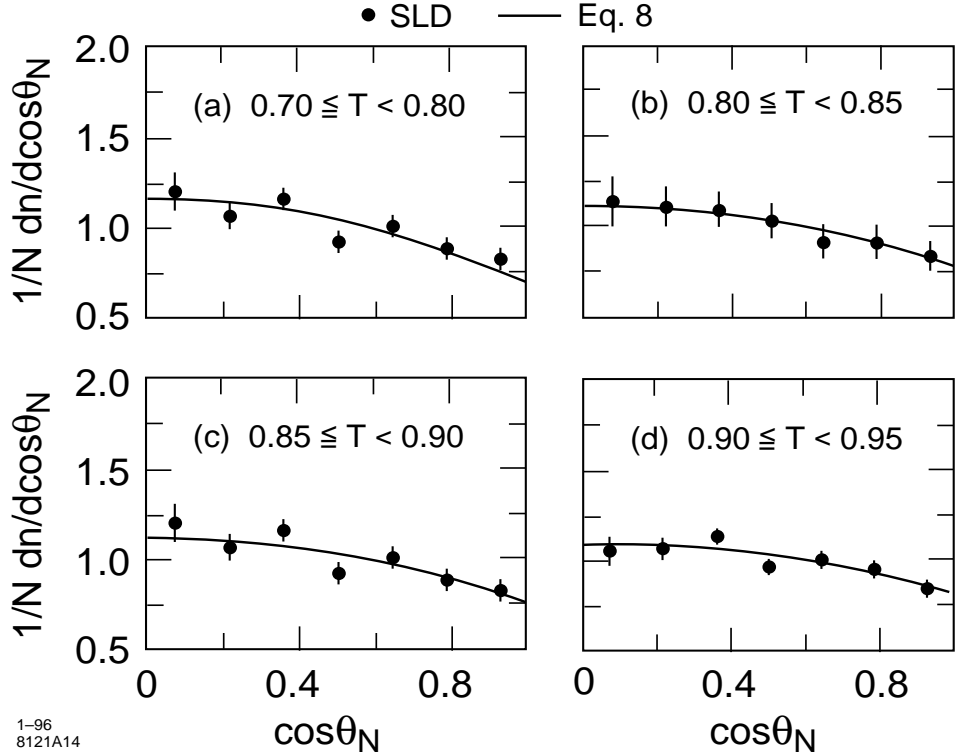}}
\end{center}
   \caption[]{
Distributions of cos$\theta_N$ as a function of event thrust from 
SLD~\cite{sldthree}.
  }
\end{figure}

%% figure 13 
\begin{figure} [tbh]
% \hspace*{5cm}
   \epsfxsize=4.0in
   \epsfysize=4.5in
   \begin{center}
    \mbox{\epsffile{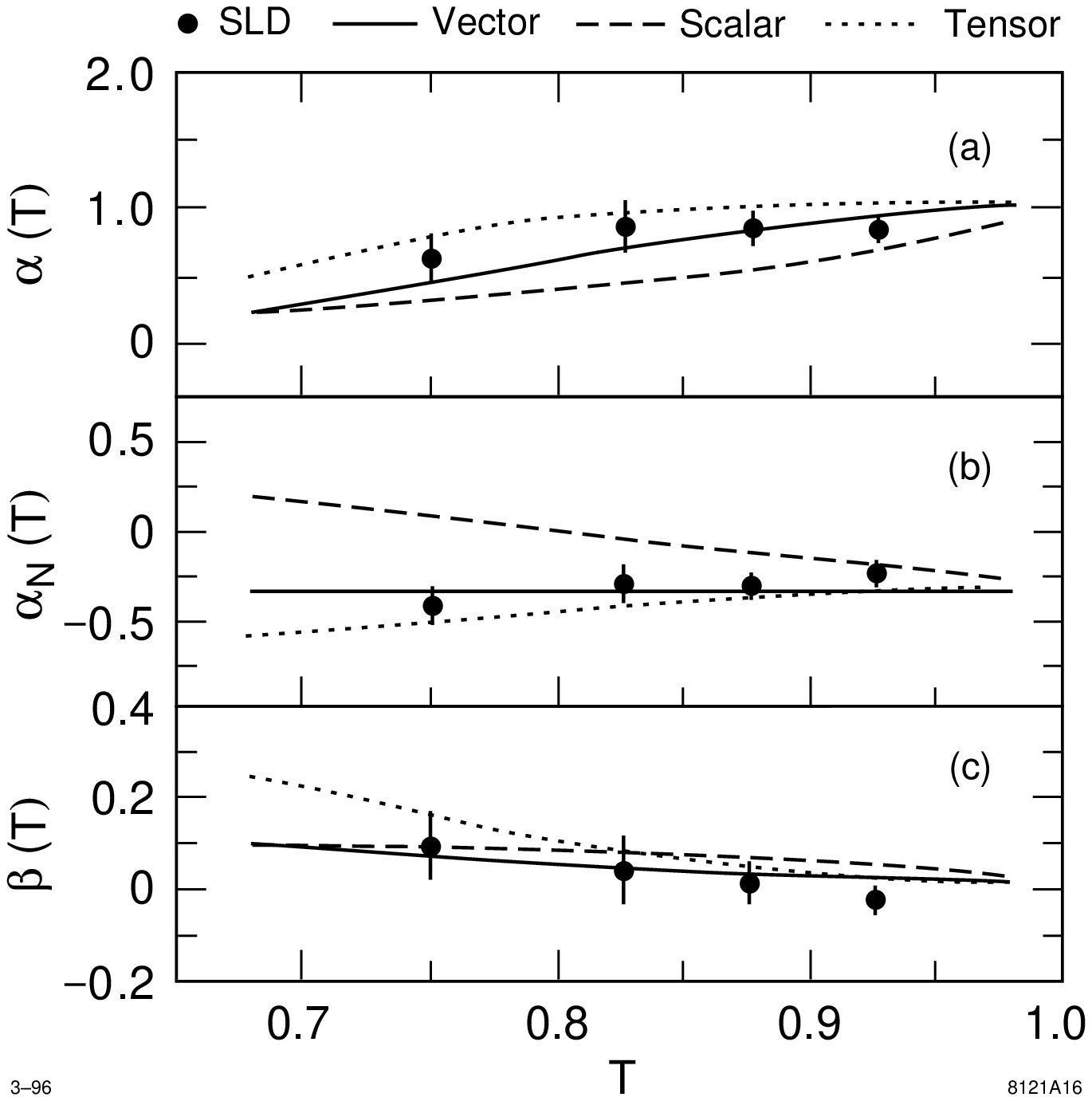}}
\end{center}
   \caption[]{
Dependence of the coefficients of the three-jet event plane orientation
angular distributions on event thrust value~\cite{sldthree}.
  }
\end{figure}

At this point it is worth pausing to take stock of what has been 
learned so far. The \ep \ra two-jet events have provided direct evidence for
\qq production, and the jet axis angular distribution indicates that the quark
and antiquark have spin-1/2. From the inclusive $R$-ratio we can confirm the 
fractional
nature of the quark charges, and know that the quarks must exist as colour
triplets since $N_C$ = 3 is the only value that brings QED + 
the quark-parton model close to the data. The value of the  
$R$-ratio also tells us that
there must be contributions to hadronic final states in addition to \qq
production, and we know that these are provided by
three-jet events, which represent direct evidence for the existence of the
gluon and its coupling to quarks and antiquarks. The distributions of jet
energies, or equivalently of jet angles within the event plane, as well as of
the event plane orientation itself, confirm that the only hypothesis that
fits the data is that the gluon has spin-1, and therefore that it 
is the vector boson of QCD. Finally, from
counting the relative rates of three- and two-jet events at $Q$ $\sim$ 30 GeV
we know that the coupling strength of the gluon to quarks is about 0.15. 
Checking the list of `essential features' of QCD we see that we have
verified about half of them! The next item in the list refers to the
triple- and quartic-gluon couplings; in order to study these we need to
examine multi-jet final-states. 

%\vskip 1truecm
\vfill
\eject
 
\noindent{\large \bf 3.3 Multi-Jet Events and Gluon Self-Couplings}

\vskip .5truecm

\noindent
Consider the Feynman diagrams for production of 4-jet final states shown in
Fig.~14. Figs.~14(a) and (b) illustrate the gluon Bremsstrahlung process,
whilst Fig.~14(d) shows the splitting of a gluon into a quark and an antiquark;
the latter may
be thought of as a QCD analogue of the QED process whereby a photon converts 
into an electron and a positron. Figs. 14(a,b,d) are sometimes referred to as 
`Abelian'
diagrams. Fig.~14(c) illustrates the lowest-order diagram for a gluon to
split into two gluons. This process has no analogue in QED since the photon
does not couple to itself, and is a consequence of the non-Abelian nature of
QCD in that the gluons, by virtue of possessing colour charge, can interact
among themselves.  
 
%% figure 14 
\begin{figure} [tbh]
% \hspace*{5cm}
   \epsfxsize=3in
   \epsfysize=3in
   \begin{center}
    \mbox{\epsffile{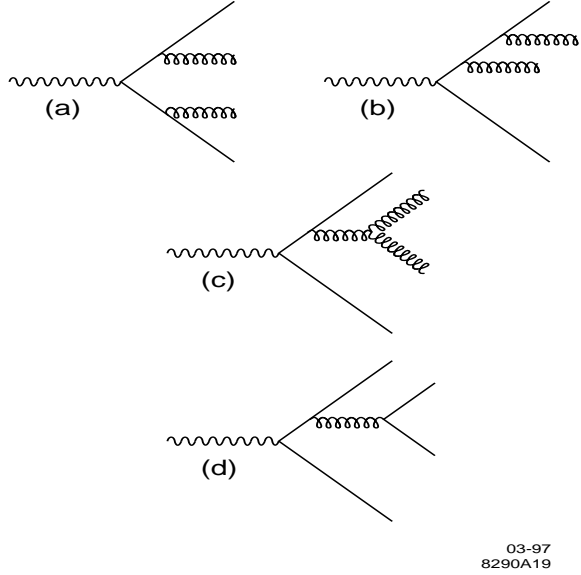}}
\end{center}
   \caption[]{
Tree-level Feynman diagrams for 4-jet production in \epa.
  }
\vskip .5truecm
\end{figure}

Now consider the formal properties of the SU(3) group. The group can be
characterised by constants known as Casimir factors that are defined by:

\vfill
\eject

\begin{eqnarray}
\Sigma_a(T^a T^a)_{ij} \qu  =  \qu \delta_{ij} C_F
\end{eqnarray}
\begin{eqnarray}
\Sigma_{a,b}(f^{abc} f^{abd}) \qu   =  \qu \delta^{cd} N_C
\end{eqnarray}
\begin{eqnarray}
Tr [T^a T^b] \qu  =  \qu \delta^{ab} T_F
\end{eqnarray}
The Casimir factors for several common groups are shown in Table~2.
We see that in the case of SU(3) $N_C$ corresponds to the now-familiar 
`number of colours' that we have already
encountered several times. The tree-level couplings appearing in Fig.~14 may be
classified in terms of the Casimir factors, as illustrated in Fig.~15.
The amplitude-squared corresponding to the Bremsstrahlung diagrams
(Fig.~14(a,b)) is proportional to $C_F$, 
that corresponding to g \ra \qq is proportional to
$T_F$, and that corresponding to the non-Abelian process g \ra gg 
is proportional to $N_C$.
 
\begin{table} [tbh]
\centering
\begin{tabular}{|l|c|c|c|} \hline
Group \qu \hfill & $N_C$ & \qu $C_F$ & \qu $T_F$      \\  \hline
U(1)  \hfill & $0$   & \qu 1            & \qu 1       \\
U(1)$_3$  \hfill & $0$   & \qu 1            & \qu 3   \\
SU(N) \hfill & $N$   & \qu $(N^2-1)/2N$ & \qu 1/2     \\
SU(3) \hfill & $3$   & \qu 4/3          & \qu 1/2     \\  \hline
\end{tabular}
\vskip .5truecm
\caption{Casimir factors for some common gauge groups.}
\end{table}
\vskip .5truecm
 
%% figure 15 
\begin{figure} [tbh]
% \hspace*{5cm}
   \epsfxsize=2.0in
   \epsfysize=3.0in
   \begin{center}
    \mbox{\epsffile{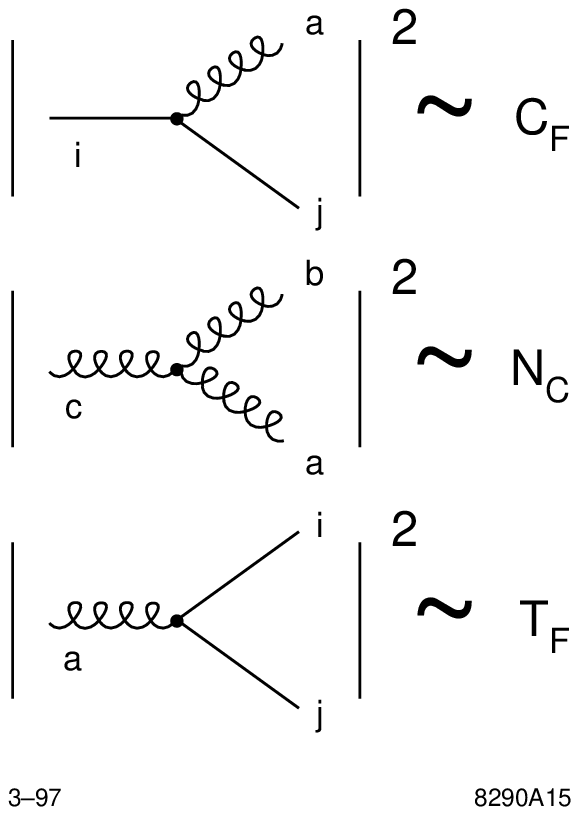}}
\end{center}
   \caption[]{
Casimir classification of tree-level QCD couplings.
  }
\end{figure}

It is interesting to consider whether the Casimir factors of SU(3) QCD
can be measured. Clearly nature does not deliver events corresponding to the
tree-level vertices shown in Fig.~15! Instead, one must write down the 
Feynman amplitudes for the 4-jet event diagrams shown in Fig.~14, add them to
those for 2- and 3-jet production at the same order of perturbation theory, 
and square them to derive the total hadronic cross section. 
The terms corresponding to 4-jet production can then be identified in a 
gauge-invariant manner, and yield a differential cross section of the form:
$$
{1\over \sigma_0} d\sigma^4\qu = \qu \left({\alp C_F \over \pi}\right)^2
\left[ F_A + \left(1-{1\over 2}{N_C\over C_F}\right) F_B +
{N_C \over C_F} F_C
\right]
$$
\begin{eqnarray}
\qu\qu\qu + \qu \left({\alp C_F \over \pi}\right)^2
\left[{T_F\over C_F} N_f F_D +
\left(1-{1\over 2}{N_C\over C_F}\right) F_E \right]
\end{eqnarray}
where $F_A$ $\dots$ $F_E$ are kinematical functions. We see that the overall
normalisation of the cross-section is proportional to $(\alp\,C_F)^2$,
and that the kinematical distribution of the four jets depends on the ratios
$N_C/C_F$ and $T_F/C_F$, which can hence in principle be measured.

The issue of jet definition will be discussed in detail in Section 4.4.
For now let us assume that 4-jet
events can be defined and measured in particle detectors, and that they can
be related meaningfully to the underlying 4-jet parton structure described by
eq.~(21).
Two important physical characteristics
underly the definition of 4-jet observables that are sensitive to 
$N_C/C_F$ and $T_F/C_F$: the first is that the two jets resulting from the
primary quark and antiquark produced in the \z0 decay tend to be more energetic
than the jets produced by the two radiated gluons or the radiated \qq; the
second is that in the `non-Abelian' process (Fig.~14c) the two gluons tend to 
be produced in the plane of the primary quark and antiquark, whereas in the
`Abelian' process (Fig.~14d) the radiated quark and antiquark tend to be 
produced along an axis normal to this plane~\cite{NR}. 

With this in mind,
a number of 4-jet observables that are potentially sensitive to the ratios of
Casimir factors have been proposed over the years. If one orders and labels 
the four jets in an event in terms of their momenta (or energies) such that
$p_1>p_2>p_3>p_4$ one can define the Bengtsson-Zerwas angle~\cite{BZ} (Fig.~16):
\begin{eqnarray}
\cos\chi_{BZ}\quad\propto\quad(\vec{p_1}\times \vec{p_2})\cdot
(\vec{p_3}\times \vec{p_4})
\end{eqnarray}
and the Nachtmann-Reiter angle~\cite{NR} (Fig.~17):
\begin{eqnarray}
\cos\theta^*_{NR}\quad\propto\quad(\vec{p_1}- \vec{p_2})\cdot
(\vec{p_3}- \vec{p_4}).
\end{eqnarray}
The sensitivity of these observables is illustrated in Fig.~18,
where the distributions of these angles are shown for SU(3) QCD, as well as for 
a straw-person U(1)$_3$ Abelian model of strong interactions, and are compared
with L3 data~\cite{lthreeNR}; the Abelian model is clearly excluded. 

%% figure 16 
\begin{figure}  [tbh]
% \hspace*{5cm}
   \epsfxsize=3.0in
   \epsfysize=3.0in
   \begin{center}
    \mbox{\epsffile{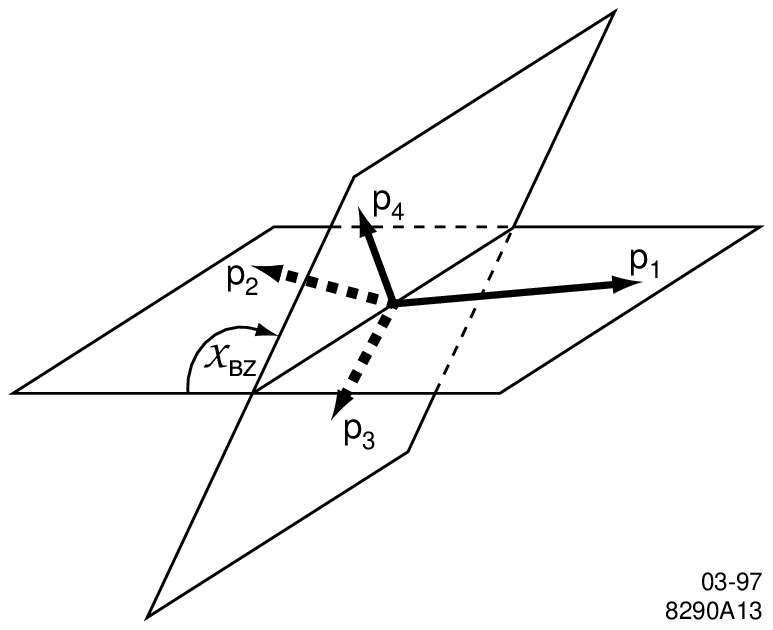}}
\end{center}
   \caption[]{
Definition of the Bengtsson-Zerwas angle.
  }
\end{figure}

%% figure 17 
\begin{figure}  [tbh]
%  \hspace*{5cm}
   \epsfxsize=3.0in
   \epsfysize=2.5in
   \begin{center}
    \mbox{\epsffile{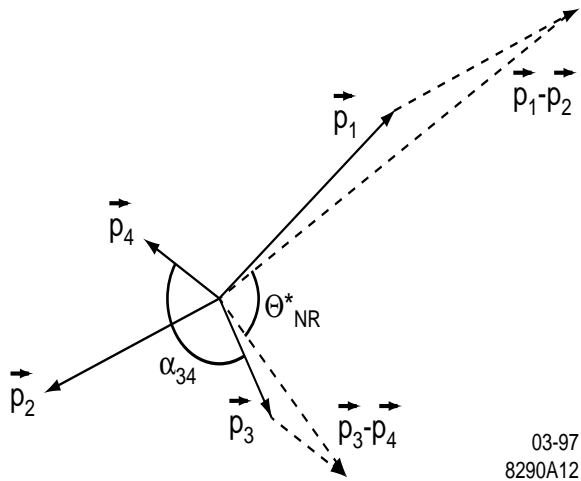}}
\end{center}
   \caption[]{
Definition of the Nachtmann-Reiter angle.
  }
\end{figure}

%% figure 18 
\begin{figure} [tbh]
% \hspace*{5cm}
   \epsfxsize=4.0in
   \epsfysize=5.5in
   \begin{center}
    \mbox{\epsffile{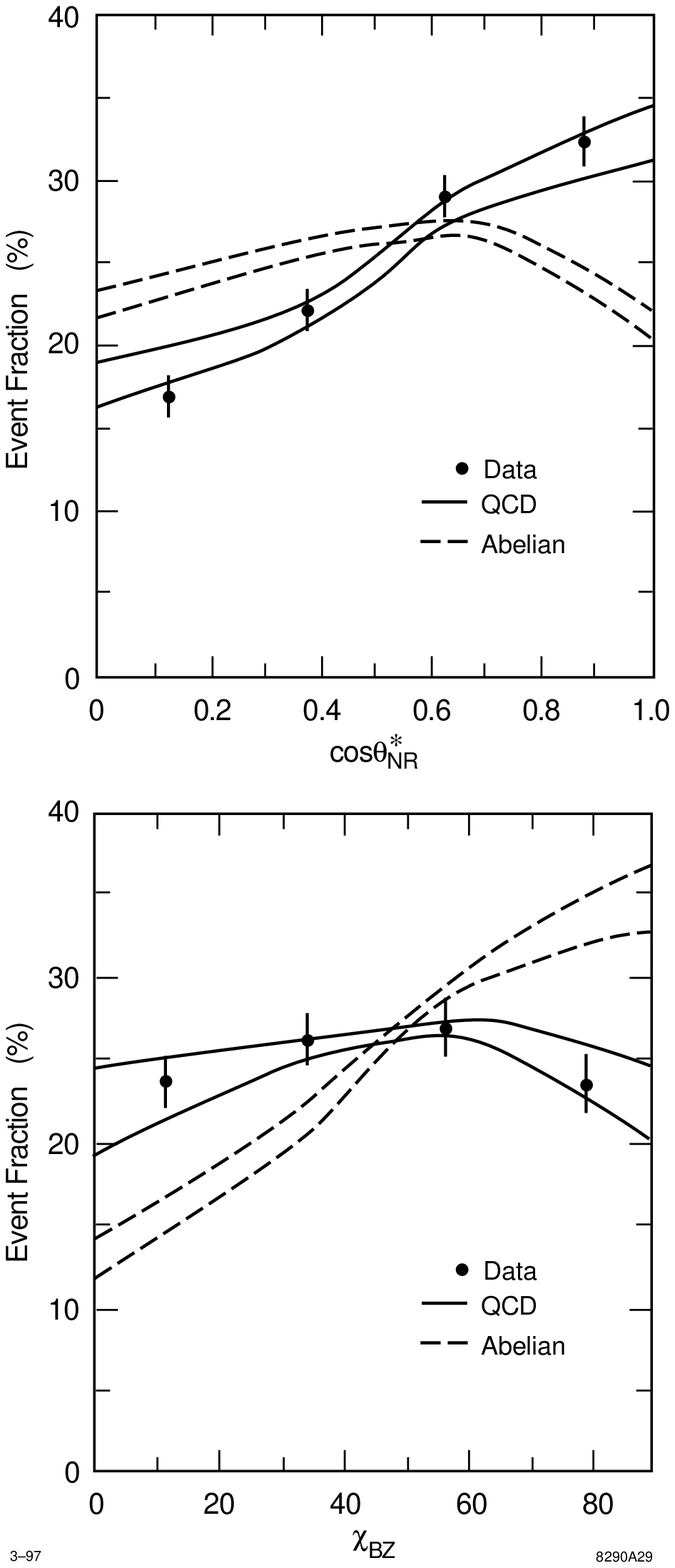}}
\end{center}
   \caption[]{
Illustration of the sensitivity of the Nachtmann-Reiter and Bengtsson-Zerwas
angles to non-Abelian effects and comparison with L3 data~\cite{lthreeNR}.
  }
\vskip .5truecm
\end{figure}

A more recent analysis by OPAL~\cite{opalnr} is summarised in Fig.~19; 
here a simultaneous fit 
was performed to the Nachtmann-Reiter and Bengtsson-Zerwas angle
distributions, as well as to the angle $\alpha_{34}$ between jets 3 and 4.
The resulting values of $N_C/C_F$ and $T_F/C_F$ are displayed in Fig.~20,
where they are compared with the expectations from numerous gauge groups.
The SU(3) QCD expectation is clearly in good agreement with the data.
The expectations from several other gauge models, such as SU(4), Sp(4) and
Sp(6), also appear to be compatible with the experimental results.
Note, however, that none of these models contains three colour degrees of
freedom for quarks, and hence all can be ruled out on that basis. Besides
SU(3), only the U(1)$_3$ and SO(3) models contain three quark colours, but
both are inconsistent with the measured values of $N_C/C_F$ and $T_F/C_F$.
The results shown in Fig.~20 hence yield the remarkable conclusion that SU(3)
is the only known viable gauge model for strong interactions.

%% figure 19 
\begin{figure} [tbh]
% \hspace*{5cm}
   \epsfxsize=5.0in
   \epsfysize=6.0in
   \begin{center}
    \mbox{\epsffile{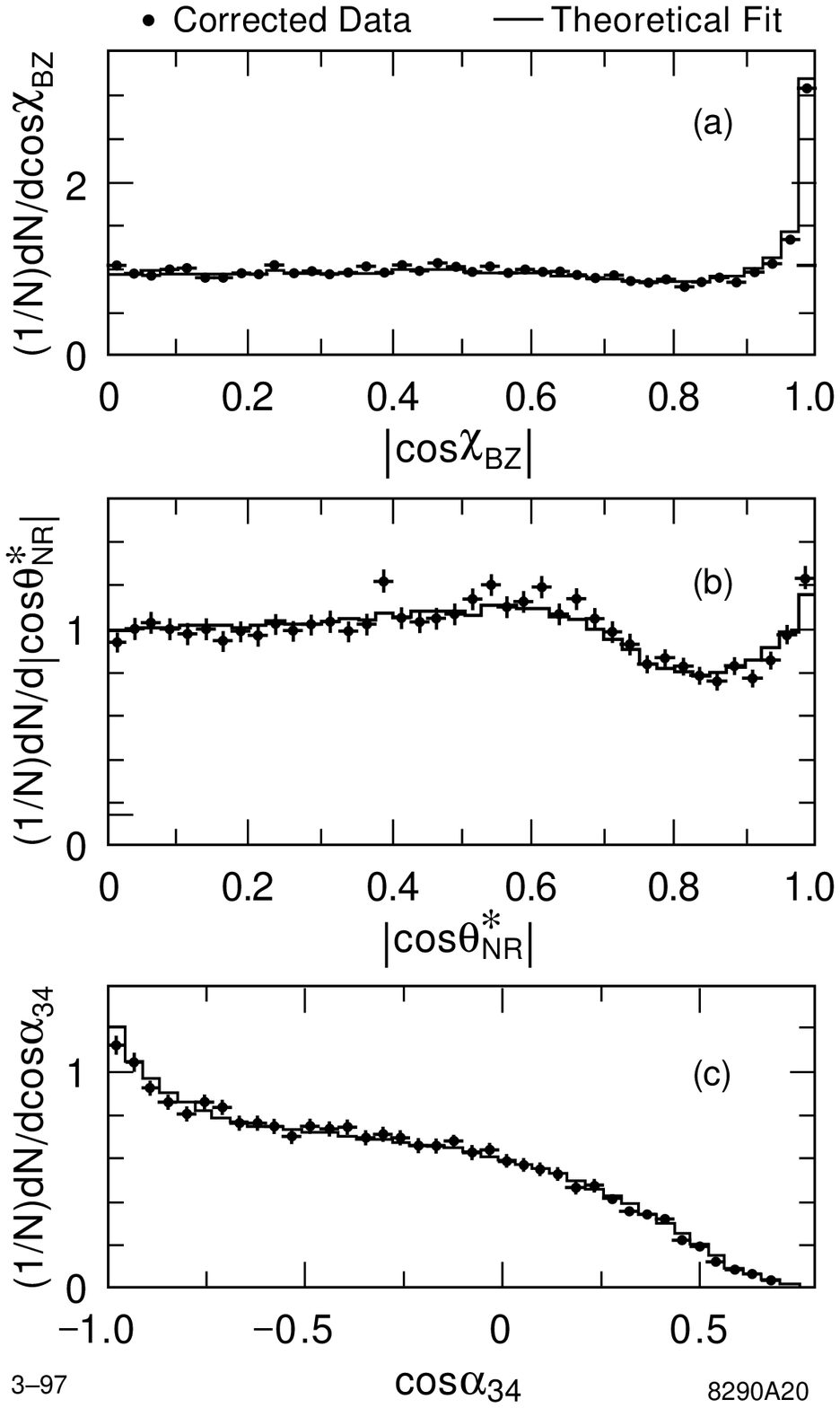}}
\end{center}
   \caption[]{
Simultaneous fit of QCD to OPAL measurements of $\chi_{BZ}$, 
$\theta_{NR}$ and $\alpha_{34}$~\cite{opalnr}.
  }
\end{figure}

%% figure 20
\begin{figure} [tbh] 
% \hspace*{5cm}
   \epsfxsize=4.0in
   \epsfysize=4.0in
   \begin{center}
    \mbox{\epsffile{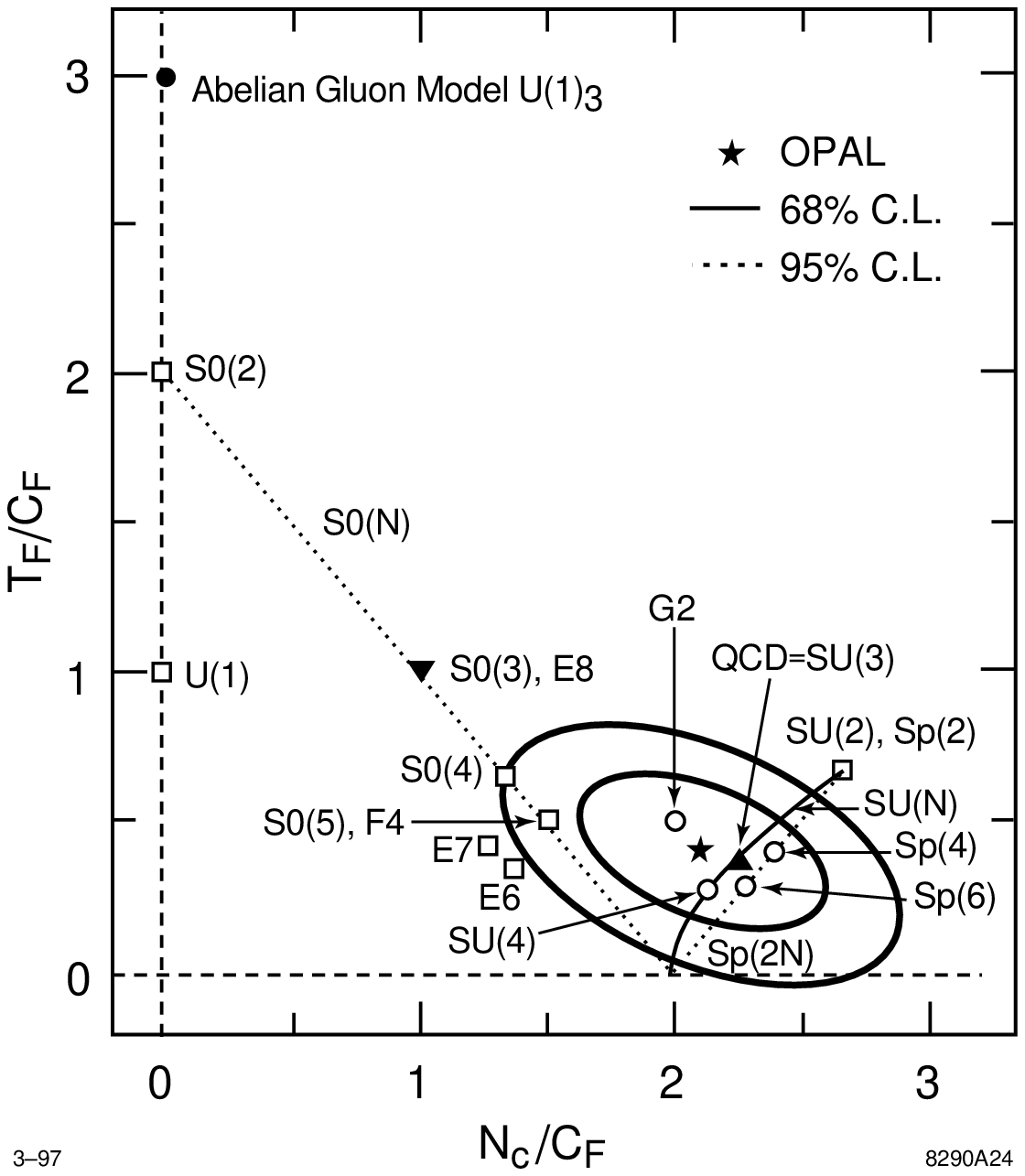}}
\end{center}
   \caption[]{
The $T_F/C_F$ vs. $N_C/C_F$ plane showing the fitted values derived 
from Fig.~19, as well as the expectations from numerous gauge 
groups~\cite{opalnr}; QCD is in good agreement with the data.
  }
\vskip .5truecm
\end{figure}

Recalling the `essential feature' of QCD that the ggg vertex must exist, we
see from Fig.~20 that the non-zero measured value of $N_C/C_F$ provides
direct evidence for its contribution to 4-jet production. Now consider the
existence of the gggg vertex; it should come as no surprise that we need to
study events of yet higher jet multiplicity in order to be sensitive to it.
The tree-level Feynman diagrams for 5-jet production in \epa are shown in
Fig.~21; the gggg vertex can be seen in the two diagrams just
left of centre on the bottom row. 

%% figure 21
\begin{figure} [tbh]
% \hspace*{5cm}
   \epsfxsize=5.0in
   \epsfysize=4.5in
   \begin{center}
    \mbox{\epsffile{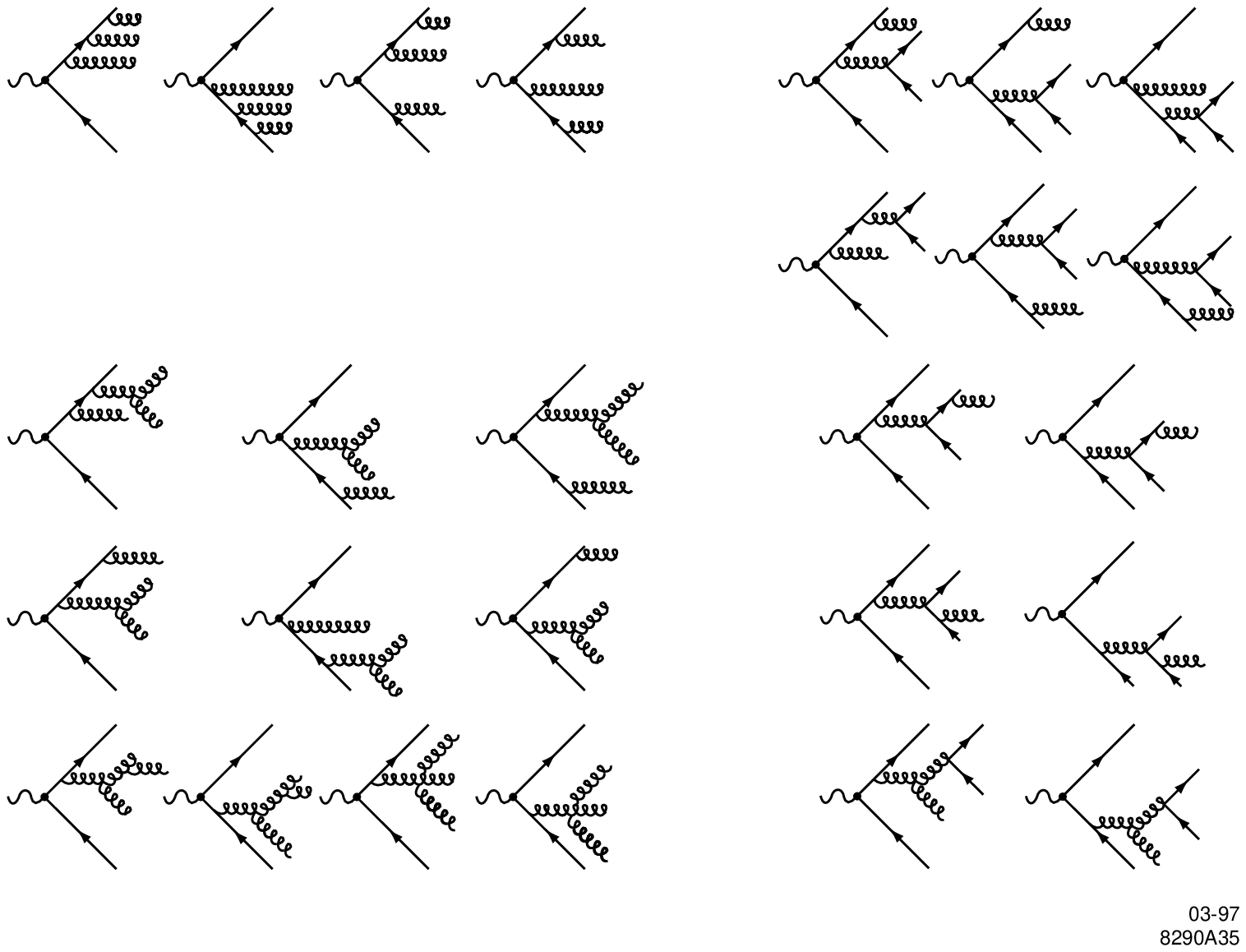}}
\end{center}
   \caption[]{
Tree-level Feynman diagrams for 5-jet production in \epa.
  }
\vskip .5truecm
\end{figure}

Performing a similar exercise to that for
the 4-jet cross section one finds:

\begin{eqnarray}
{1\over \sigma_0} d\sigma^5\qu = \qu {1\over \sigma_0} d\sigma^{2q3g}\qu
+ \qu {1\over \sigma_0} d\sigma^{4q1g},
\end{eqnarray}
The first term contributes about 85\% of the 5-jet cross section, and may be
written:
\begin{eqnarray}
{1\over \sigma_0} d\sigma^{2q3g}\qu = \qu
\left({\alp C_F \over \pi}\right)^3
\left[\;G_A \qu + \qu {N_C\over C_F} G_B \qu + \qu
\left({N_C\over C_F}\right)^2 G_C \;\right]
\end{eqnarray}
where 
$G_A$, $G_B$ and $G_C$ are kinematical functions. The contribution of the gggg
vertex is represented by the last term in eq.~(25), 
which is proportional to $(N_C/C_F)^2$.
We have just seen that $N_C/C_F$ must be non-vanishing
in order to describe the 4-jet data, so that the existence of the gggg vertex
is absolutely required in QCD in order for the theory to be gauge-invariant and
self-consistent.
Pushing pedagogy to its limits, however, one can still ask if the data actually
{\it require} the existence of the gggg vertex, from a phenomenological
point-of-view. One can therefore define a set of \adhoc 5-jet correlation
observables, such as those illustrated in Fig.~22~\cite{opalfive}. The
measured distributions of the five of these observables that are 
most sensitive to the $(N_C/C_F)^2$ term are shown in Fig.~23, from the OPAL
Collaboration~\cite{opalfive}.

%% figure 22
\begin{figure} [tbh]
% \hspace*{5cm}
   \epsfxsize=5.0in
   \epsfysize=5.0in
   \begin{center}
    \mbox{\epsffile{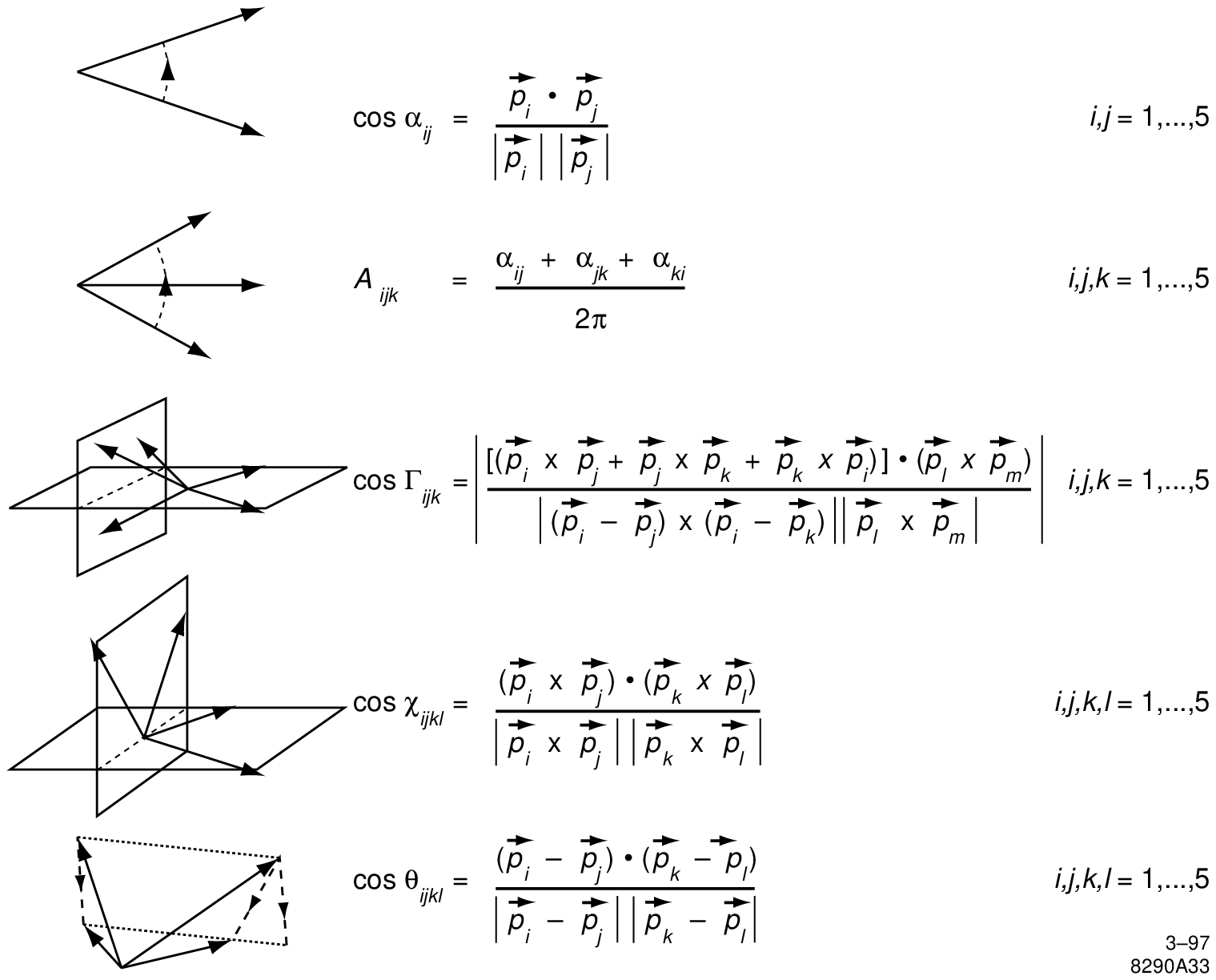}}
\end{center}
   \caption[]{
Illustration of correlation observables among the jets in 5-jet 
events~\cite{opalfive}.
  }
\vskip .5truecm
\end{figure}

%% figure 23
\begin{figure} [tbh]
% \hspace*{5cm}
   \epsfxsize=4.0in
   \epsfysize=5.0in
   \begin{center}
    \mbox{\epsffile{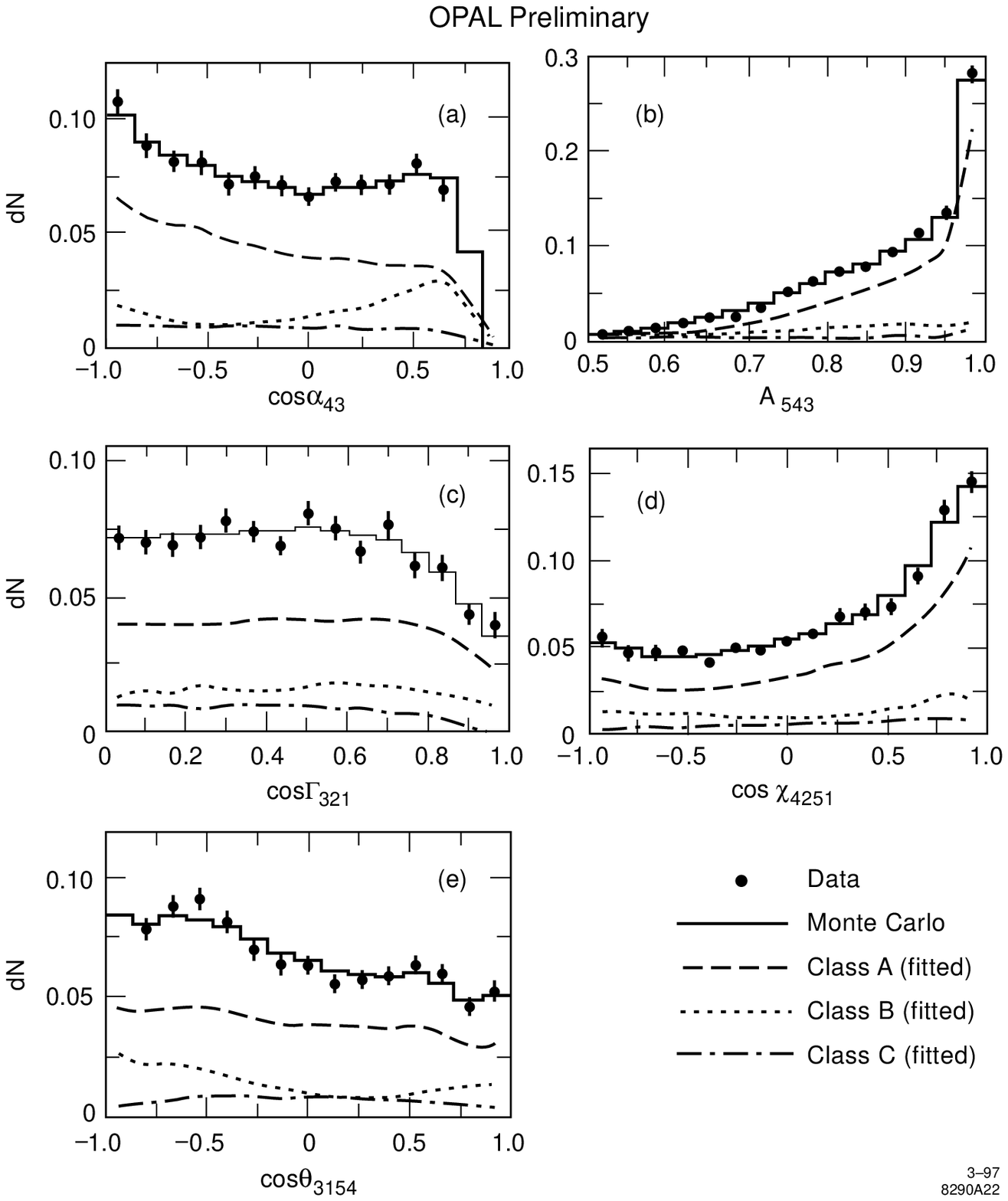}}
\end{center}
   \caption[]{
The five correlation observables in 5-jet events that are most sensitive to 
contributions from the gggg vertex~\cite{opalfive}.
  }
\vskip .5truecm
\end{figure}

Two possible strategies now present themselves for testing the
self-consistencey of QCD. One could fit inclusively the quantity $N_C/C_F$ to
the 5-jet data shown in Fig.~23 and compare it with the value determined 
from 4-jet
events; the results of such a comparison are shown in Fig.~24a; the 4-jet and
5-jet events clearly yield consistent results. A second possibility is to
fit phenomenologically only the gggg contribution proportional to
$(N_C/C_F)^2$; the results are shown in Fig.~24b. In the latter case the
error bars are large due to the small number of 5-jet events, as well as to
the large uncertainties on multijet production that arise from hadronisation
effects (see Section 4.4). The measured value of $(N_C/C_F)^2$ is clearly 
consistent with the QCD expectation 
of $(9/4)^2$ $\approx$ 5, but it is also consistent with zero, so that
the existence of the gggg vertex has not yet been established from a 
phenomenological point-of-view.

%% figure 24
\begin{figure} [tbh]
% \hspace*{5cm}
   \epsfxsize=3.0in
   \epsfysize=4.0in
   \begin{center}
    \mbox{\epsffile{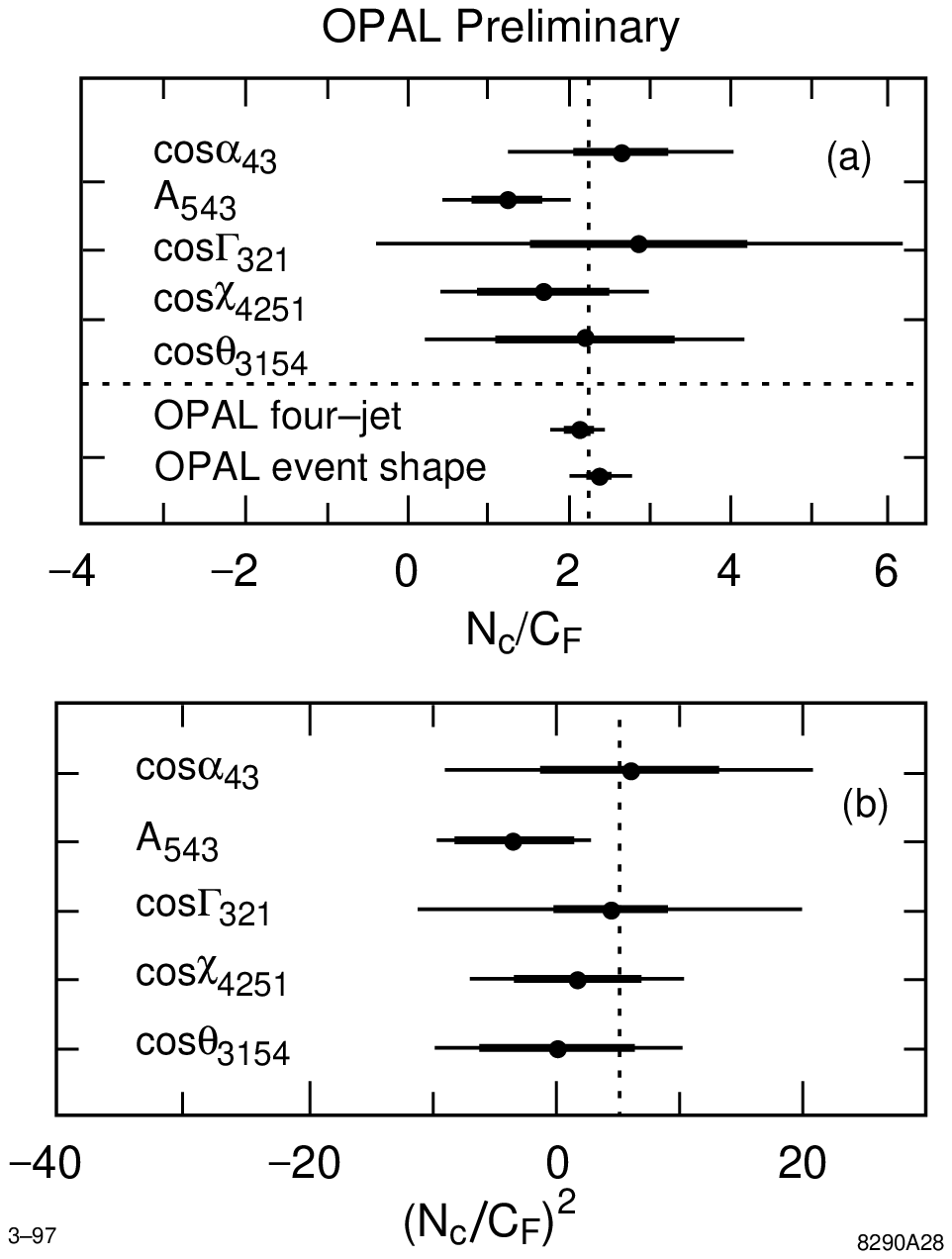}}
\end{center}
   \caption[]{
Measurements of (a) $(N_C/C_F)$ and (b) $(N_C/C_F)^2$ made using 5-jet events
at OPAL~\cite{opalfive}.
  }
\vskip .5truecm
\end{figure}
 
%\vskip 1truecm
\vfill
\eject
 
\noindent{\large \bf 3.4 Review of Strategy for QCD Tests}

\vskip .5truecm

\noindent
At this point we have seen that most of the
`essential features' of QCD have been established empirically, with the 
possible exception of the gggg coupling. Even in this case, 
given the existence of the ggg vertex, the gggg vertex must exist in QCD in
order for the theory to be gauge-invariant. The last 20 years of
hadronic-event studies at \ep colliders have hence established, in a  
qualitative sense, that the QCD Lagrangian is the correct one to describe 
strong interactions. At this point it therefore seems sensible to revise the
strategy for testing QCD.

Since the theory contains in principle only one free parameter, the
strong coupling \alp, QCD can be tested in a quantitative fashion
by measuring \alp in different processes
and at different hard scales $Q$. The precision of these measurements, and
the resulting degree of consistency among them, 
determine quantitatively the precision with which the theory has been tested.
This philosophy is directly analogous to that used to test the electroweak
theory by measuring a large number of observables that are sensitive to a few
key unknown parameters of the theory.
In addition to testing QCD, the
precise measurement of \alp allows constraints on
possible extensions to the Standard Model (SM) of elementary particles;
see \eg \cite{kane}. 
Measurements of \alp have been performed in \epa, hadron-hadron
collisions, and deep-inelastic lepton-hadron scattering,
covering a range of $Q^2$ from roughly 1 to $10^5$ GeV$^2$.
In the next section I shall describe the \ep measurements, 
and compare them with those made in other hard processes; for a review
of this field see~\cite{cracow}.

\vskip 1truecm
 
\noindent{\large \bf 4. Measurements of \alp in \ep Annihilation}

\vskip .7truecm
 
\noindent{\large \bf 4.1 Theoretical Considerations}

\vskip .5truecm

\noindent
An inclusive observable $X$ may be written schematically:
\begin{eqnarray}
X \qu=\qu X^{EW}\;(1\;+\;\delta^{QCD})
\end{eqnarray}
where $X^{EW}$ represents the electroweak contribution.
Since, with observables of this type,
\alp enters via the small QCD radiative correction, $\delta^{QCD}$,
a precise measurement of \alp generally requires a large data sample.
Observables can also be defined that
are directly proportional to $\delta^{QCD}$ and hence
potentially more sensitive to \alp. In either case
$\delta^{QCD}$ can be separated into perturbative and non-perturbative
contributions:
\begin{eqnarray}
\delta^{QCD}\qu=\qu \delta^{pert} + \delta^{non-pert}.
\end{eqnarray}
The perturbative contribution can in principle be calculated as a power
series in \alp, though in practice
the large number of Feynman diagrams involved renders a complete calculation
beyond the first few orders intractable. The non-perturbative contribution,
often called a `hadronisation correction' in \epa
or a `higher twist effect' in lepton-hadron scattering, is expected
to have the form of a series of inverse powers
of the physical scale (see section 5). 

In practice most QCD calculations of
observables are performed using finite-order perturbation theory, and
calculations beyond leading order depend on the
{\it renormalisation scheme} employed, implying a scheme-dependent
strong-interaction scale $\Lambda$. It is conventional to work in
the modified minimal subtraction scheme ($\overline{MS}$
scheme) \cite{msbar}, and to use the strong interaction scale
\lam for five active quark flavours.
If one knows \lam one may calculate the strong coupling \alp($Q^2$) from
the solution of the QCD renormalisation group equation \cite{ian}:

\begin{eqnarray}
\alp(Q^2)\quad=\quad{4\pi\over{\beta_0{\rm ln}(Q^2/\lam^2)}}
\lbrace \;1\;-\;{2\beta_1\over \beta_0^2}\;{{\rm ln(ln}(Q^2/\lam^2))
\over {\rm ln} (Q^2/\lam^2)}\;+\ldots\;\rbrace
\end{eqnarray}

\noindent
Because of the large data samples taken in \epa at the \z0 resonance,
it has become conventional to use as a yardstick \alpmzsq,
where $M_Z$ is the mass of the \z0 boson; $M_Z$ $\approx$ 91.2 GeV
\cite{mz}.
Tests of QCD can therefore be quantified in terms of the consistency
of the values of \alpmzsq measured in different experiments. The 
`QCD-challenged' reader may like to think of \alpmzsq as being `the
sin$^2\theta_W$ of strong interactions'.
 
In \epa \alpmzsq has been measured from inclusive observables relating to the
\z0 lineshape and to hadronic decays of the $\tau$ lepton, as well as from
jet-related hadronic event shape observables, and scaling violations in
inclusive hadron fragmentation functions.

\vskip 1truecm
 
\noindent{\large \bf 4.2 $R$ and the \z0 Lineshape}

\vskip .5truecm

\noindent
For the inclusive ratio $R$ =
$\sigma$(\ep \ra hadrons)/$\sigma$(\ep \ra $\mu^+\mu^-$),
the SM electroweak
contributions are well understood theoretically and the perturbative
QCD series has been calculated
up to \oalpc \cite{thirdo} for massless quarks and up to \oalpsq
including quark mass effects \cite{kuhn}; the large
size of the \oalpc term is potentially a cause for concern about the
degree of convergence of the series.
Closely-related observables at the \z0 resonance are:
 
%\vskip .2truecm
\vfill
\eject
 
$\bullet$
the \z0 total width, $\Gamma_Z$
 
$\bullet$
the pole cross section, $\sigma_h^0\;\equiv\;
12\pi\Gamma_{ee}\Gamma_{had}/{M_Z^2\Gamma_Z^2}$
 
$\bullet$
the ratio of hadronic to leptonic \z0 decay branching
ratios $R_l\;\equiv\;\Gamma_{had}/\Gamma_{ll}$
 
\vskip .2truecm

\noindent
which all depend on the \z0 hadronic width:

\begin{eqnarray}
\Gamma_{had}\qu=\qu1.671\left(1\;+\;a_1
\left({\alp\over\pi}\right)
\;+\;a_2\left({\alp\over\pi}\right)^2
\;+\;a_3\left({\alp\over\pi}\right)^3\;+\qu\ldots\right)
\end{eqnarray}

where: $a_1$ = 1, $a_2$ = 0.75 and $a_3$ = $-15.3$.
In these cases
the non-perturbative contributions are expected to be O(1/$M_Z$) and
are usually ignored.
A concern is that recent measurements of observables that probe
the electroweak couplings of the \z0 to b and c quarks deviate
slightly from SM expectations \cite{blondel}. Since these couplings
must be known in order to extract \alpmzsq,
this effect, whatever its origin,
is a potential source of bias \cite{ian}. Further analysis
is in progress from the SLC and LEP experiments and the situation is not
yet resolved.
 
Proceeding nonetheless, the procedure adopted \cite{blondel}
is to perform a global SM fit
to a panoply of electroweak data that includes the W and top quark masses
as well as the \z0 observables relating to the lineshape,
left-right production asymmetry,
decay fermion forward-backward asymmetries, branching ratios
to heavy quarks, and $\tau$ polarisation. The free parameters are
the Higgs mass, $M_{Higgs}$, which contributes to $X^{EW}$, and \alpmzsq. 
Data presented at the 1996 summer conferences yield
the results shown in Fig.~25~\cite{blondel}, from which
the positively-correlated results $M_{Higgs}=149^{+190}_{-82}$ GeV and
\begin{eqnarray}
\alpmzsq\qu=\qu0.1202\pm0.0033\;\; ({\rm exp.})
\end{eqnarray}
are obtained. The \alpmzsq value is lower than the corresponding results
presented at the 1995 conferences \cite{renton}, \alpmzsq =
$0.123\pm0.005$, and at the
1994 conferences, \alpmzsq $0.125\pm0.005$ \cite{schaile}, whose large
central values were partly responsible for a supposed discrepancy
between `low-$Q^2$' and `high-$Q^2$' \alpmzsq measurements
\cite{shifman}.
The change between 1995 and 1996 is due to a combination of
shifts in the values of the \z0 lineshape parameters, redetermined in
light of the recalibration of the LEP beam energy due to the
`TGV effect' \cite{blondel}, and a change in the central value of
$M_{Higgs}$ at which \alpmzsq is quoted, from 300 GeV (1995)
to the fitted value 149 GeV (1996). A detailed study of
theoretical uncertainties implies \cite{chyla} that they contribute at
a level substantially below $\pm0.001$.
Since data-taking at the \z0 resonance has now been completed at the
LEP collider the precision of this result is not expected to improve
further.

%% figure 25
\begin{figure} [tbh]
% \hspace*{5cm}
   \epsfxsize=4.0in
   \epsfysize=4.0in
   \begin{center}
    \mbox{\epsffile{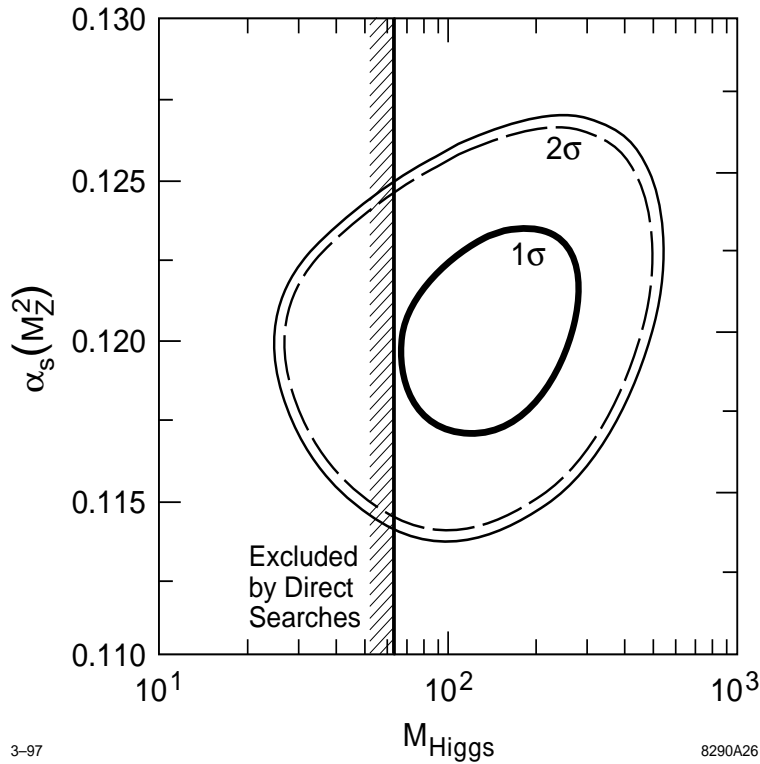}}
\end{center}
   \caption[]{
Results of a global fit of the Standard Model to electroweak 
observables~\cite{blondel};
the 1- and 2-standard deviation contours are shown in the \alpmzsq vs.
$M_{Higgs}$ plane.
  }
\vskip .5truecm
\end{figure}

%\vskip 1truecm
\vfill
\eject
 
\noindent{\large \bf 4.3 Hadronic $\tau$ Decays}

\vskip .5truecm

\noindent
 An inclusive quantity similar to $R$ is the ratio $R_{\tau}$ of
hadronic to leptonic decay branching ratios, $B_h$ and $B_l$
respectively, of the $\tau$ lepton:
\begin{eqnarray}
R_{\tau}\qu \qu \equiv \qu {B_h \over B_l} \qu
= \qu {1-B_e-B_{\mu} \over B_e}
\end{eqnarray}
where $B_e$ and $B_{\mu}$ can either be
measured directly, or deduced from a measurement of the $\tau$
lifetime $\tau_{\tau}$.
In addition, a family of observables known as `spectral
moments' of the invariant mass-squared
$s$ of the hadronic system has been proposed \cite{spectral}:
\begin{eqnarray}
R_{\tau}^{kl} \qu
\equiv \qu {1\over B_e} \int_0^{M_{\tau}^2} ds\left(
1-{s\over M_{\tau}^2}\right)^k \left({s\over M_{\tau}^2}\right)^l
{d B_h \over ds}
\end{eqnarray}
where $M_{\tau}$ is the $\tau$ mass. In this case the integrand
can be measured independently of $B_e$. It is easily seen that
$R_{\tau}$ = $R_{\tau}^{00}$.
 
$R_{\tau}$ and $R_{\tau}^{kl}$
have been calculated perturbatively up to \oalpc.
However, because  $M_{\tau}$ $\sim$ 1 GeV one expects (eq.~(28))
\alp($M_{\tau}$) $\sim$ 0.3 and
it is not \apriori obvious that the perturbative calculation can be
expected to be reliable, or that the non-perturbative contributions of 
O($1/M_{\tau}$) will be small.
In recent years a large theoretical effort has been devoted to this
subject; see \eg \cite{spectral,bnplp,neubert}.
 
The ALEPH Collaboration derived $R_{\tau}$ from its measurements of
$B_e$, $B_{\mu}$, and $\tau_{\tau}$, and also
measured the (10), (11), (12), and (13) spectral moments.
A combined fit yielded~\cite{alconf} \alpmzsq =
$0.124\pm0.0022\pm0.001$, where the first error receives equal
contributions from experiment and theory, and the second derives from
uncertainties in evolving \alp across the c and b thresholds.
The OPAL Collaboration measured $R_{\tau}$ from
$B_e$, $B_{\mu}$, and $\tau_{\tau}$, and derived \cite{optau}
\alpmzsq=0.1229$^{+0.0016}_{-0.0017}$ (exp.) $^{+0.0025}_{-0.0021}$
(theor.). The CLEO Collaboration measured the same four spectral
moments as ALEPH and also derived $R_{\tau}$ using 1994 Particle
Data Group values for $B_e$, $B_{\mu}$ and $\tau_{\tau}$.
A combined fit yielded \cite{cltau}
\alpmzsq = $0.114\pm0.003$. This central value is slightly lower than
the ALEPH and OPAL values. If more recent
world average values of $B_e$ and $B_{\mu}$ are used CLEO obtains
a higher central \alpmzsq value~\cite{cltau}.
Averaging the second CLEO result and the ALEPH and OPAL
results by weighting with the experimental
errors, assuming they are uncorrelated,
yields:
\begin{eqnarray}
 \alpmzsq\quad =\quad 0.122 \pm 0.001\;\; ({\rm exp.})\;\; \pm0.002\;\; 
({\rm theor.}).
\end{eqnarray}
This is nominally a very precise measurement, although recent studies have 
ruggested that additional theoretical uncertainties 
may be as large as $\pm0.006$~\cite{alttau}.

\vskip 1truecm
 
\noindent{\large \bf 4.4 Hadronic Event Shape Observables}

\vskip .5truecm

\noindent 
As discussed in Section 3.2, in \epa the rate of 3-jet production:
\begin{eqnarray}
R_3\quad\equiv\quad\frac{\sigma_{3-jet}}{\sigma_{had}} 
\end{eqnarray}
is directly
proportional to \alp and can hence be used to determine \alp. In order to make
a meaningful measurement that can be compared with those just discussed one
must calculate $R_3$ to at least next-to-leading order in \alp, \ie to
\oalpsq. The relevant contributing Feynman diagrams are shown in Fig.~26; these
form the basis of the \oalpsq calculation of $R_3$~\cite{ert,kl,kn}.

%% figure 26
\begin{figure} [tbh]
% \hspace*{5cm}
   \epsfxsize=4.0in
   \epsfysize=5.0in
   \begin{center}
    \mbox{\epsffile{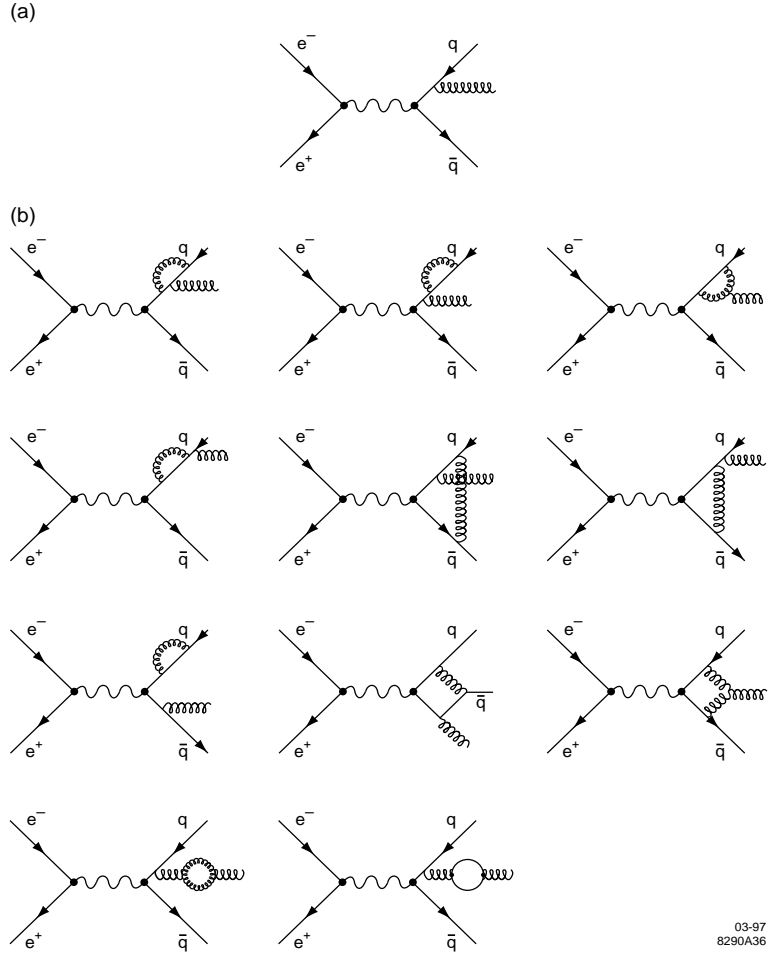}}
\end{center}
   \caption[]{
Tree-level and one-loop Feynman diagrams that contribute to 3-jet observables up
to \oalpsq in QCD perturbation theory.
  }
\end{figure}

\vskip 1truecm
 
\noindent{\large \bf 4.4.1 Definition of Jets and Event Shape Measures}

\vskip .5truecm

\noindent 
The task is, in principle, straightforward. One must count the number of
3-jet events and divide by the total number of hadronic events to obtain
$R_3$, then compare with the theoretical prediction to obtain \alp.
However, it is immediately apparent that one cannot simply define the jet
multiplicity of events on the basis of a visual inspection! On the experimental
side, the classic `Mercedes-Benz' 3-jet event measured in a detector is rather 
rare; many events contain broad
particle flows that might be classified as a single jet by one observer but as
two or more jets by another observer. Moreover, in QCD the Bremsstrahlung
spectrum of parton radiation peaks at small angles and is continuous. Hence
even theoretically the issue of when a radiated parton is sufficiently 
energetic, and at a sufficiently wide angle relative to its parent, so as to
be resolved as a separate jet is not without ambiguity. 
After due Cartesian deliberation one pragmatically concludes that one needs 
an algorithmic definition of a jet that can be applied to hadrons recorded in
a detector, as well as to partons in perturbative QCD calculations, and a
sensible recipe to translate between the two.

A convenient solution is provided by iterative clustering algorithms in which
a measure $y_{ij}$, such as invariant mass-squared/$Q^2$, is calculated
for all pairs of particles $i$ and $j$ in an event, 
and the pair with the smallest $y_{ij}$
is combined into a single `particle'. This process is repeated until all pairs
have $y_{ij}$ exceeding a value
$y_c$, and the jet multiplicity of the event is defined as the
number of particles remaining. For a sample of events the $n$-jet rate $R_n$ 
is then defined as the number of $n$-jet events divided by the total number of
events. This number is not a constant, but rather depends on the choice of
algorithm and on the $y_c$ value. The $y_c$-dependence is illustrated in 
Fig.~27 for jets defined using the JADE algorithm~\cite{JADE} applied to SLD
data~\cite{sldjet}. 
One can think of $y_c$ as the `jet resolution' scale. Large $y_c$ values
correspond to poor eyesight, most events look 2-jet-like, and hence $R_2$
$\approx$ 1.  Small $y_c$ values
correspond to good eyesight, a richer jet structure is discernible, and $R_3$
and $R_4$ are non-zero. It should be noted, however, that from an operational
point-of-view the data points shown in Fig.~27 are awkward to handle in that
they are correlated between different $y_c$ values. A more convenient observable
is the {\it differential} 2-jet rate:
\begin{eqnarray}
D_2(y_c)\;\equiv\;(R_2(y_c) - R_2(y_c-\Delta y_c))/\Delta y_c
\end{eqnarray}
which is a measure of the rate of events that {\it change} their 
classification between 2-jet-like and $\geq3$-jet-like as $y_c$ is varied
across the range $\Delta y_c$. $D_2(y_c)$ is illustrated in Fig.~28.
                      
%% figure 27
\begin{figure} [tbh] 
% \hspace*{5cm}
   \epsfxsize=5.0in
   \epsfysize=4.0in
   \begin{center}
    \mbox{\epsffile{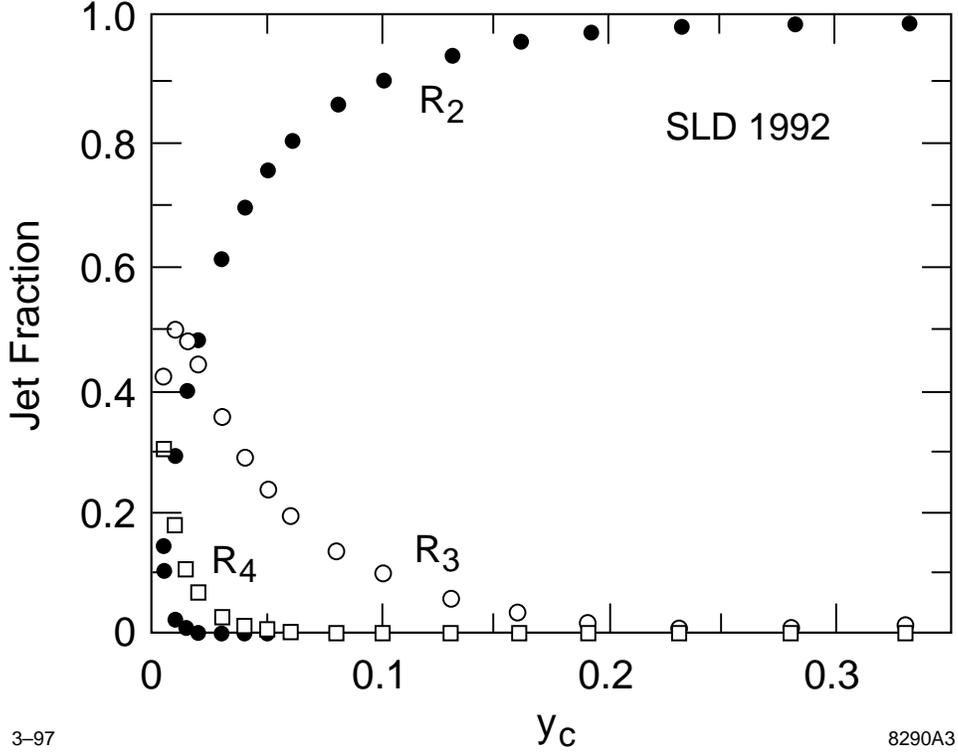}}
\end{center}
   \caption[]{
Dependence on the jet resolution parameter $y_c$ of the $n$-jet rates $R_n$
measured using SLD data~\cite{sldjet} with the JADE algorithm.
  }
\vskip .5truecm
\end{figure}
                      
%% figure 28
\begin{figure} [tbh] 
% \hspace*{5cm}
   \epsfxsize=5.0in
   \epsfysize=4.0in
   \begin{center}
    \mbox{\epsffile{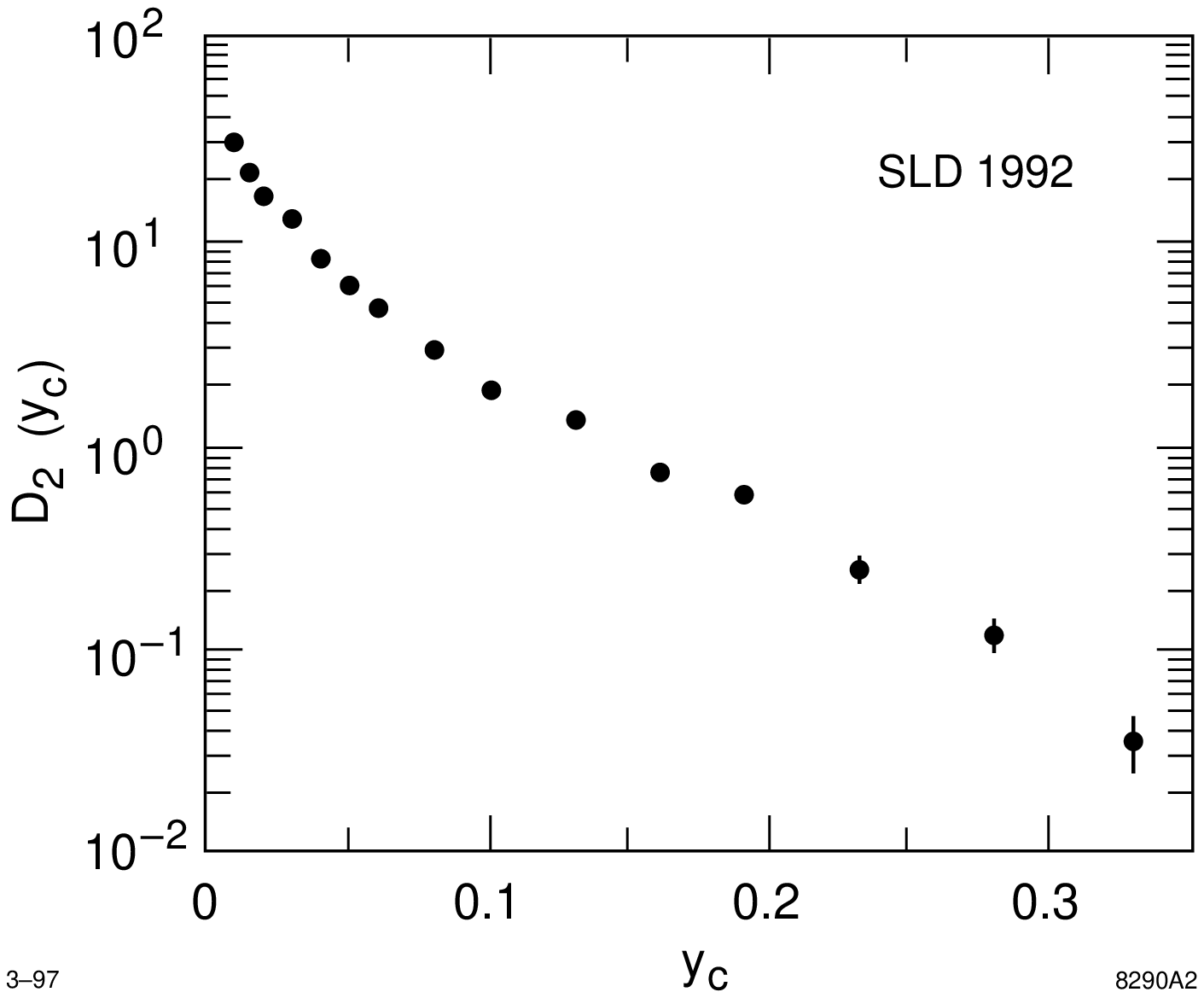}}
\end{center}
   \caption[]{
Dependence on the jet resolution parameter $y_c$ of the differential 2-jet 
rate $D_2$~\cite{sldjet}.
  }
\vskip .5truecm
\end{figure}

In fact several variations of the JADE algorithm have been 
suggested~\cite{bkssjet}; these differ in the definition of the resolution
measure $y_{ij}$, and/or in the
`recombination scheme' prescription for combining two particles that are
unresolvable. A full discussion is beyond the scope of these lectures, but it
is important to note that the `E', `E0', `P' and `P0' variations of the
JADE algorithm, as well as the `Durham' (`D') and
`Geneva' (`G') algorithms, are all {\it collinear- and infra-red-safe} 
observables,
which, for our purposes, means that they can be calculated in perturbative
QCD~\cite{soper}.

More generally one can define other infra-red- and collinear-safe measures of 
the topology of hadronic final states; a list of 15 such observables 
is given in Table~3. Thrust has already been encountered in Section~3 and is
related to the longitudinal momentum flow in events:
\begin{eqnarray}
T = \mbox{max}\frac{\sum_i\mid \vec{p}_i \cdot \vec{n}_T \mid}
{\sum_i\mid \vec{p}_i \mid},
\label{eqthrust}
\end{eqnarray}
where $\vec{p}_i$ is the momentum vector of particle $i$,
and $\vec{n}_T$ is the thrust axis to be determined.
It is useful to define $\tau \equiv 1 - T$.
For back-to-back two-parton final states $\tau$ is zero,
while $0 \leq \tau \leq \frac{1}{3}$ for planar three-parton final states.
Spherical events have $\tau = \frac{1}{2}$.
An axis $\vec{n}_{maj}$ can be found to maximize the momentum sum transverse
to $\vec{n}_T$, and
an axis $\vec{n}_{min}$ is defined to be perpendicular to the two
axes $\vec{n}_T$ and $\vec{n}_{maj}$. The variables thrust-major $T_{maj}$
and thrust-minor $T_{min}$
are obtained by replacing $\vec{n}_T$ in Eq.~(\ref{eqthrust})
by $\vec{n}_{maj}$ or $\vec{n}_{min}$, respectively. The oblateness $O$ is then
defined by \cite{oblate}
\begin{eqnarray}
O = T_{maj} - T_{min}.
\label{oblate}
\end{eqnarray}
Other measures are related to jet masses, 
and energy-energy correlations between particles;
for a discussion see \eg \cite{sldalp}.

\begin{table}[tbh]
\centering
\begin{tabular}{|l|c|} \hline
Observable & symbol \\ \hline
1 -- Thrust \hfill & $\tau$      \\
Heavy jet mass \hfill & $\rho$   \\
Jet broadening: \hfill &              \\
$\quad$ Total           \hfill & $B_T$  \\
$\quad$ Wide \hfill            & $ B_W$  \\
Oblateness \hfill      & $ O$    \\
C-parameter \hfill     & $ C$    \\
Differential jet rates: \hfill &      \\
$\quad D_2(y_c)$ = ${\Delta R_2(y_c)\over{\Delta y_c}}$ &  E \\
                       &  E0   \\
                       &  P    \\
                       &  P0   \\
                       &  D    \\
                       &  G    \\
Energy-energy correlations \hfill & $ EEC$ \\
Asymmetry of EEC \hfill    & $ AEEC$       \\
Jet cone energy fraction \hfill  & $JCEF$ \\ \hline
\end{tabular}
\vskip .5truecm
\caption{
Fifteen infra-red- and collinear-safe measures of the topology of \ep hadronic
final states.
}
\end{table}
\vskip .5truecm

The observables are all constructed to be
directly proportional to \alp at leading order, and
so are potentially sensitive measures of the strong coupling.
The \oalpsq QCD prediction for each of these observables $X$ can be 
written~\cite{kn}:
\begin{eqnarray}
{1\over\sigma_0} {{\rm d}\sigma\over{\rm d}X}
\quad=\quad A(X)\;\left({\alp\over 2\pi}\right)
\quad+\quad B(X)\left(
{\alp\over 2\pi}\right)^2
\end{eqnarray}
so that \alp can be determined from each. Though these observables are
intrinsically highly correlated, by using all 15 to study \alp one is
attempting to maximise the use of the information in complicated multi-hadron
events, and in some sense is making a more demanding test of QCD than by using
only one or two observables. Moreover, it will be seen that the study of many
observables is essential, as it may expose systematic effects. Finally, the \alp
determination from hadronic event shape observables is based on the 
information content within 3-jet-like events, and is essentially uncorrelated 
with the measurements from the \z0 lineshape which are based on event-counting
of predominantly 2-jet-like final states.

The technology of this approach has been developed
over the past 15 years of analysis at the PETRA, PEP, TRISTAN, SLC and
LEP colliders, so that the method is considered to be well understood
both experimentally and theoretically. Note, however, that before they can be
compared with perturbative QCD predictions,
it is necessary to correct the measured distributions for any
bias effects originating from the detector
acceptance, resolution, and inefficiency, as well
as for the effects of initial-state radiation and
hadronisation, to yield `parton-level' distributions.
                      
%% figure 29
\begin{figure} [tbh]
% \hspace*{5cm}
   \epsfxsize=5.0in
   \epsfysize=4.0in
   \begin{center}
    \mbox{\epsffile{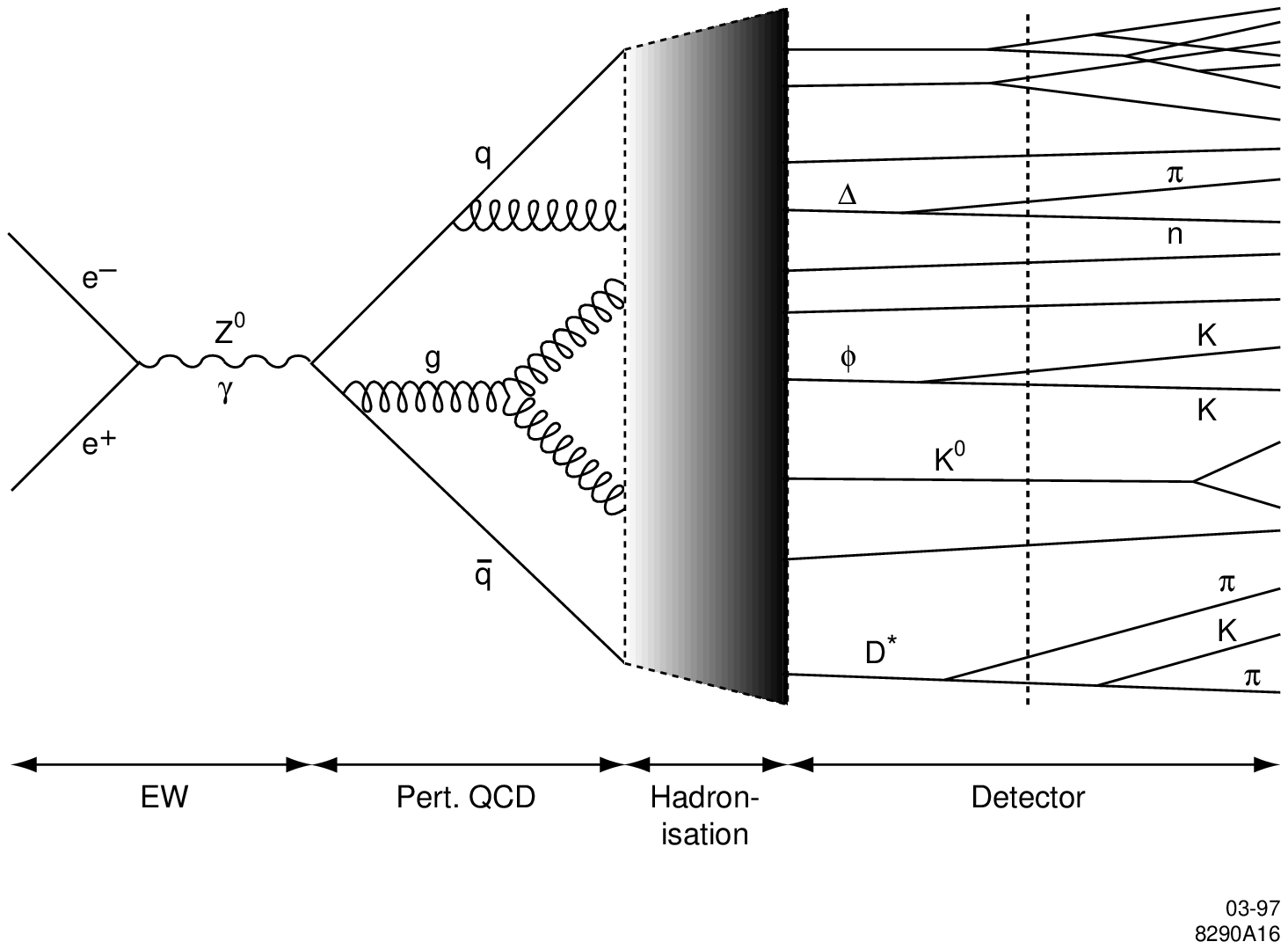}}
\end{center}
   \caption[]{
Schematic of hadron production in \epa.
  }
\end{figure}

%\vskip 1truecm
\vfill
\eject

\noindent{\large \bf 4.4.2 Hadronisation and Monte Carlo Models}

\vskip .5truecm

\noindent
A schematic of hadron production in \epa is shown in Fig.~29. 
One may divide this process into several phases:

\vskip .5truecm

\noindent 
1. A hard electroweak process in which the primary quark and antiquark may be
produced off mass-shell:  

\ep \qu \ra \qq

\vskip .2truecm
 
\noindent 
2. Perturbative QCD evolution of the primary \qq via parton Bremsstrahlung:
 
\qq \qu \ra \qu several q, \qbar, g

\vskip .2truecm
 
\noindent 
3. Hadronisation of partonic system:
 
(q, \qbar, g)s \qu \ra \qu primary resonances

\vskip .2truecm
 
\noindent 
4. Decays of primary resonances into `stable' particles:
 
B, C, \K0, $\phi$, $\Delta$, $\rho$ $\ldots$ \ra $\pi^{\pm}$, K$^{\pm}$,
p,\pbar $\ldots$ \qu (e$^{\pm}$, $\mu^{\pm}$, $\tau^{\pm}$, $\nu$)

\vskip .5truecm

\noindent 
Phases 1 and 2 are generally agreed to be calculable `respectably' using 
perturbative techniques applied to the electroweak theory and QCD,
respectively. Phases 3 and 4 are more problematic in that they are 
intrinsically non-perturbative processes that cannot in general be calculated
from first principles. 
In the absence of non-perturbative calculations we are forced 
to rely on phenomenological models. 

Since it is also necessary in phase 4 to simulate the interaction of particles 
with detectors, which can only be done in a deterministic fashion, 
Monte Carlo event generators have been developed for the complete simulation
of hadronic event production in \epa and are now essential components of 
data analysis. I shall discuss only the two most widely used generators
JETSET~\cite{jetset} and HERWIG~\cite{herwig}; other generators are described
in~\cite{knowles}, and will be discussed later by Buchanan~\cite{chuck}.
I shall not discuss at all the GEANT program~\cite{geant}, which is widely used
for the simulation of the geometry and material response of particle 
detectors. The philosophy here is to outline the main features of these
generators in the context of their use as tools in understanding and 
correcting the data; no attempt will be made to justify these models 
on phenomenological grounds, and the outline will necessarily be brief.

Both JETSET and HERWIG implement electroweak matrix elements for
the production of a primary \qq, as well as a perturbative QCD 
`parton shower' evolution of
the system into a set of low-virtual-mass quarks and gluons. More formally,
the latter is based on a probabilistic parton branching process that is 
derived from a
leading + partial next-to-leading logarithmic resummation of the QCD matrix 
elements~\cite{mw}. JETSET and HERWIG implement the parton branching process 
slightly
differently, a discussion of which is beyond the scope of these 
lectures, but
both generators have a parameter $\Lambda$ that characterises the
scale of strong interactions, as well as a parameter $Q_0$ that
characterises the minimum virtual-mass scale of the parton evolution.
                      
%% figure 30
\begin{figure} [tbh]
% \hspace*{5cm}
   \epsfxsize=5.0in
   \epsfysize=4.0in
   \begin{center}
    \mbox{\epsffile{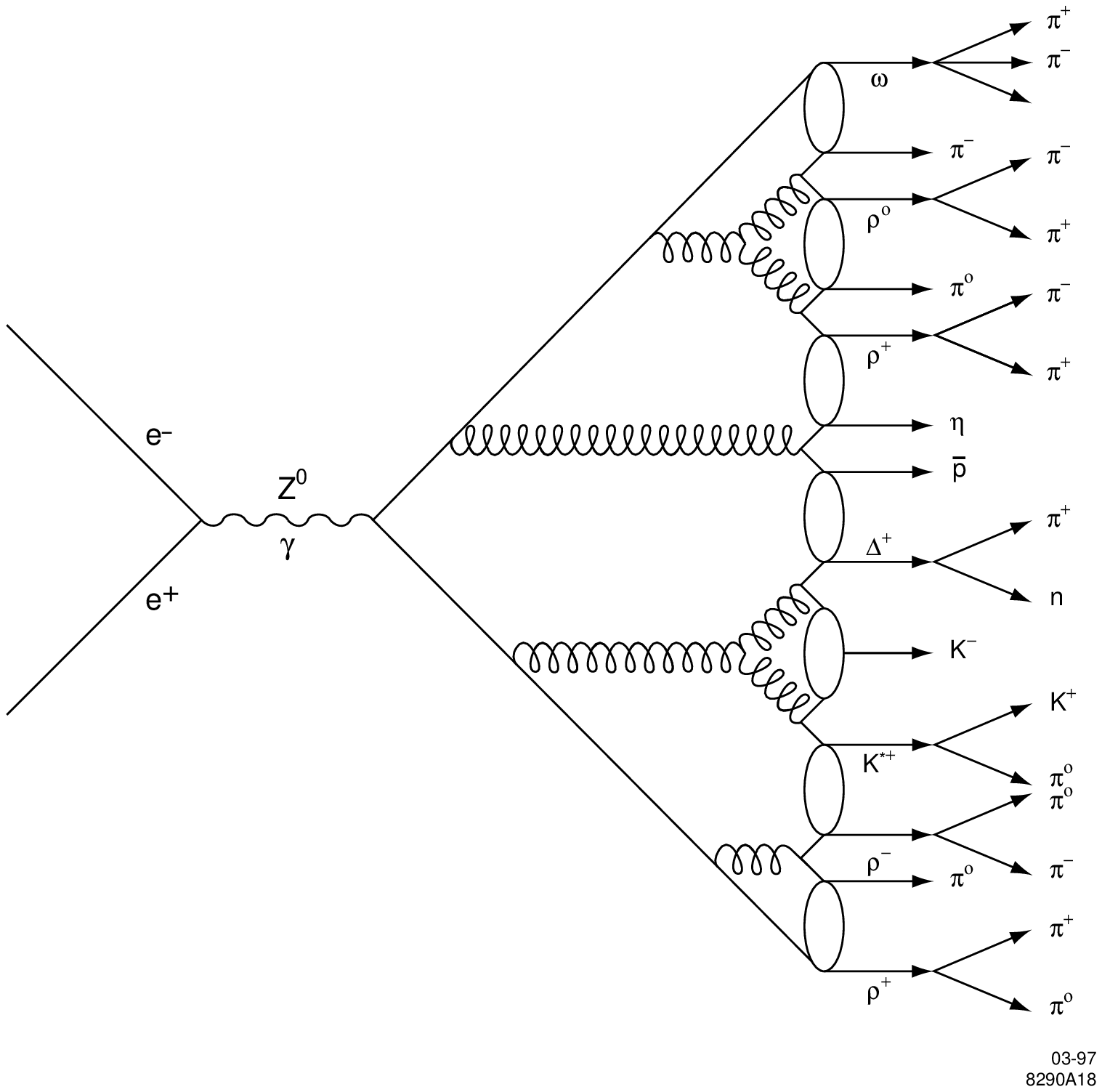}}
\end{center}
   \caption[]{
Schematic of hadronisation in HERWIG.
  }
\end{figure}

A schematic of the hadronisation process as implemented in HERWIG is shown in
Fig.~30. At the termination of the parton shower pairs of partons are 
associateed into colourless clusters; these then undergo phase-space decay to
produce stable pions, kaons and baryons. Clusters with mass larger than 
a parameter $M_{cl}$ are split into two before the phase-space decay. Additional
parameters control the properties of heavy (B or C) hadron decay~\cite{herwig}.
 
JETSET implements the `Lund string model' of jet fragmentation~\cite{Lund}, 
illustrated in
Fig.~31. In this case the colour field between partons at the end of the
parton shower is represented as a one-dimensional massless 
relativistic string. String
pieces terminate at quarks and antiquarks, and gluons are represented by
momentum-carrying `kinks' in the string. The string is fragmented iteratively
according to the recipe:
\begin{eqnarray}
f(z)\quad\propto\quad \frac{1}{z} (1-z)^a {\rm exp}({-b\frac{m_{\perp}^2}{z}}) 
\end{eqnarray} 
where $z$ is the fraction of the quantity $E+p_{\parallel}$ of a parent string
piece taken by the daughter, $m_{\perp}$ = $\sqrt{p_{\perp}^2+m^2}$, `$\perp$'
and $`\parallel$' refer to the string axis, and $a$ and $b$ are parameters.
Momentum transverse to the string axis, $p_{\perp}$, is introduced in an \adhoc
fashion using a Gaussian probability distribution. A large number of additional
parameters is used to fine-tune the relative production of particles such as
strange, pseudoscalar, and vector mesons, as well as strange and non-strange, 
octet and decuplet baryons~\cite{jetset}.
                      
%% figure 31
\begin{figure} [tbh]
% \hspace*{5cm}
   \epsfxsize=3.5in
   \epsfysize=3.0in
   \begin{center}
    \mbox{\epsffile{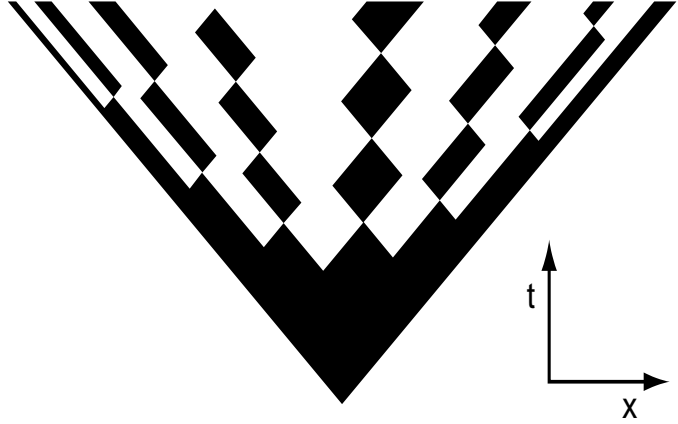}}
\end{center}
   \caption[]{
Schematic of hadronisation in JETSET.
  }
\end{figure}

\vskip 1truecm

\noindent{\large \bf 4.4.3 Data Correction}

\vskip .5truecm

\noindent 
For the \alp analysis we wish to use these event generators to understand the
effect of the hadronisation process on the hadronic momentum flow in 
events, and to correct for any bias, as well as to understand the influence 
of the response of the detector. One conventional approach involves
using a sample of Monte Carlo events to calculate bin-by-bin correction 
factors, and then applying these to the measured distribution.  
For a distribution $D(X)$, the correction for detector effects is defined: 
\begin{eqnarray}
C_{DET}^{MC}(X)\qu=\qu{D_{HAD}^{MC}(X)\over D_{DET}^{MC}(X)},
\end{eqnarray}
where $HAD$ and $DET$ refer to the simulated distribution at the hadron-level
and detector-level phases, respectively.
The correction for hadronisation effects is analogously defined: 
\begin{eqnarray}
C_{HAD}^{MC}(X)\qu=\qu{D_{PART}^{MC}(X)\over D_{HAD}^{MC}(X)},
\end{eqnarray}
where $PART$ refers to the simulated distribution at the parton-level.
The data distribution corrected back to the parton-level is given by: 
\begin{eqnarray}
{D^{Data}}^{\prime}(X)\qu=\qu C_{HAD}^{MC}(X)\cdot C_{DET}^{MC}(X)\cdot 
D^{Data}(X),
\end{eqnarray}
and ${D^{Data}}^{\prime}(X)$ can be compared with perturbative QCD.
More sophisticated correction procedures can also be defined; see 
\eg~\cite{blobel}.

As with any correction procedure one must take care not to introduce bias from
implicit model-dependence, and must estimate the systematic uncertainties
involved. A prerequisite is that the simulation describe the
distribution measured in the detector! The parameters of the detector
simulation, as well as of the event generator itself, should then be varied, 
the stability of the correction factors examined, and systematic errors assigned
accordingly. An example of a raw measured $D_2$ distribution from 
SLD~\cite{sldalp}, compared with
simulations based on JETSET and HERWIG, is shown in Fig.~32. The corresponding
corrected distribution, and the correction factors, are shown in Fig.~33.
                      
%% figure 32
\begin{figure} [tbh]
% \hspace*{5cm}
   \epsfxsize=3.0in
   \epsfysize=4.0in
   \begin{center}
    \mbox{\epsffile{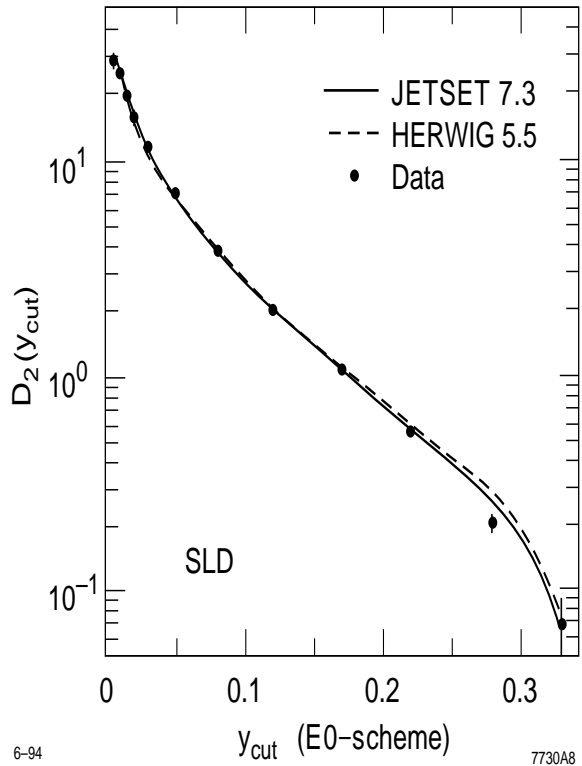}}
\end{center}
   \caption[]{
Measured $D_2$ distribution~\cite{sldalp}
compared with JETSET and HERWIG predictions.
  }
\vskip .5truecm
\end{figure}
                      
%% figure 33
\begin{figure} [tbh]
% \hspace*{5cm}
   \epsfxsize=3.0in
   \epsfysize=4.0in
   \begin{center}
    \mbox{\epsffile{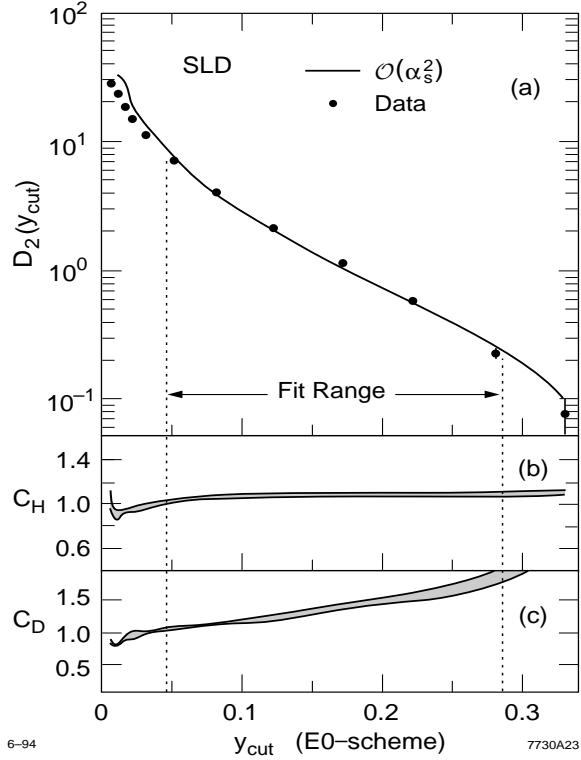}}
\end{center}
   \caption[]{
(a) Corrected $D_2$ distribution~\cite{sldalp}; 
(b) hadronisation correction factor; (c)
detector effects correction factor. In (a) the line shows a fit to \oalpsq
QCD.
  }
\vskip .5truecm
\end{figure}

%\vskip 1truecm
\vfill
\eject

\noindent{\large \bf 4.4.4 Comparison with Perturbative QCD}

\vskip .5truecm

\noindent 
A fit of \oalpsq perturbative QCD to the $D_2$ distribution is shown in
Fig.~33; it yields \alpmzsq = 0.1175 $\pm0.0007$ (stat.) $\pm0.0027$ 
(syst.)~\cite{sldalp}. One can repeat this procedure for
all 15 observables listed in Table~3 and derive in each case a fitted value of
\alpmzsq; these are shown in Fig.~34a. The distressing result of this exercise
is that the \alpmzsq values so determined are not internally consistent with
one another! A measure of the scatter among the results is given by the r.m.s.
deviation of $\pm0.008$, which is much larger than the experimental error of
$\pm0.003$ on a typical observable.
An exciting, though remote, possibility is that we have observed a spectacular
breakdown of QCD! A more likely explanation is that some systematic effect 
that we have not yet considered is at work. In fact an implicit assumption
was made in deriving the results shown in Fig.~34a that relates to the arcane
issue of choosing the {\it renormalisation scale} in QCD.
                      
%% figure 34
\begin{figure} [tbh]
% \hspace*{5cm}
   \epsfxsize=5.0in
   \epsfysize=4.0in
   \begin{center}
    \mbox{\epsffile{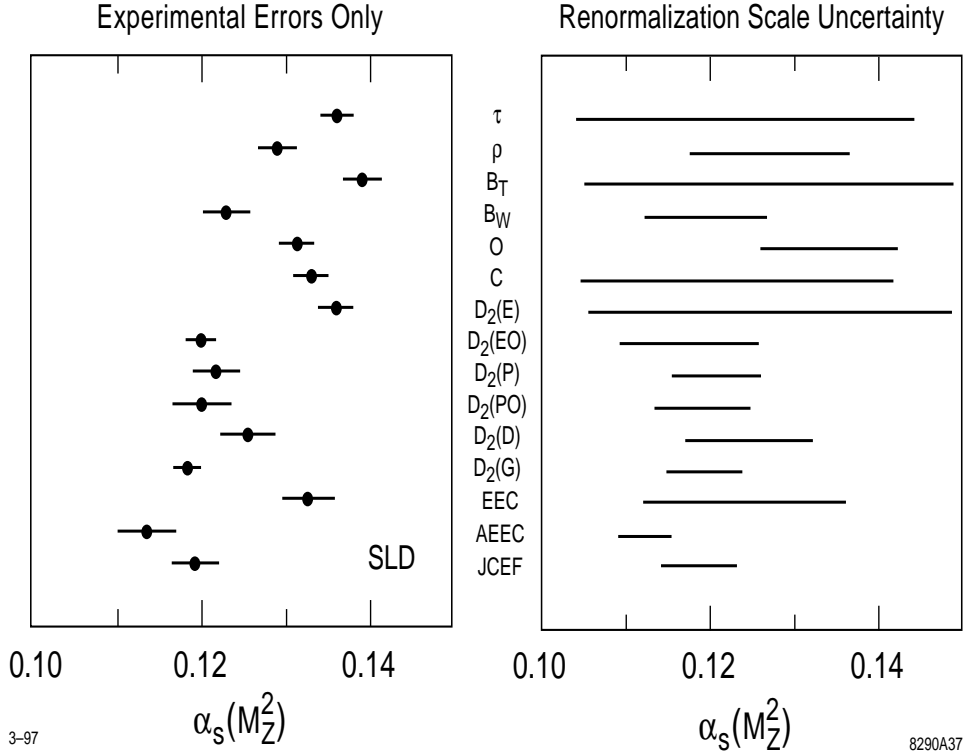}}
\end{center}
   \caption[]{
(a) Values of \alpmzsq determined~\cite{sldalp} 
by fitting \oalpsq QCD predictions to 15 hadronic
event shape observables using a fixed value of the renormalisation scale
$\mu$ = $Q$; the results are clearly inconsistent within the
experimental errors. (b) Renormalisation scale uncertainties.
  }
\vskip .5truecm
\end{figure}

\vskip 1truecm

\noindent{\large \bf 4.4.5 Renormalisation Scale Uncertainty}

\vskip .5truecm

\noindent 
For any observable, truncation of the QCD perturbation series at finite
order causes a
residual dependence on the (scheme-dependent) {\it renormalisation
scale} $\mu$. This parameter is formally unphysical and
should not enter at all into an exact infinite-order calculation,
and its value is arbitrary. For the event shape observables an explicit
$\mu$-dependence enters the next-to-leading coefficient:

\begin{eqnarray}
{1\over\sigma_0} {{\rm d}\sigma\over{\rm d}X}
\;=\; A(X)\;\left({\alp(\mu^2)\over 2\pi}\right)
\;+\; (\;B(X)\quad+\;A(X) 2\pi b_0\;{\rm ln}\mu^2/Q^2\;)\left(
{\alp(\mu^2)\over 2\pi}\right)^2
\end{eqnarray}

\noindent
so that a measurement of \alp must be in the context of some chosen 
value of $\mu$. This is illustrated in Fig.~35, where the value of \lam 
from fits to $D_2$ is
shown as a function of the choice of $\mu$; there is 
clearly a strong $\mu$-dependence. The top portion of Fig.~35 shows the
corresponding $\chi^2_{dof}$ for each fit; amusingly the data show no
preference for any particular value of $\mu$ provided it is larger than 
$\sqrt{0.001Q^2}$. A full discussion of the form of the $\mu$-dependence is
beyond the scope of these lectures; see~\cite{bm}. Figures of the 
$\mu$-dependence for the other observables can be found in~\cite{sldalp}.
                      
%% figure 35
\begin{figure} [tbh]
% \hspace*{5cm}
   \epsfxsize=4.0in
   \epsfysize=3.0in
   \begin{center}
    \mbox{\epsffile{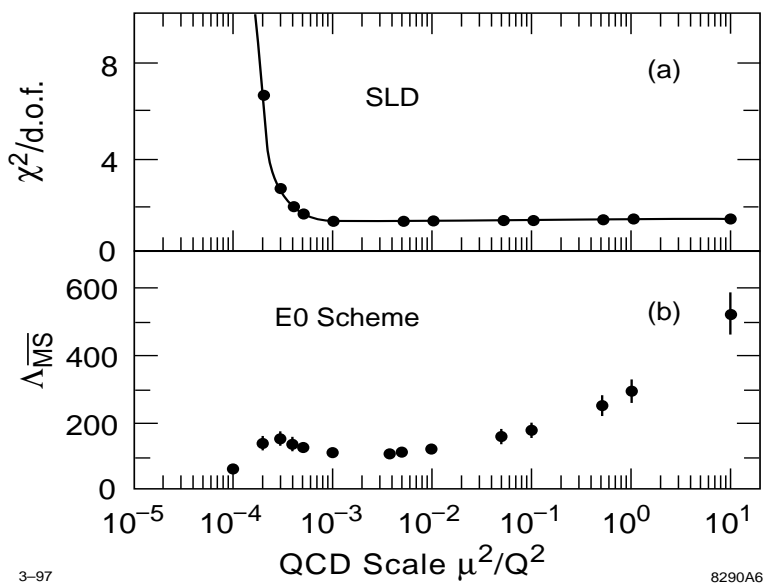}}
\end{center}
   \caption[]{
Dependence of (a) $\chi^2_{dof}$ and (b) \lam on the value of $\mu$ chosen 
in fits to the SLD $D_2$ distribution~\cite{sldjet}.
  }
\vskip .5truecm
\end{figure}

A consensus has arisen among experimentalists 
that the effect of missing higher-order terms can hence be
estimated from the dependence of \alpmzsq on the value of $\mu$
assumed in fits of the calculations to the data, and a
{\it renormalisation scale uncertainty} is often quoted.
This procedure, well-motivated in that the $\mu$-dependence
caused by the truncation of the perturbation series
would be cancelled by addition of the higher-order terms, is, however,
arbitrary, and is not equivalent to knowledge of the size of the
\apriori unknown terms. In cases where scale uncertainties are considered
this arbitrariness is manifested in the
wide variation among the ranges and central values of
$\mu$ chosen by different experimental groups, see \eg \cite{bmmo};
in other cases this source of uncertainty is not included in the errors.
Different \alpmzsq results with similar experimental
precision can hence be quoted with different total errors
depending on the procedure adopted for assigning the
theoretical uncertainties.
The interpretation of the central values and errors on \alpmzsq
measurements is hence not always straightforward. 
The SLD estimate of the renormalisation scale uncertainty for each observable
is shown in Fig.~34b. It is apparent that the scale uncertainty is
much larger than the experimental error, and that the \alpmzsq values are
consistent within these uncertainties.
Though this is comforting, in that it indicates that QCD is self-consistent, 
the necessary addition of large theoretical uncertainties to otherwise precise
experimental measurements is frustrating, at least to experimentalists!
 
The best resolution of the scale ambiguity would be to reduce its
effect by calculating
observables to higher order in perturbation theory. Though this is in principle
possible, the large number of Feynman diagrams involved renders the task
difficult and unattractive. In \epa only the $R$-related observables
and the $\tau$ hadronic decay ratio $R_{\tau}$, have been calculated exactly
up to \oalpc. For the hadronic
event shape observables \oalpc contributions have not yet been
calculated completely. However, for six observables 
(indicated in Table~3) improved calculations can be
formulated that incorporate the resummation \cite{resum}
of leading and next-to-leading
logarithmic terms matched to the \oalpsq results. The matched
calculations are expected {\it a priori} both
to describe the data in a larger
region of phase space than the fixed-order results, and to yield a
reduced dependence of \alp on the renormalization scale. 
This is illustrated in Fig.~36 for the case of thrust ($\tau$). Though not well
described by the \oalpsq calculation, the low-$\tau$
region is well reproduced when resummed contributions are included.

Application of other approaches to circumvent the scale ambiguity in $\alpha_s$
measurement, involving the use of `optimised' perturbation theory'~\cite{opt}
and Pad${\acute{\rm e}}$ Approximants~\cite{pade}, can be found 
in~\cite{bmmo,bs} respectively.
                      
%% figure 36
\begin{figure} [tbh]
% \hspace*{5cm}
   \epsfxsize=3.0in
   \epsfysize=4.0in
   \begin{center}
    \mbox{\epsffile{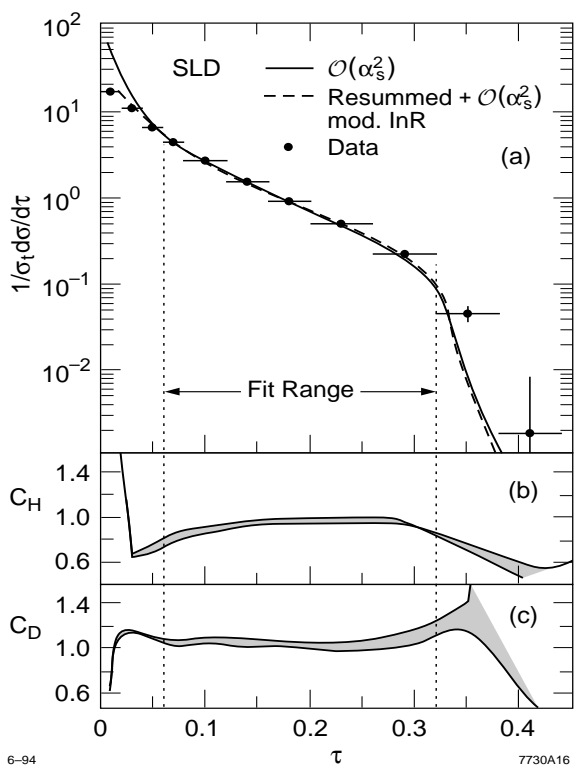}}
\end{center}
   \caption[]{
Illustration of the need for resummed contributions: the $\tau$ distribution
measured by SLD~\cite{sldalp}.  At low $\tau$ the \oalpsq calculation is
unable to describe the data unless resummed terms are considered.
  }
\end{figure}

%\vskip 1truecm
\vfill
\eject

\noindent{\large \bf 4.4.6 Summary of \alp Measurements}

\vskip .5truecm

\noindent 
Hinchliffe has reviewed the various hadronic event shapes-based 
measurements from experiments
performed in the c.m. energy range $10\leq Q\leq 91$ GeV, utilising both
\oalpsq and resummed calculations, and quotes an average value of
\alpmzsq = $0.122\pm0.007$ \cite{ian},
where the large error is dominated by the renormalisation 
scale uncertainty, which far
exceeds the experimental error of about $\pm0.002$.
Schmelling has also compiled the measurements, including the recent
results from the LEP-II run at $Q$ $\sim$ 133 GeV \cite{lephi}, and
quotes a global average \cite{schmelling} \alpmzsq = $0.121 \pm 0.005$,
in agreement with \cite{ian}, but assuming
a more aggressive scale uncertainty. 

\vskip 1truecm
 
\noindent
{\large \bf 4.5 Scaling Violations in Fragmentation Functions}

\vskip .5truecm
 
\noindent
Though distributions of final-state hadrons are not, in general, calculable
in perturbative QCD, the $Q^2$-evolution of the
scaled energy ($x_p = 2E/Q$) distributions of hadrons, or
`fragmentation functions', can be calculated and used to determine \alp.
In addition to the usual renormalisation scale $\mu$,
a {\it factorisation
scale} $\mu_F$ must be defined that delineates the boundary between the
calculable perturbative, and incalculable non-perturbative, domains.
Additional complications
arise from the changing composition of the underlying event
flavour with $Q$ due to the different $Q$-dependence of the $\gamma$
and \z0 exchange processes. Since B and C hadrons typically
carry a large fraction of the beam momentum, and contribute a large
multiplicity from their decays, it is necessary to consider the scaling
violations separately in b, c, and light quark events, as well as in
gluon jet fragmentation.
 
In an early analysis \cite{delfrag} the DELPHI Collaboration
parametrised the fragmentation functions using the \oalpsq matrix
elements and the string fragmentation model implemented in JETSET
\cite{jetset}. They fitted data in the range $14\leq Q\leq 91$ GeV
to determine \alpmzsq = $0.118\pm0.005$, where the error is
dominated by varying $\mu$ in the range $0.1\leq\mu/Q\leq 1$.
The ALEPH Collaboration used its \z0 data to constrain
flavour-dependent effects by tagging event
samples enriched in light, c, and b quarks, as well as a sample of
gluon jets \cite{alfrag}.
The fragmentation functions for the different flavours and the gluon
were parametrised at a reference energy, evolved with $Q$ according to
the perturbative DGLAP formalism calculated at next-to-leading order
\cite{dglap}, in conjunction with a parametrisation
proportional to $1/Q$ to represent non-perturbative effects (Section 5), 
and fitted to
data in the range $22\leq Q\leq 91$ GeV (Fig.~37). They derived
\alpmzsq = $0.126\pm0.007$ (exp.) $\pm0.006$ (theor.), where the
theoretical uncertainty is dominated by variation of the
factorisation scale $\mu_F$
in the range $-1\leq {\rm ln}\mu_F^2/Q^2\leq 1$;
variation of the renormalisation scale in the same range contributed
only $\pm0.002$.
DELPHI has recently reported a similar analysis \cite{newdel} yielding
\alpmzsq = $0.121^{+0.006}_{-0.007}$ (exp.) $\pm0.010$ (theor.).
Curiously, although a similar range as ALEPH,
$0.3\leq \mu/Q\leq 3$, was used to examine variation of the
renormalisation and factorisation scales, here the
renormalisation scale dominates the
theoretical uncertainty, with a contribution of $\pm0.009$, in
contrast to $\pm0.002$ from factorisation.
Combining the ALEPH and later DELPHI results, assuming uncorrelated
experimental errors, yields~\cite{cracow}:
\begin{eqnarray}
\alpmzsq \qu = \qu 0.124 \pm 0.005 ({\rm exp.})\pm0.010({\rm theor.})
\end{eqnarray}
                      
%% figure 37
\begin{figure} [tbh]
% \hspace*{5cm}
   \epsfxsize=4.0in
   \epsfysize=5.0in
   \begin{center}
    \mbox{\epsffile{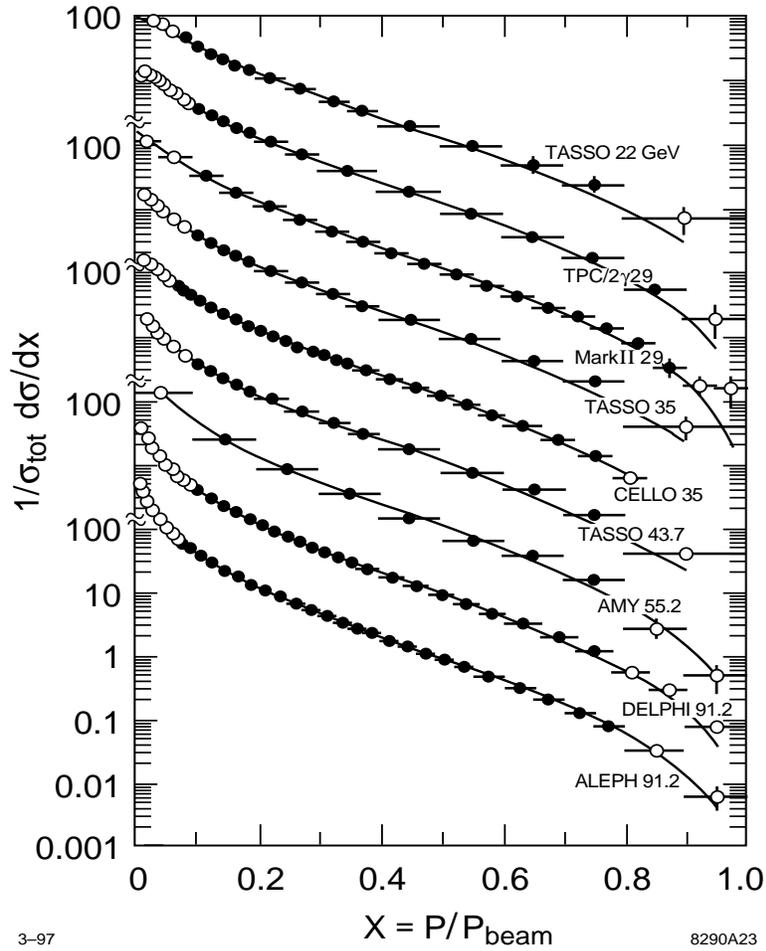}}
\end{center}
   \caption[]{
Illustration of scaling violations in \ep fragmentation functions.
  }
\end{figure}
 
\vskip 1truecm
  
\noindent
{\large \bf 4.6 Comparison with Other Measurements of \alpmzsq}
 
\vskip .5truecm

\begin{figure} [tbh]
   \epsfxsize=3.0in
   \epsfysize=5.0in
%   \begin{center}
\includegraphics{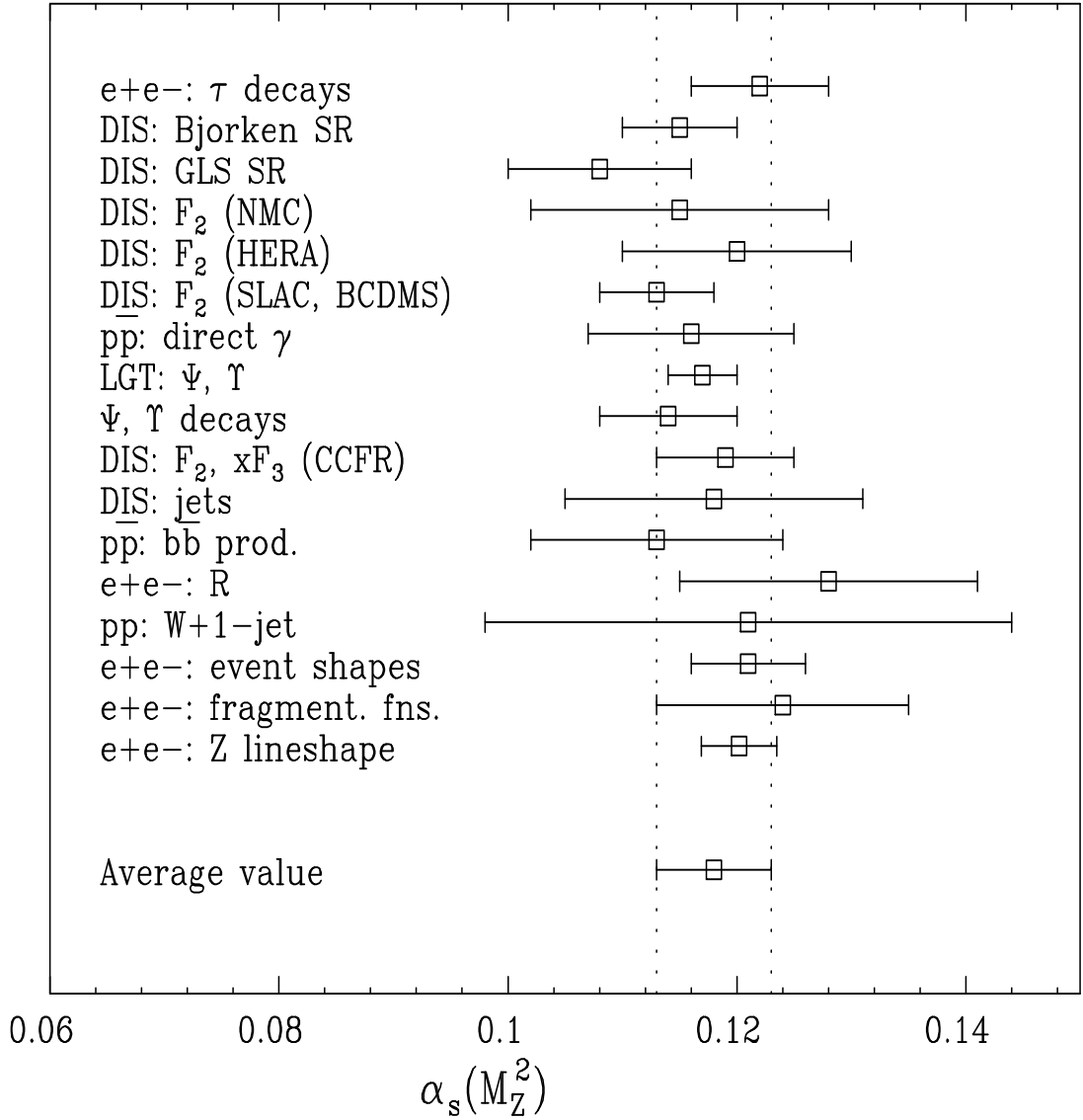}
%   \end{center}
\vskip 20truecm
   \caption[]{
Summary of world \alpmzsq measurements~\cite{cracow}.
The results are ordered vertically in terms of the hard scale $Q$ of the 
experiment.
  }
\label{figalpsum}
\end{figure}

\noindent
A summary of world \alp measurements, all evolved to $Q$ = $M_Z$, is shown in
Fig.~\ref{figalpsum}~\cite{cracow}. These are drawn from lepton-hadron
scattering, hadron-hadron collisions, heavy quarkonia decays and lattice
gauge theory, as well as \epa. In addition to being relatively precise, 
the \ep results have the invaluable feature
that they bracket the $Q$-range of the experiments, from around 1 GeV for
$\tau$ decays to around 100 GeV for \z0 production, providing the largest
lever-arm for tests of consistency of \alpmzsq measured at different energy
scales. It is clear that, within the uncertainties, all results are consistent
with one another. 

\clearpage
                      
%% figure 38
%\begin{figure} [tbh]
%   \epsfxsize=4.0in
%   \epsfysize=5.0in
%   \begin{center}
%    \mbox{\special
%{epsfile{$sld_axp_fac:[000000.sldqcd.ps]allalpp.ps[angle=90]}}}
%\end{center}
%   \caption[]{
%Summary of world \alpmzsq measurements~\cite{cracow}.
%The results are ordered vertically in terms of the hard scale $Q$ of the 
%experiment.
%  }
%\label{figalpsum}
%\end{figure}
                      
%% figure 38

\clearpage
 
Taking an average over all 17 measurements
{\it assuming} they are independent, by weighting each by its {\it total}
error, yields \alpmzsq = 0.118 with a $\chi^2$ of 6.4; the low $\chi^2$
value reflects the fact that most of the
measurements are theoretical-systematics-limited. Taking an unweighted average,
which in some sense corresponds to the assumption that all 17
measurements are completely correlated, yields the same result. 
The r.m.s. deviation of the 17
measurements w.r.t. the average value characterises the
dispersion, and is $\pm0.005$. In a quantitative sense, therefore, QCD has
been tested to a level of about 5\%.
 
If further progress is to be made in testing QCD, future
measurements of \alpmzsq should aim for substantially improved
precision. The prospects for
achieving 1\%-level measurements are discussed in detail 
elsewhere \cite{snowmass}. Lattice QCD determinations may reach this precision
within the next few years. A precise \alpmzsq
measurement has yet to emerge from the TeVatron, but feasibility
studies are in progress and appear promising.
Deep-inelastic scattering and \epa
will probably require higher-energy facilities, as well as significant
theoretical effort with regard to \oalpc perturbative contributions. An
{\alpmzsq} measurement at a high-energy \ep collider will be discussed in 
Section 6.2.

\vskip 1truecm
  
\noindent
{\large \bf 5. Towards a Theory of Hadronisation}
 
\vskip .5truecm
 
\noindent
We expect that
the strong coupling becomes large in long-distance (low-$Q$) q$-$q 
interactions such that finite-order perturbation theory is no longer valid.
Lattice gauge theory~\cite{degrand} is the only practical 
non-perturbative calculational tool available today. It is presently limited 
in applicability to static properties of hadrons, such as masses and decay 
constants, although in principle it might eventually
be applied to the dynamical process of hadronisation.

\begin{figure} [tbh]
% \hspace*{5cm}
   \epsfxsize=4.0in
   \epsfysize=4.0in
   \begin{center}
    \mbox{\epsffile{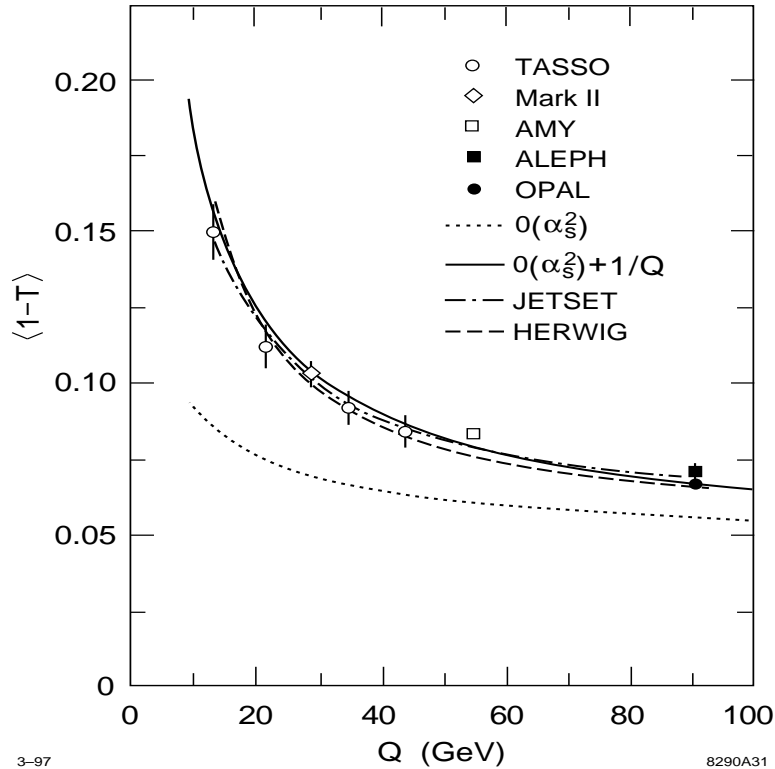}}
\end{center}
   \caption[]{
Energy-dependence of the mean value of $1-T$. The \adhoc addition of a $1/Q$
term to the \oalpsq QCD prediction describes the data~\cite{bryan}. 
  }
\label{fighadt}
\end{figure}

From the 
operator product expansion (OPE) one expects (see \eg~\cite{beneke}) that the 
expectation value of an observable $O$ may be written: 
\begin{eqnarray}
<O> \qu = \qu \Sigma{\;a_i
\left({\alp\over\pi}\right)^i}\qu+\qu\Sigma{\;b_j\over Q^j}
\end{eqnarray}
During the past 15 years 
much theoretical effort has been focussed on the perturbative component
represented by the first term in this equation.
More recently attention has turned to the `power corrections' represented by
the second term, whose origin is intrinsically non-perturbative.
In particular, attempts have been made to evaluate power corrections for 
\ep observables. An illustration of the potential of such an
approach is provided by Fig.~\ref{fighadt}~\cite{bryan}. 
The \adhoc addition of a $1/Q$
term to the \oalpsq QCD prediction describes the energy-dependence of 
$<1-T>$ remarkably well. It will be seen in Section 6.2 that the inverse
power-law behaviour of hadronisation effects has important consequences for a
precise \alp measurement at a high energy \ep collider. The explicit 
calculation of leading power corrections~\cite{beneke} 
hence represents our first tentative step
towards a consistent theoretical treatment of hadronisation.
                      
%% figure 39

\vskip 1 truecm

\noindent{\large \bf 6. QCD at a High Energy \ep Collider}
 
\vskip .5truecm

\noindent{\large \bf 6.1 Introduction}

\vskip .5truecm

\noindent
Since QCD {\it is}
our theory of strong interactions it would be
irresponsible not to test it at the highest energy scales available in
different hard scattering processes. For this reason testing QCD at a
0.5--1.5 TeV \ep collider (`XLC') is mandatory. For a detailed discussion 
see~\cite{morioka}.

Precise determination of the strong 
coupling \alp is key to a better understanding of high
energy physics. The current precision of \alpmzsq measurements, limited to
about 5\% (Section 4.6), results in the dominant uncertainty on our 
prediction of the energy scale at
which grand unification of the strong, weak and electromagnetic forces
takes place. An \alpmzsq measurement of 1\% precision may be possible at a
high energy \ep collider. Such a measurement would also allow improved
determination of the mass and width of the top quark from the
threshold behaviour of the \tt cross-section.
Measurements of hadronic event properties at high energies, combined
with existing lower energy data, would allow one to test further the gauge
structure of QCD by searching for anomalous `running' of observables,
such as the rate of production of events containing three jets,
and to set limits on models which predict
such effects, for example those involving light gluinos which are
difficult to exclude by other means.

Gluon radiation in \tt events is expected to be strongly regulated by
the large mass and width of the top quark; \ttg events will hence
provide an exciting new domain for QCD studies. As a corollary,
measurements of
gluon radiation patterns in \ttg events may provide valuable
additional constraints on the top quark decay width. Furthermore,
searches could be made for anomalous chromo-electric and chromo-magnetic 
moments of quarks~\cite{rizzochrom}, 
which effectively modify the rate and pattern of gluon
radiation, and for which the phase space increases as the c.m. energy is
raised. Finally, polarised electron beams will be exploited at high energy
\ep colliders and will allow tests of symmetries using multi-jet
final states~\cite{todd}.

\vskip 1 truecm

\noindent{\large \bf 6.2 Is a 1\%--level Measurement of \alpmzsq Possible?}
 
\vskip .5truecm
 
\noindent
It is interesting to consider whether a measurement of \alpmzsq at the
1\%--level of precision is possible at the XLC. Consider the SLD \alpmzsq
measurement, discussed in Section 4.4, 
based on 15 hadronic event shape observables measured with
a data sample comprising approximately 50,000 hadronic events~\cite{sldalp}:
\begin{eqnarray}
\alpmzsq\quad=\quad 0.1200\pm 0.0025 \;{\rm (exp.)} \pm 0.0078\;
{\rm (theor.)}
\end{eqnarray}
where the experimental error is composed of statistical and systematic
components of about $\pm 0.001$ and $\pm 0.002$ respectively, and the
theoretical uncertainty has components of $\pm 0.003$ and $\pm 0.007$
arising from hadronisation and missing higher order terms, respectively.
Now consider `scaling' this result to estimate the precision of a
similar measurement at $Q$ = 500 GeV.
 
\vskip .5truecm
 
\noindent
$\bullet$
{\bf Statistical error}: At design luminosity the 500 GeV
XLC would deliver roughly 100,000 \qq$\;$ (q=u,d,s,c,b) events per year
(Section 6.4), implying that a statistical error on \alpmzsq well
below $\pm$ 0.001 could be obtained.
 
\vskip .5truecm
 
\noindent
$\bullet$
{\bf Systematic error}: This results primarily from the uncertainty in
modelling the jet resolution of the detector. The situation
may be improved at the XLC by a combination of building
better detectors and benefitting from improved calorimeter energy
resolution for higher energy jets.
It is not unreasonable to suppose that the current systematic
error of roughly $\pm0.002$ could be reduced by a factor of two.
 
\vskip .5truecm
 
\begin{figure} [tbh] 
% \hspace*{5cm}
   \epsfxsize=4.0in
   \epsfysize=4.0in
   \begin{center}
    \mbox{\epsffile{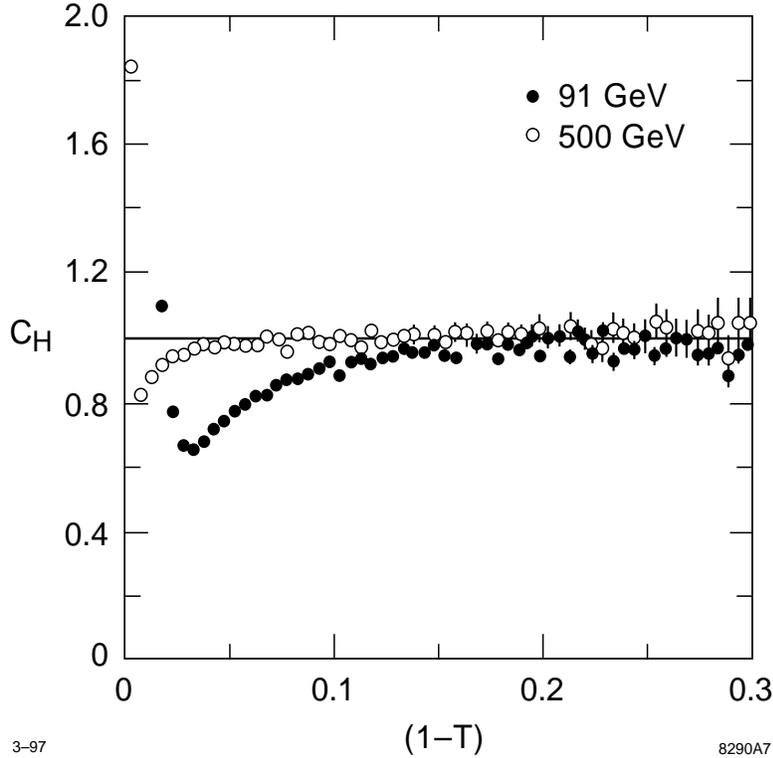}}
\end{center}
   \caption[]{
Estimate of the hadronisation correction factor (Section 4.4.2) for thrust at 
$Q$ = 91 GeV and 500 GeV~\cite{bethke}. At 500 GeV the factor barely deviates
from unity for most of the kinematic range.
  }
\label{figxlct}
\end{figure}

\noindent
$\bullet$
{\bf Hadronisation uncertainty}: 
From the discussion in Section 5 it can be seen that non-perturbative 
corrections to
jet final states in \epa can be parametrised in terms of inverse powers
of the hard scale $Q$. At leading order, perturbative evolution is
proportional to ${1/{\rm ln} Q}$. Hence for a generic observable $X$
the ratio of non-perturbative to perturbative QCD contributions
is dominated by a term of the form:
\begin{eqnarray}
{\delta X^{\rm non-pert} \over
X^{\rm pert}}\quad\sim\quad {{\rm ln}Q \over Q}.
\end{eqnarray}
Increasing $Q$ from 91 GeV to 500 GeV causes this ratio to
decrease by a factor of 5, implying that hadronisation
corrections in the `3-jet region' of observables 
should be of order 2\% at XLC. The conclusion of this analysis is 
reinforced by explicit simulation of hadronisation effects, illustrated in
Fig.~\ref{figxlct}~\cite{bethke} for thrust.  Assuming that these
corrections can be estimated to better than $\pm50$\%, the
hadronisation {\it uncertainty} should contribute less than 1\% to the error
on \alpmzsq.

%% figure 40
 
\vskip .5truecm
 
\noindent
$\bullet$
{\bf Uncertainty due to missing higher orders}:
Currently perturbative QCD calculations of hadronic event shapes are
available complete up to \oalpsq. Since the data contain knowledge of all
orders one must estimate the possible bias inherent in measuring
\alpmzsq using the truncated QCD series (Section 4.4.5).
Since the missing perturbative terms are \oalpc, and since at $Q$ =
500 GeV \alp is expected to be about 25\% smaller than its value at
the \z0, one naively expects the uncalculated terms to be
almost a factor of two smaller at the higher energy, leading to an
estimated uncertainty of $\pm0.004$ on \alp(500 GeV). However,
translating to the yardstick \alpmzsq yields an uncertainty of
$\pm0.006$, only slightly reduced compared with the current uncertainty.
 
\vskip .5truecm
 
From this simple analysis it seems reasonable to conclude that
achievement of the
luminosity necessary for `discovery potential' at the XLC will result
in a \qq event sample of sufficient size to measure \alpmzsq with a
statistical uncertainty of better than 1\%. Construction of detectors
superior in performance to those in operation today at SLC and LEP
may be necessary in order to reduce systematic errors to the 1\% level.
Hadronisation effects should be significantly smaller, implying
a sub--1\% uncertainty. However, unless \oalpc contributions are
calculated, \alpmzsq measurements at 500 GeV will be limited by
theoretical uncertainties to a
precision of $\pm 0.006$, only marginally better than that achieved at
present.

\vskip 1 truecm

\noindent{\large \bf 6.3 Top Quark Mass Determination and \alp}
          
\vskip .5truecm
 
\noindent
It is clear that the value of \alp controls the shape
of the strong potential that binds quarkonia resonances. In the case of
\tt production near threshold, the large top mass $m_t$, and hence large
decay width $\Gamma$, ensure that the top quarks decay in a time comparable with
the classical period of rotation of the bound system, making the
toponium resonance a very short-lived phenomenon, and washing out most
of the resonant structure in the cross-section. The shape of the
\tt cross-section near threshold
hence depends strongly not only on the top mass, but also on \alp.

%% figure 41
\begin{figure} [tbh]
% \hspace*{5cm}
   \epsfxsize=4.0in
   \epsfysize=4.0in
   \begin{center}
    \mbox{\epsffile{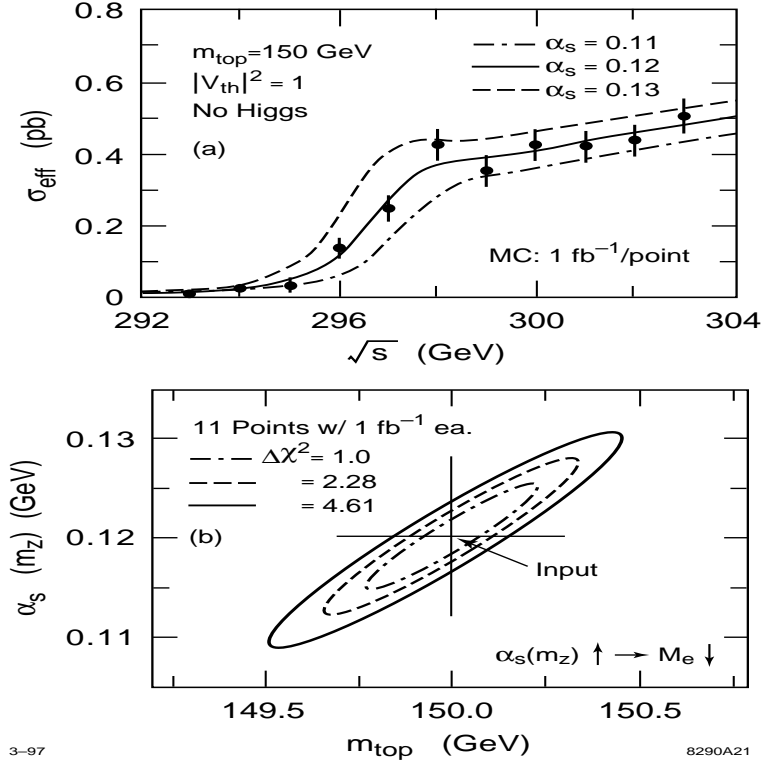}}
\end{center}
   \caption[]{
Simulation of a simultaneous measurement of \alpmzsq and $m_t$ at a high-energy
\ep collider~\cite{top}: (a) \tt production cross section; (b) 
correlation between fitted values.
  }
\label{figtopjlc}
\end{figure}
 
Fits to simulations of measurements of this cross-section have
shown~\cite{top} that the top mass so determined is strongly correlated with
the assumed value of \alpmzsq. This is illustrated in Fig.~\ref{figtopjlc}.
The European Top Quark Working Group has
updated these simulations for the latest measured values of the top
mass and has shown~\cite{eurotop} that a simultaneous determination
of $m_t$ and \alpmzsq by fitting to the threshold
cross-section measured with one design-year of luminosity yields
statistical precisions of $\pm$250 MeV/$c^2$ and $\pm0.006$ on $m_t$ and
\alpmzsq, respectively. Fixing \alpmzsq to 0.120 reduces the error on
$m_t$ by a factor of 2. Since this technique would yield a
measurement of \alpmzsq no more precise
than those made today, and since systematic uncertainties may be
large and have not yet been considered, a more sensible strategy would be
to measure \alpmzsq as precisely as possible, as described in the
previous section, and to use this value to allow better
determination of the top quark parameters.

\vskip 1 truecm

\noindent{\large \bf 6.4 Energy Evolution Studies}
          
\vskip .5truecm
 
\noindent
The non-Abelian gauge structure of QCD implies that as the hard
scattering scale $Q$ increases, the strong coupling decreases
roughly as 1/ln$Q$. Existing hadronic final states data from \epa
at the PETRA, PEP, TRISTAN, SLC and LEP colliders span the range
$14\leq Q\leq 170$ GeV, although hadronisation uncertainties are large
on the data below 25 GeV. A 1.5 TeV \ep collider would increase the
lever-arm in 1/ln$Q$ by almost a factor of two, hence allowing detailed
study of the energy evolution of QCD observables that are proportional
to \alp, such as the rate of
production of final states containing three hadronic jets, $R_3$.
This would provide not only a test of the fundamental structure of
SU(3) QCD, but also a search-ground
for new physics that might produce `anomalous' running.
 
\begin{figure} [tbh]
% \hspace*{5cm}
   \epsfxsize=4.0in
   \epsfysize=4.0in
   \begin{center}
    \mbox{\epsffile{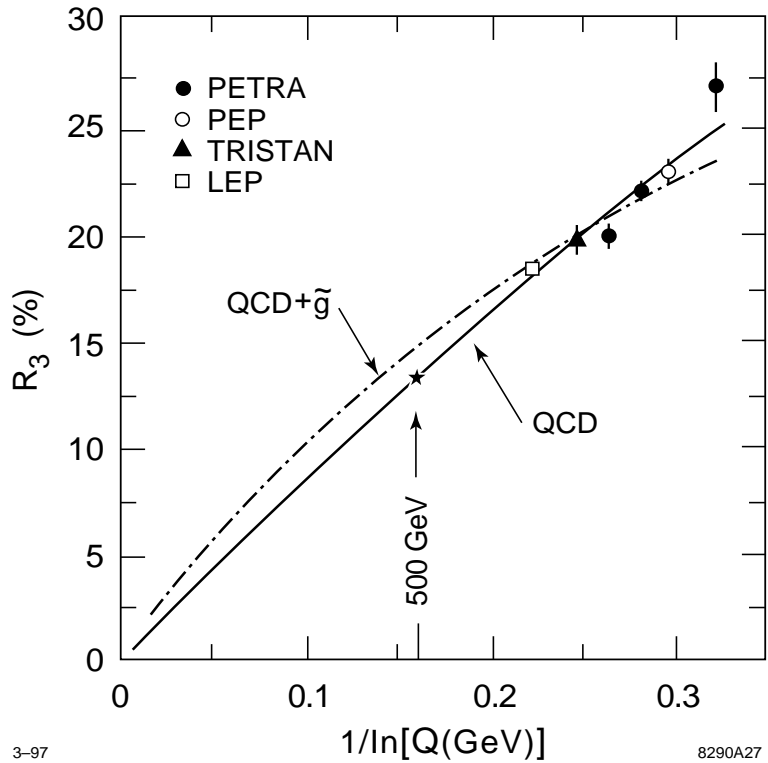}}
\end{center}
   \caption[]{
Energy-evolution of the 3-jet rate $R_3$~\cite{bethke}. For illustration an
\oalpsq QCD fit, as well as a fit allowing the possibility of one family of
light gluinos, is shown. The simulated data point at $Q$ = 500 GeV would add
considerable lever-arm.
  }
\label{figrunning}
\end{figure}

One such possibility is the existence of a light, electrically neutral,
coloured fermion that couples to gluons, often called a `light gluino'
and denoted by \gluino. The existence of such a particle would
manifest itself via a modification of gluon vacuum polarisation
contributions involving fermion loops, effectively increasing the
number of light fermions entering into the QCD $\beta$-function. At
one-loop level the
effective number of flavours would change from $N_F$ to $N_F + 3
N_{\gluino}$, where $N_{\gluino}$ is the number of families of light
gluinos, causing a decrease in the running of \alp as a function
of $Q$. The existence of a light gluino of mass between 2 and 5
GeV/$c^2$ has not been excluded by searches with current data~\cite{bethke}.
A simulated measurement of $R_3$ at $Q$ = 500 GeV,
corresponding to one design-luminosity-year, is shown in
Fig.~\ref{figrunning}~\cite{bethke}, together with existing measurements, 
plotted as a
function of 1/ln$Q$. The presence of one family of light gluinos of mass
2 GeV/$c^2$ would cause an
increase in the predicted value of $R_3$ at 500 GeV by 10\%. A
1\%-level measurement of $R_3$, as discussed in the previous section,
would allow this difference to be measured with a significance of
many standard deviations.

%% figure 42
  
It should be noted, however, that data from a number of experiments at
different \ep colliders contribute to Fig.~\ref{figrunning}. 
Some of these data were
recorded more than 10 years ago, were treated differently by the
various experimental groups, and have relatively large systematic
errors that are at least partly uncorrelated from point to point.
Furthermore,
the sophistication and performance of particle detectors constructed
in the last decade has improved significantly, and it is reasonable to
assume that future detectors will be even better. In addition,
our understanding of the modelling of hadronisation effects and
theoretical uncertainties has improved enormously as a result of
studies at the \z0. Therefore, the precision of searches for anomalous
running of QCD observables at XLC would be improved significantly if new
data were taken at the lower c.m. energies with the {\it same}
detector and analysis procedures.
 
In fact, if the luminosity of the 500 GeV
XLC could be preserved at lower c.m. energies, very large data samples
would be recorded. Table~\ref{tabxlcev}~\cite{bethke}
shows the number of \qq events delivered per
day at various c.m. energies by the XLC operating at the design
luminosity of $5\times10^{33}$ cm$^{-2}s^{-1}$. At each energy more
luminosity would be delivered {\it per day} than was recorded in total
by the original dedicated colliders! This argument is of course naive,
in that a collider designed to operate at a luminosity of
$5\times10^{33}$ cm$^{-2}s^{-1}$ at 500 GeV would not automatically be
operable at the same luminosity at energies a factor of 5 or 10 lower;
such capability would have to be designed from the outset. Furthermore,
the requirements on the triggering and data processing capabilities of
the detector are extreme by the standards of \epa, and this would also
have to be designed from the start. Nevertheless, the prospect of
running the XLC at the \z0 resonance, or at even lower energies,
for QCD studies, not
to mention high-statistics electroweak physics measurements, is
very attractive.

\begin{table}[t]
\centering
\begin{tabular}{|c|c|} \hline
c.m. energy $Q$ (GeV) & \qq events/day              \\ \hline
 500 \hfill & $\quad$ 1750   \hfill     \\
 91  \hfill & $\quad$ 20,000,000 \hfill \\
 60  \hfill & $\quad$ 75,000  \hfill    \\
 35  \hfill & $\quad$ 150,000 \hfill    \\ \hline
\end{tabular}
\caption{Number of \qq events per day delivered by an
\ep collider operating at a luminosity of $5\times10^{33}$ cm$^{-2}s^{-1}$.}
\label{tabxlcev}
\end{table}

\vskip 1 truecm

\noindent{\large \bf 6.5 Gluon Radiation in \tt Events}
        
\vskip .5truecm
 
\noindent
The large mass and decay width of the top quark serve to make the
study of gluon radiation in \tt events a new arena for testing QCD.
The large mass acts as a cutoff for collinear gluon radiation,
and the large decay width acts as a cutoff for soft gluon radiation,
allowing reliable perturbative QCD calculations to be performed;
these effects are of course correlated. The
latter case is particularly interesting. If the top width were
infinite, top quarks would decay immediately to bottom quarks, and any
gluons would be radiated from the secondary b's.
If the top width were zero, top quarks would live forever and all
radiation would be from the primary t's. In the case of a large but
finite width, expected to be around 2 GeV for a top mass of 180
GeV/$c^2$, gluon radiation in \tt events will be a coherent sum of
contributions from these two limiting cases, with a degree of
coherence regulated by the top width itself.
 
%% figure 43
\begin{figure} [tbh]
% \hspace*{5cm}
   \epsfxsize=5.0in
   \epsfysize=4.0in
   \begin{center}
    \mbox{\epsffile{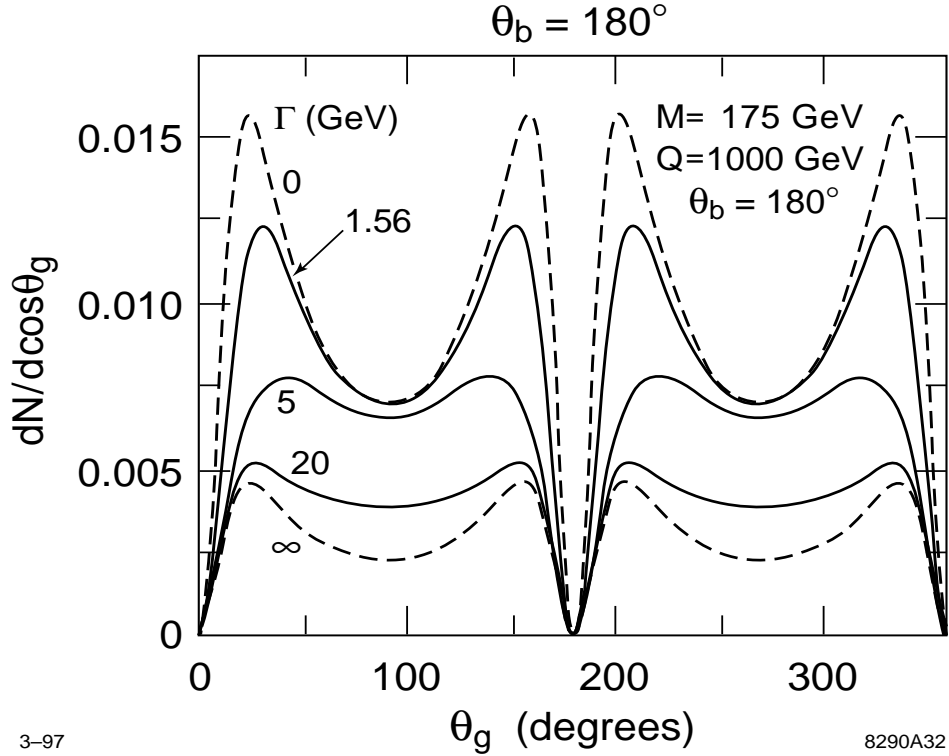}}
\end{center}
   \caption[]{
Angular distribution of 5 GeV gluons w.r.t. \tt axis, at $Q$ = 1 TeV, 
illustrating the dependence on the top width $\Gamma$~\cite{lynnenew}.
  }
\label{figlynne}
\end{figure}
 
A theoretical study of \tt production above threshold,
assuming $m_t$ = 175 GeV/$c^2$ at $Q$ = 1 TeV, is
illustrated in Fig.~\ref{figlynne}~\cite{lynnenew}. This shows the angular
distribution of 5 GeV gluons w.r.t. the \tt axis for the kinematic
configuration in which the decay b-quark travels backwards w.r.t. the t
flight direction. The dependence of the radiation pattern on the top
decay width is strong. Similar effects are predicted in the spectrum
of gluon radiation in \tt events around threshold~\cite{lynneold}.
Measurement of such effects would yield not only a dramatic
demonstration of quantum interference in strong interactions,
but might also provide an essential cross-check on the value of the
top quark decay width, which may prove difficult to disentangle from
measurements of the \tt threshold cross-section and top momentum
distributions, which also depend on \alp and $m_t$ (section 6.3), as
well as on the beam energy distribution. 
   
\vskip 1 truecm

\noindent{\large \bf 7. Concluding Remarks}
 
\vskip .5truecm

\noindent 
We have seen that \ep annihilation is an ideal laboratory for
precise studies of QCD. One observes jets indicating the primary production 
of quarks and gluons, and one can measure precisely the quark and gluon spins.
Multijet events allow the very gauge structure of QCD to be tested via 
measurement of the Casimir factors $N_C$, $T_F$, and $C_F$, leading us to the
conclusion that QCD {\it is} the theory of strong interactions.
One can then measure the single parameter of QCD, the coupling \alp, 
from inclusive observables such as $R$, or equivalently the \z0 lineshape
parameters, and from hadronic $\tau$ decays, as well as from 
event shape measures and scaling violations in inclusive single-particle 
fragmentation functions. These \alpmzsq measurements are internally consistent,
and agree with results from lepton-nucleon scattering, hadron-hadron
collisions, and lattice gauge theory determined across a wide range of energy
scales.
 
There was no time to cover many interesting topics, including: 
differences between quark and gluon jets, tests
of the flavour-independence of strong interactions, polarisation phenomena,
particle multiplicities and correlations,
production of B and C mesons and baryons, and
production of identified hadrons such as 
$\pi^{\pm}$, \K0, K$^{\pm}$, p/\pbar, $\Lambda$, $\phi$, K$^*$ \etc
Some of these topics are discussed in other contributions to these 
proceedings~\cite{chuck,willocq}.

Looking towards the future,
tests of QCD will provide an important component of the physics programme at a
future high energy \ep collider operating in the c.m. energy range 
$0.5\leq Q \leq 1.5$ TeV. Measurement of \alpmzsq at the 1\% level of
precision appears feasible experimentally, but will require
considerable theoretical effort to calculate \oalpc contributions in
QCD perturbation theory. A search for anomalous running of \alp($Q^2$),
by operating the collider at different c.m. energies, is an attractive
prospect. Quantum coherence is expected to give rise to interesting gluon
radiation patterns in \tt events, which could be used to constrain the
top quark decay width, and measurement of the gluon radiation spectrum
would also constrain anomalous top quark chromomagnetic couplings.

More immediately, 
the next generation of {\it low} energy \ep colliders, known as B factories, 
also has the potential to make a precise \alp measurement from the
$R$-ratio at $Q$ $\approx$ 10 GeV, as well as from hadronic
$\tau$ decays. Even more precise tests of QCD in \epa will hence continue to 
enhance our confidence in the theory, and may even yield surprises$\ldots$
   
%\vskip 1 truecm
\vfill
\eject

\noindent{\large \bf Acknowledgements}
 
\vskip .5truecm

\noindent 
I am grateful for the support of my colleagues in the SLD Collaboration and
the SLAC Theory Group. I thank Lance Dixon and David Muller for careful
reading of this manuscript.

\vfill\eject

\end{document}